\documentclass[aps,prd,reprint,nofootinbib,superscriptaddress,longbibliography,floatfix]{revtex4-2}

\usepackage{amsmath,amssymb,mathtools,bm,bbm}
\usepackage{graphicx}
\usepackage{booktabs}
\usepackage{array}
\usepackage{tabularx}
\usepackage[hidelinks]{hyperref}
\usepackage[capitalize]{cleveref}
\usepackage{xcolor}
\usepackage{placeins}
\usepackage{multirow}
\usepackage{microtype}

\usepackage{tikz}
\usetikzlibrary{arrows.meta,calc,decorations.pathreplacing}
\usetikzlibrary{decorations.pathmorphing}
\usetikzlibrary{decorations.markings}
\usetikzlibrary{shapes.geometric}
\usepackage{pgfplots}
\pgfplotsset{compat=1.17}

\hypersetup{
  pdftitle={Momentum-projected hadron entanglement from lattice-QCD replica correlators},
  pdfauthor={Kiminad A. Mamo},
  pdfsubject={Finite-volume QCD, hadron entanglement, replica correlators, lattice QCD},
  pdfkeywords={QCD, lattice QCD, entanglement entropy, Renyi entropy, finite volume, hadron structure, replica method}
}

\newif\ifusetikzfigures
\usetikzfigurestrue

\newcommand{\diff}{\mathrm{d}}
\newcommand{\e}{\mathrm{e}}

\newcommand{\be}{\begin{equation}}
\newcommand{\ee}{\end{equation}}
\newcommand{\ba}{\begin{aligned}}
\newcommand{\ea}{\end{aligned}}
\newcommand{\lab}[1]{\label{#1}}

\newcommand{\Tr}{\mathop{\mathrm{Tr}}\nolimits}

\newcommand{\ket}[1]{\left|#1\right\rangle}
\newcommand{\bra}[1]{\left\langle#1\right|}

\newcommand{\cutC}{\mathcal C}

\newcommand{\regA}{\mathcal{A}}
\newcommand{\regB}{\mathcal{B}}

\newcommand{\Sig}{\Sigma}

\newcommand{\Reff}{R_{\rm eff}}
\newcommand{\Rfit}{R_{\rm fit}}

\newcommand{\DeltaS}{\Delta S}
\newcommand{\Fone}{\mathcal F}

\newcommand{\Vol}{\mathcal V}
\newcommand{\Ccp}{\mathcal C_{\rm cp}}

\newcommand{\Sh}{\mathcal S}

\newenvironment{centralidentity}[1]{%
  \par\smallskip\noindent\textbf{#1.}\ }{%
  \par\smallskip
}

\DeclareMathSizes{11}{11}{8}{6}
\begin{document}
\makeatletter
\let\ArxivNormalHangFrom\@hangfrom@section
\let\ArxivNormalSectionCntFormat\@sectioncntformat
\makeatother

\title{Momentum-projected hadron entanglement from lattice-QCD replica correlators}

\author{Kiminad A.~Mamo}
\affiliation{Department of Physics, University of Connecticut, Storrs, CT 06269-3046, USA}
\email{ska25005@uconn.edu}

\date{October 1, 2026}

\begin{abstract}
We define a finite-volume lattice-QCD observable for the vacuum-subtracted spatial R\'enyi response of a source--sink-prepared, momentum-projected hadron. At fixed regulator, integer replica index $n>1$, spatial region and gauge-theory cut prescription, double-sided state projection expresses this response as a ratio of replicated and ordinary hadron correlators. The first numerical target is the $n=2$ ratio; its proposed $L^{-3}$ scaling at fixed physical region is to be tested with matched volumes. We implement the measurement in quenched SU(3) for a charged pseudoscalar meson with heavy degenerate Wilson valence quarks, $am_{\rm eff}\simeq2.07$, on two $16^3\times32$ sheets with a central-plaquette cut. Five finite balls and empty/full controls are measured at nine source/sink placements. Their finite-ball means increase with radius at every placement. At the reference placement the three largest responses are 1.6--2.5 times an ideal single-quasiparticle occupancy reference. A full-covariance fit $\Delta_{\rm cp}=\kappa_{\rm occ}S_2^{\rm occ}$ gives $\kappa_{\rm occ}=1.68\pm0.12$; radius-window and full-control comparisons give values spanning $1.49$--$1.73$. An exact pure-state quark--antiquark pair model interpolates, in the dilute limit, between the unit-occupancy reference ($\kappa_{\rm occ}=1$) for an unresolved meson and the double-occupancy reference ($\kappa_{\rm occ}=2$) for a separately resolved quark and antiquark, and yields a boundary-area correction governed by their separation.
\end{abstract}

\maketitle

\section{Introduction}
\lab{sec:introduction}

Hadron structure is usually described through matrix elements of local or light-ray operators.  Energy-momentum-tensor matrix elements, for example, are parameterized by gravitational form factors and encode mass, spin, pressure, shear, and mechanical information~\cite{Ji:1996ek,Polyakov:2002yz,Polyakov:2018zvc,Hackett:2023rif}.  Such matrix elements probe local or light-ray operator densities in a one-hadron state.  Spatial entanglement asks a different question: how the reduced density matrix of a spatial region changes when the QCD state is changed from the vacuum to a one-hadron state.  These two questions are not equivalent, even though both may encode spatial information about the hadron.  Since the vacuum entropy is UV dominated, the useful finite-volume object is a vacuum-subtracted, prescription-matched R\'enyi response.

The physical content of this response is the radius dependence of the hadron-induced change in the reduced density matrix.  It is sensitive not only to how a delocalized one-particle state occupies a spatial region, but also to how color, sea-quark, and gauge-field correlations across the entangling surface are modified by hadron-state preparation.  Thus the response function \(\Fone_n^{\cutC}(R;h)\) is not another density profile and is not reconstructible from gravitational form factors alone.  Its \(R\)-dependence can distinguish smooth volume-like contributions from cut-localized QCD correlations, and it provides a first-principles target for questions usually discussed in more approximate language, such as partonic entanglement, small-\(x\) saturation, rapidity-space entropy, and the effective liberation of color degrees of freedom in high-energy processes.

This hierarchy is important.  The hadron spatial-entanglement-entropy program has often been developed in partonic, lower-dimensional, or high-energy effective descriptions.  Deep inelastic scattering has been proposed as a probe of entanglement in a partonic picture~\cite{Kharzeev:2017qzs}, universal rapidity scaling has been studied using lower-dimensional and conformal arguments~\cite{Gursoy:2023hge}, and meson entanglement has been studied in two-dimensional QCD~\cite{Goykhman:2015sga,Liu:2022qqf}.  More recently, an infinite-momentum-frame and wave-packet interpretation of hadron spatial entanglement was developed in Ref.~\cite{Mamo:2025}.  These approaches are physically important, but they are downstream of a more basic question: what is the QCD density-matrix observable whose limits, state preparations, or effective descriptions they approximate?  The purpose of the present work is to identify that first-principles starting point in finite-volume lattice QCD.  Once the finite-volume response is measured, effective descriptions of DIS, rapidity-space entropy, or infinite-momentum-frame wave packets can be constrained by matching their reduced-state information to the same QCD observable or to controlled limits of it.  This paper defines the Euclidean finite-volume target; it does not attempt that phenomenological matching.

We use the Euclidean replica cut-and-glue definition of reduced density matrices in gauge theory.  On the lattice this is the Buividovich-Polikarpov construction and its successors: one chooses a spatial region at a Euclidean time cut, glues the complement across the cut, leaves independent boundary data on the region, and evaluates \(\Tr\rho_{\regA}^{\,n}\) through a path integral on an \(n\)-sheeted cut geometry~\cite{Buividovich:0806.3376,Buividovich:08:2,Velytsky:2008sv,Nakagawa:2009jk,Nakagawa:2010kjk,Itou:2015cyu,Rabenstein:2018bri,Rindlisbacher:2022,Rindlisbacher:2023}.  Concretely, in the replica trace the fields on \(\regB\) are glued across the time cut within each sheet, while the two open \(\regA\) edges are glued cyclically between adjacent sheets; after \(n\) such gluings the final sheet is identified with the first.  The gauge-theory observable is defined only after choosing a cut prescription \(\cutC\): center choice, sheet map, boundary-link convention, fermion cut convention, ball discretization, and radius assignment.  The ball \(B_R\) is used here because its radius gives a convenient hadronic response variable; the construction applies equally to slabs, strips, and general lattice regions.  Related gauge-theory entanglement constructions include Refs.~\cite{Bombelli:1986rw,Srednicki:1993im,Callan:1994py,Calabrese:2004eu,Ryu:2006bv,Ryu:2006ef,Klebanov:2007ws,Solodukhin:2008dh,Casini:2011kv,CasiniHuertaRosabal:2013,DonnellyWall:2014edge,Huang:2014pfa,Goykhman:2015sga,Agarwal:2016cir,Donnelly:2019zde,Anegawa:2021osi,Panizza:2022gvd,Liu:2022qqf,Bulgarelli:2023ofi}.  Existing non-Abelian lattice entanglement studies have mostly focused on the vacuum, slab or strip geometries, entropic \(C\)-functions, external static sources, static-quark color correlations, flux tubes, lower-dimensional systems, or simplified-model constructions~\cite{Buividovich:08:2,Itou:2015cyu,Rabenstein:2018bri,Rindlisbacher:2022,Rindlisbacher:2023,Takahashi:2019xjv,Takahashi:2020vct,Amorosso:2024glf,Amorosso:2024leg,Amorosso:2025tgg,Grieninger:2025rdi,Horn:2026bfs}.  They do not provide the source-sink replica observable for the spatial entanglement response of a dynamical, momentum-projected finite-volume hadron in full lattice QCD.

A particularly close lattice precedent is the flux-tube entanglement construction of Ref.~\cite{Amorosso:2024leg}, where the state is prepared by a static \(Q\bar Q\) pair and the vacuum-subtracted R\'enyi response is expressed as a ratio of Polyakov-loop correlators on replicated and ordinary lattices.  That calculation provides an important feasibility demonstration of the replicated-ratio strategy in a pure-gauge setting with static sources.  The present work replaces the Polyakov-loop preparation of a static flux tube by ordinary hadron source-sink operators and therefore targets a dynamical, momentum-projected finite-volume hadron. Multi-slab flux-tube tomography in $(2+1)$-dimensional Yang--Mills theory also relates the response to whether the entangling region severs an extended excitation~\cite{Amorosso2026Tomography}. It provides a useful geometric precedent for the pair-model boundary analysis below.

The state used here is the standard finite-volume hadron state of lattice spectroscopy: a color-singlet finite-volume level isolated by a spatially summed source and sink, projected to rest.  For resonant or strongly mixed channels, \(h\) denotes the selected finite-volume eigenstate.  Momentum projection removes a preferred hadron center; a fixed physical region \(B_R\) probes a delocalized, box-normalized one-particle state.  The main novelty of this work is a source-sink replica identity for precisely this momentum-projected finite-volume hadron state.

The central result is the projected source-sink replica identity
\begin{align}
&\DeltaS^{\cutC}_{n,L}(B_R;h,\lambda)
\nonumber\\
&\quad=
\lim_{\substack{\tau_f\to+\infty\\-\tau_i\to+\infty}}
\frac{1}{1-n}
\ln
\frac{
\left\langle
\prod_{k=1}^{n}C_{h,\lambda}^{(k)}
\right\rangle^{\cutC}_{n,B_R,L}
}{
\langle C_{h,\lambda}\rangle_{1,L}^{\,n}
}.
\lab{eq:intro-correlator-ratio}
\end{align}
The numerator is evaluated on the \(n\)-sheeted cut geometry associated with \(B_R\), with the same hadron source-sink functional inserted on every sheet.  The denominator is the \(n\)th power of the ordinary single-sheet source-sink correlator.  The sheet insertions are the \(n\) replica copies needed to compute \(\Tr\rho_{B_R,h}^{\,n}\); they are not an \(n\)-hadron physical state.  Absolute partition functions cancel in this ratio.  This cancellation removes the need to compute the vacuum replicated partition-function ratio itself, but it does not remove the need to evaluate the numerator in the replicated cut measure.

For a rest-frame momentum eigenstate normalized in a periodic volume \(L^3\), a fixed region samples the hadron with inverse-volume weight.  The expected leading behavior is therefore
\begin{align}
\DeltaS^{\cutC}_{n,L}(B_R;h,\lambda)
=
\frac{1}{L^3}\,
\Fone^{\cutC}_{n}(R;h,\lambda)
+
\cdots ,
\lab{eq:intro-scaling}
\end{align}
for fixed physical \(R\), fixed integer \(n>1\), and fixed \(\cutC\), when the leading coefficient exists.  Thus the primary lattice target is the response function
\begin{align}
\Fone^{\cutC}_{n}(R;h,\lambda)
=
\lim_{L\to\infty}
L^3\DeltaS^{\cutC}_{n,L}(B_R;h,\lambda).
\lab{eq:intro-F-limit}
\end{align}
Equations~\eqref{eq:intro-scaling} and \eqref{eq:intro-F-limit} should be read as the expected one-particle finite-volume ansatz, not as a theorem derived here from the nonlinear reduced density matrix.  The corresponding assumption is that, at fixed regulator and fixed \(\cutC\), the hadron-induced change of the reduced state begins at order \(L^{-3}\).  Gauge-theory cut-localized structures can be subtle, so the exponent should be measured in the finite-volume data; the \(L^3\) rescaling is the expected one-particle scaling for a box-normalized momentum eigenstate.

At leading planar order, the light-front von Neumann meson response is proportional to \(L_{\rm box}^{-1}\) at fixed physical momentum and interval, with a coefficient controlled by quark and antiquark PDF moments~\cite{Liu:2022qqf}.

The status of the main statements is therefore as follows.  The source-sink identity in \cref{eq:intro-correlator-ratio} is an exact fixed-prescription replica identity after the standard double-sided source-sink projection.  The \(L^{-3}\) behavior in \cref{eq:intro-scaling} is a one-particle finite-volume ansatz for the four-dimensional lattice observable, motivated by box normalization and by the linearized reduced-density-matrix expansion.  The finite-volume exponent is part of the lattice output.

The observable in \cref{eq:intro-correlator-ratio} probes how tracing out the complement of $B_R$ changes the normalized hadron density matrix relative to the vacuum. It therefore complements the local and light-ray operator information encoded in conventional form factors. Other state preparations, such as localized wave packets, boosted states, or infinite-momentum-frame limits, define their own entanglement observables and effective descriptions~\cite{Mamo:2025}. The present construction specifies the rest-frame, momentum-projected finite-volume target and its lattice measurement formula.

The numerical study in \cref{sec:numerical} implements the proposal on five finite balls and empty/full controls throughout a nine-placement source/sink grid. An explicit central-plaquette prescription $\mathcal C_{\rm cp}$ realizes the Euclidean cut. We compare the radius response with the occupancy model and evaluate regional contrasts with a joint jackknife covariance. \Cref{pair:lattice} fits the overall occupancy normalization and relates it to the unresolved-particle and resolved-pair limits derived in Appendix~\ref{hr:appendix}. The Supplemental Material (SM)~\cite{Mamo:SupplementalMaterial} gives the complete placement scan, cut and contraction definitions, essential uncertainty analysis, temporal controls, and finite-mask model calculations.

The paper is organized as follows.  \Cref{sec:definition} defines the finite-volume state and the fixed-prescription R\'enyi entropy.  \Cref{sec:identity} derives the source-sink replica identity.  \Cref{sec:scaling} gives the finite-volume scaling logic and the occupancy reference for the response coefficient.  \Cref{sec:lattice} gives the implementation strategy; \cref{sec:numerical} presents the exploratory quenched results.  \Cref{pair:lattice} discusses their occupancy normalization and pair-model interpretation, followed by the conclusions in \cref{sec:conclusion}.  The appendices give the kernel derivation, sheet maps, finite-cutoff occupancy checks, and correlated volume-fitting methods. Appendix~\ref{hr:appendix} derives the exact pair-model interpolation and boundary-area correction and retains the measured-radius fit comparisons.

\section{Finite-volume definition}
\lab{sec:definition}

\begin{figure*}[t]
\centering
\ifusetikzfigures
\resizebox{0.68\textwidth}{!}{%
\begin{tikzpicture}[x=1cm,y=1cm,>=Latex,font=\small]
  \tikzset{
    whitebox/.style={fill=white,fill opacity=0.94,text opacity=1,inner sep=1.6pt},
    paneltitle/.style={anchor=north,align=center},
    regionlabel/.style={whitebox,align=center}
  }
  \path[use as bounding box] (-5.45,-2.78) rectangle (6.50,2.45);

  \begin{scope}[shift={(-3.35,0)}]
    \def\Rsmall{1.00}
    \def\Rbig{2.05}
    \fill[blue!15,opacity=0.42] (0,0) circle (\Rsmall);
    \draw[very thick] (0,0) circle (\Rsmall);
    \draw[thick] (0,0) circle (\Rbig);
    \filldraw[black] (0,0) circle (0.75pt);
    \draw[->,thick] (0,0) -- (\Rsmall,0) node[midway,below=1pt] {$R$};
    \node[regionlabel] at (-0.22,0.35) {$\regA$};
    \node[regionlabel] at (1.37,0.50) {$\regB$};
    \node[regionlabel] at (0,\Rsmall+0.28) {$\Sigma$};
    \node[paneltitle] at (0,-2.36) {(a) spatial region $\regA=B_R$};
  \end{scope}

  \begin{scope}[shift={(3.05,0)},scale=1.05]
    \def\xL{-3.05}
    \def\xR{ 3.05}
    \def\xa{-1.12}
    \def\xb{ 1.12}
    \def\eps{0.22}
    \def\tT{ 2.0}
    \def\tB{-2.0}

    \fill[gray!7] (\xL,0) rectangle (\xR,\tT);
    \fill[gray!4] (\xL,0) rectangle (\xR,\tB);
    \draw[->,thick] (\xL-0.20,0) -- (\xR+0.28,0) node[right] {$x$};
    \draw[->,thick] (0,\tB-0.26) -- (0,\tT+0.28) node[above] {$\tau$};
    \draw[very thick,black!75] (\xL,0) -- (\xa,0);
    \draw[very thick,black!75] (\xb,0) -- (\xR,0);
    \draw[very thick] (\xa,\eps) -- (\xb,\eps);
    \draw[very thick] (\xa,-\eps) -- (\xb,-\eps);
    \draw[densely dashed,black!55] (\xa,\tB) -- (\xa,\tT);
    \draw[densely dashed,black!55] (\xb,\tB) -- (\xb,\tT);
    \filldraw[black] (\xa,0) circle (0.95pt);
    \filldraw[black] (\xb,0) circle (0.95pt);

    \node[whitebox] at (0,1.08) {$\regA$ open};
    \node[whitebox] at (-2.15,-0.66) {$\regB$ glued};
    \node[whitebox] at ( 2.15,-0.66) {$\regB$ glued};
    \node[whitebox] at (0,0.54) {$\Phi_{\regA,+}$};
    \node[whitebox] at (0,-0.54) {$\Phi_{\regA,-}$};
    \node[whitebox,align=center] at (-2.22,0.88)
      {glue on $\regB$:\\[-1pt]$\Phi(0^+,\regB)=\Phi(0^-,\regB)$};
    \node[paneltitle] at (0,-2.44) {(b) Euclidean cut-and-glue kernel};
  \end{scope}
\end{tikzpicture}}
\else
\includegraphics[width=0.45\textwidth]{ballAB.pdf}\hfill
\includegraphics[width=0.50\textwidth]{cut_glue.pdf}
\fi
\caption{\textbf{Spatial region and cut-and-glue construction.}
The ball \(\regA=B_R\) has complement \(\regB\) and boundary \(\Sigma=\partial\regA\).  At the Euclidean time cut, \(\regB\) is glued across the cut, while \(\regA\) is left open and later cyclically glued between replicas.}
\lab{fig:region-cut}
\end{figure*}

\subsection{Region, prescription, and entropy}

Let the spatial volume be a periodic box of size \(L^3\), and let \(\regA=B_R\) be a ball at Euclidean time \(\tau=0\), with complement \(\regB\) and entangling surface \(\Sig=\partial\regA\). Figure~\ref{fig:region-cut} illustrates the region and the cut-and-glue construction. All expressions are defined at fixed gauge-theory cut prescription \(\cutC\).

The vacuum reduced density matrix kernel is obtained by gluing the fields on \(\regB\) across the cut while keeping independent boundary data on \(\regA\):
\begin{align}
&\rho_{\regA,0,L}^{\cutC}
[\Phi_{\regA,+},\Phi_{\regA,-}]
\nonumber\\
&\quad=
\frac{1}{\mathcal N_0}
\left.
\int \mathcal D\Phi\;e^{-S_E[\Phi]}
\right|_{\substack{
\Phi(0^+,\regA)=\Phi_{\regA,+}\\
\Phi(0^-,\regA)=\Phi_{\regA,-}\\
\Phi(0^+,\regB)=\Phi(0^-,\regB)}} .
\lab{eq:rhoA-vac}
\end{align}
The fixed-prescription finite-volume R\'enyi entropy is
\begin{align}
S^{\cutC}_{n,L}(B_R;\rho)
=
\frac{1}{1-n}
\ln
\Tr\left[
\left(\rho^{\cutC}_{B_R,L}\right)^n
\right],
\qquad n=2,3,\ldots .
\lab{eq:Sn-def}
\end{align}
The hadron observable is the vacuum-subtracted entropy
\begin{align}
\DeltaS^{\cutC}_{n,L}(B_R;h,\lambda)
=
S^{\cutC}_{n,L}(B_R;\rho_{h,\lambda,L})
-
S^{\cutC}_{n,L}(B_R;\rho_{0,L}).
\lab{eq:DeltaS-def-main}
\end{align}
This is not a relative entropy; its sign is not fixed.  A positive value means that the hadron-prepared reduced state has a larger R\'enyi entropy than the vacuum for the chosen \(n\), region, and prescription.  A negative value is also physically meaningful: it would mean that the hadron preparation reduces the relevant R\'enyi entropy of \(B_R\) relative to the vacuum, for example through a more constrained reduced-state spectrum or through prescription-dependent cut correlations.  Simple delocalized occupancy models illustrate a positive leading response.

The prescription \(\cutC\) is part of the proposal and need not be the electric center. One possible choice is the link-based extended-Hilbert-space, or electric-center, construction. A second, explicit Euclidean realization is the central-plaquette prescription \(\Ccp\) used in \cref{sec:cp-prescription}. Both use within-sheet gluing on \(\regB\) and cyclic gluing on \(\regA\), but their boundary action and state-space interpretation must be specified rather than inferred from connectivity alone. The numerical study reports the finite-time \(\Ccp\) response without assuming equality to the electric-center answer.

\subsection{Continuum status and prescription dependence}
\lab{sec:continuum-prescription}

The superscript \(\cutC\) is part of the definition of the observable.  Vacuum subtraction cancels the matched vacuum R\'enyi entropy computed with the same regulator, region, and cut prescription, but it does not by itself erase all prescription dependence.  At nonzero lattice spacing one should write
\begin{align}
\DeltaS^{\cutC}_{n,L,a}(B_R;h,\lambda)
\end{align}
when the regulator dependence is relevant.  The finite-volume response at fixed cutoff is
\begin{align}
\Fone^{\cutC}_{n,a}(R;h,\lambda)
=
\lim_{L\to\infty}L^3\DeltaS^{\cutC}_{n,L,a}(B_R;h,\lambda),
\lab{eq:F-a-def}
\end{align}
if the infinite-volume limit exists.  A continuum-extrapolated target is then
\begin{align}
\Fone^{\cutC}_{n}(R;h,\lambda)
=
\lim_{a\to0}\Fone^{\cutC}_{n,a}(R;h,\lambda),
\lab{eq:F-continuum-def}
\end{align}
provided the region, radius assignment, center choice, boundary-link convention, fermion cut convention, and action are matched along the extrapolation.  Different continuum center choices or edge-mode prescriptions may define different gauge-theory entanglement observables rather than different estimates of one universal number.  This is not a defect of the construction; it is a scheme dependence that should be stated, tested, and, where possible, compared across prescriptions.

\subsection{Momentum-projected source-sink state}

For rest-frame kinematics, the finite-volume interpolating operator is spatially summed,
\begin{align}
\mathcal O_{H,\lambda}(\tau;\bm0)
=
\int \diff^3\bm x\;\mathcal O_{H,\lambda}(\tau,\bm x),
\lab{eq:OH-rest}
\end{align}
with spin projection \(\lambda\).  On the lattice this means either \(a^3\sum_{\bm x}\mathcal O_{H,\lambda}(\tau,\bm x)\) or the conventionally normalized spatial sum \(\sum_{\bm x}\mathcal O_{H,\lambda}(\tau,\bm x)\).  The choice changes overlap factors such as \(Z_{h,\lambda,L}\), but these factors cancel in the ratio below when the same convention is used in numerator and denominator.  With \(\tau_i<0<\tau_f\), define
\begin{align}
\Sh_{0,\lambda}^{\dagger}
=
\mathcal O_{H,\lambda}^{\dagger}(\tau_i;\bm0),
\qquad
\Sh_{0,\lambda}
=
\mathcal O_{H,\lambda}(\tau_f;\bm0),
\lab{eq:source-sink-def}
\end{align}
and
\begin{align}
C_{h,\lambda}(T_{\rm sep})
=
\Sh_{0,\lambda}\Sh_{0,\lambda}^{\dagger},
\qquad
T_{\rm sep}=\tau_f-\tau_i.
\lab{eq:C-source-sink-def}
\end{align}
With exterior-vacuum projection, the ordinary correlator has the spectral form
\begin{align}
\langle C_{h,\lambda}(T_{\rm sep})\rangle_{1,L}
&=
\sum_{\alpha}
\left|
\langle0|\mathcal O_{H,\lambda}(0;\bm0)
|\alpha,L,\bm0\rangle
\right|^2
\nonumber\\
&\quad\times
\e^{-E_{\alpha,L}T_{\rm sep}}
\nonumber\\
&=
|Z_{h,\lambda,L}|^2\e^{-M_{h,L}T_{\rm sep}}
\nonumber\\
&\quad\times
\left[1+O(\e^{-\Delta E_L T_{\rm sep}})\right].
\lab{eq:spectral-projection}
\end{align}
On a finite temporal circle of extent $\mathcal T$, the leading charged-pseudoscalar contribution instead contains both paths,
\[
 C_\pi(t)\simeq |Z_\pi|^2
 \left[e^{-M_\pi t}+e^{-M_\pi(\mathcal T-t)}\right],
\]
with additional excited and thermal terms. The temporal scan in Sec.~S4 of the Supplemental Material~\cite{Mamo:SupplementalMaterial} examines the crossover between these propagation directions.

The large-time projected state is
\begin{align}
\rho_{h,\lambda,L}
=
\ket{h,L,\bm0,\lambda}\bra{h,L,\bm0,\lambda}.
\lab{eq:finite-volume-state}
\end{align}
Because the entanglement cut is at \(\tau=0\), the projection limit is double-sided:
\begin{align}
\tau_f\to +\infty,
\qquad
-\tau_i\to +\infty,
\lab{eq:double-sided-projection}
\end{align}
with \(R,L,a,n\), and \(\cutC\) held fixed.  At finite \(T_{\rm sep}\), the same ratio defines a source-sink-prepared estimator with excited-state contamination. On a finite temporal circle, the exterior propagation must also project the vacuum. Thus the formal limit requires a sufficiently large or extrapolated temporal extent, not simply increasing \(T_{\rm sep}\) toward half a fixed period. The temporal diagnostics in Sec.~S4 of the Supplemental Material~\cite{Mamo:SupplementalMaterial} illustrate this distinction.

For a spinful hadron, fixed-\(\lambda\) responses are defined first.  A convenient spin-averaged response is
\begin{align}
\DeltaS^{\cutC}_{n,L}(B_R;h)
&=
\frac{1}{2S+1}
\sum_{\lambda=-S}^{S}
\DeltaS^{\cutC}_{n,L}(B_R;h,\lambda),
\lab{eq:spin-avg-DeltaS}
\\
\Fone^{\cutC}_{n,L}(R;h)
&=
\frac{1}{2S+1}
\sum_{\lambda=-S}^{S}
\Fone^{\cutC}_{n,L}(R;h,\lambda).
\lab{eq:spin-avg-F}
\end{align}
This arithmetic average is not, in general, the R\'enyi entropy of the mixed unpolarized density matrix \((2S+1)^{-1}\sum_{\lambda}\rho_{h,\lambda,L}\), because R\'enyi entropies are nonlinear in the density matrix.  At finite lattice spacing, spin projection is implemented through the appropriate cubic-group irreps and projectors; the continuum spin labels are recovered after the usual continuum matching or extrapolation.

\section{Source-sink replica identity}
\lab{sec:identity}

\subsection{Replica traces}

Let \(Z_{1,L}[0]\) be the ordinary vacuum partition function and \(Z_{1,L}[h_\lambda]\) the single-sheet path integral with the hadron source-sink functional inserted:
\begin{align}
Z_{1,L}[h_\lambda]
=
Z_{1,L}[0]\,
\langle C_{h,\lambda}\rangle_{1,L}.
\lab{eq:Z1-h}
\end{align}
Let \(Z^{\cutC}_{n,L}[0;B_R]\) be the vacuum path integral on the \(n\)-sheeted cut geometry and \(Z^{\cutC}_{n,L}[h_\lambda;B_R]\) the same replicated path integral with the source-sink functional inserted on every sheet.

The hadron and vacuum replica traces are
\begin{align}
\Tr\rho_{B_R,h,\lambda,L}^{\,n}
&=
\frac{Z^{\cutC}_{n,L}[h_\lambda;B_R]}
{Z_{1,L}[h_\lambda]^n},
\nonumber\\
\Tr\rho_{B_R,0,L}^{\,n}
&=
\frac{Z^{\cutC}_{n,L}[0;B_R]}
{Z_{1,L}[0]^n}.
\lab{eq:replica-traces}
\end{align}
Therefore
\begin{align}
\DeltaS^{\cutC}_{n,L}(B_R;h,\lambda)
&=
\frac{1}{1-n}
\ln
\frac{
\Tr\rho_{B_R,h,\lambda,L}^{\,n}
}{
\Tr\rho_{B_R,0,L}^{\,n}
}
\nonumber\\
&=
\frac{1}{1-n}
\ln\Bigg[
\frac{Z^{\cutC}_{n,L}[h_\lambda;B_R]}
{Z^{\cutC}_{n,L}[0;B_R]}
\nonumber\\
&\qquad\qquad\times
\left(
\frac{Z_{1,L}[0]}{Z_{1,L}[h_\lambda]}
\right)^n
\Bigg].
\lab{eq:DeltaSn-replica-ratio}
\end{align}
The absolute vacuum partition functions cancel.  This cancellation removes the need to compute \(Z^{\cutC}_{n,L}[0;B_R]/Z_{1,L}[0]^n\) as an independent normalization factor, but it does not eliminate the need to generate, reweight, or otherwise sample the replicated cut measure that defines the expectation value in the numerator.

\subsection{Correlator representation}

The replicated hadron functional divided by the replicated vacuum functional is
\begin{align}
\frac{Z^{\cutC}_{n,L}[h_\lambda;B_R]}{Z^{\cutC}_{n,L}[0;B_R]}
=
\left\langle
\prod_{k=1}^{n} C_{h,\lambda}^{(k)}
\right\rangle^{\cutC}_{n,B_R,L},
\lab{eq:replicated-correlator-def}
\end{align}
where
\begin{align}
C_{h,\lambda}^{(k)}
=
\Sh_{0,\lambda}^{(k)}
\Sh_{0,\lambda}^{\dagger(k)}
\lab{eq:C-sheet-def}
\end{align}
is inserted on sheet \(k\).  The superscript labels the sheet of the source and sink.  In full QCD the contractions are performed with the replicated Dirac operator on the cut graph; they do not factorize into ordinary single-sheet contractions.  Also,
\begin{align}
\frac{Z_{1,L}[h_\lambda]}{Z_{1,L}[0]}
=
\langle C_{h,\lambda}\rangle_{1,L}.
\lab{eq:ordinary-correlator-def}
\end{align}

\begin{centralidentity}{Source-sink replica identity}
At fixed regulator, cut prescription \(\cutC\), integer \(n>1\), and after the double-sided projection in \cref{eq:double-sided-projection},
\begin{align}
\boxed{
\DeltaS^{\cutC}_{n,L}(B_R;h,\lambda)
=
\frac{1}{1-n}
\ln
\frac{
\left\langle
\prod_{k=1}^{n}C_{h,\lambda}^{(k)}
\right\rangle^{\cutC}_{n,B_R,L}
}{
\langle C_{h,\lambda}\rangle_{1,L}^{\,n}
}}
.
\lab{eq:DeltaSn-correlator-ratio}
\end{align}
\end{centralidentity}

Equation~\eqref{eq:DeltaSn-correlator-ratio} is the measurement formula.  The product over sheets represents the nonlinear trace \(\Tr\rho_{B_R,h}^{\,n}\), not a physical \(n\)-hadron state.  The overlap factors and ordinary Euclidean propagation cancel between numerator and denominator in the projected limit.  The cancellation of absolute partition functions should therefore be interpreted as a normalization cancellation inside a replicated path-integral ratio, not as a statement that the full-QCD numerator can be measured by post-processing ordinary single-sheet two-point ensembles.

A closely related vacuum-subtracted replica ratio was derived and numerically implemented in Ref.~\cite{Amorosso:2024leg} for a static heavy-quark pair in pure Yang-Mills theory.  In that case the state preparation is by Polyakov-loop insertions, and the excess flux-tube R\'enyi entropy is obtained from a replicated Polyakov-loop correlator divided by the corresponding power of the ordinary Polyakov-loop correlator.  That result can be viewed as a first feasibility demonstration of the replicated-ratio method before tackling the more demanding source-sink hadron correlators in full QCD.

\subsection{Transfer-matrix form}

Abbreviating \(\ket h_L=\ket{h,\bm0,\lambda}_L\), the double-sided large-time and exterior-vacuum limits give
\begin{align}
\left\langle\prod_{k=1}^{n}C_{h,\lambda}^{(k)}\right\rangle^{\cutC}_{n,B_R,L}
&\to |Z_{h,\lambda,L}|^{2n}\e^{-nM_{h,L}T_{\rm sep}}\nonumber\\
&\quad\times
\frac{{}_{L}\!\bra h^{\otimes n}\widehat{\mathcal R}^{\cutC}_{n}(B_R)\ket h_{L}^{\otimes n}}
{{}_{L}\!\bra0^{\otimes n}\widehat{\mathcal R}^{\cutC}_{n}(B_R)\ket0_{L}^{\otimes n}}
+\cdots .
\lab{eq:transfer-replica-operator}
\end{align}
Here \(\widehat{\mathcal R}^{\cutC}_{n}(B_R)\) is the cyclic cut-and-glue insertion: it glues \(\regB\) within each sheet and permutes the open \(B_R\) edges. Its vacuum matrix element is the vacuum regional purity in the same prescription. The division in \cref{eq:transfer-replica-operator} is already implicit in the normalized replicated measure; it does not require an additional partition-function measurement.  The ordinary denominator contains
\begin{align}
\langle C_{h,\lambda}\rangle_{1,L}^{\,n}
=
|Z_{h,\lambda,L}|^{2n}
\e^{-nM_{h,L}T_{\rm sep}}
\left[1+O(\e^{-\Delta E_L T_{\rm sep}})\right].
\lab{eq:denominator-asymptotic}
\end{align}
Thus the ratio isolates the hadron replica matrix element divided by its matched vacuum matrix element.  The cut \(B_R\) fixes a spatial region and can exchange momentum internally, while the external source and sink still project the hadron to \(\bm p=\bm0\).

\section{Finite-volume scaling and response coefficient}
\lab{sec:scaling}

\subsection{One-particle scaling}

For a normalized rest-frame momentum eigenstate in a periodic box, a fixed physical region \(B_R\) samples an inverse-volume fraction of the state,
\begin{align}
\frac{\Vol_R}{\Vol_L}
=
\frac{4\pi R^3/3}{L^3}.
\lab{eq:volume-fraction}
\end{align}
The same normalization appears in fixed-subregion matrix elements.  With infinite-volume relativistic normalization,
\begin{align}
{}_\infty\langle h,\bm p',\lambda'|h,\bm p,\lambda\rangle_\infty
=
2E_{\bm p}(2\pi)^3\delta^{(3)}(\bm p'-\bm p)\delta_{\lambda'\lambda},
\end{align}
the finite-volume rest state gives, for a fixed-subregion operator \(\mathcal O_{B_R}\),
\begin{align}
&
\langle h,L,\bm0,\lambda|
\mathcal O_{B_R}
|h,L,\bm0,\lambda\rangle
-
\langle0,L|\mathcal O_{B_R}|0,L\rangle
\nonumber\\
&=
\frac{1}{2E_h L^3}
{}_\infty\langle h,\bm0,\lambda|
\mathcal O_{B_R}
|h,\bm0,\lambda\rangle_{\infty,{\rm conn}}
+
\cdots .
\lab{eq:local-operator-FV-scaling}
\end{align}
This motivates, but does not prove, the corresponding statement for the nonlinear fixed-cutoff R\'enyi functional.  The operative one-particle ansatz is the following finite-cutoff expansion, which must be tested against the volume dependence of the data.

\begin{centralidentity}{Finite-cutoff consequence of the one-particle ansatz}
At fixed lattice spacing, fixed physical \(R\), fixed integer \(n>1\), and fixed cut prescription \(\cutC\), suppose the prescription-defined reduced state has the expansion
\begin{align}
\rho_{B_R,h,\lambda,L}^{\cutC}
&=
\rho_{B_R,0,L}^{\cutC}
+
\frac{1}{L^3}
\delta\rho_{B_R,h,\lambda}^{\cutC}
\nonumber\\
&\quad+
O(L^{-3-\omega}),
\qquad \omega>0 .
\lab{eq:rho-expansion-operational}
\end{align}
Then the fixed-\(n\) R\'enyi response satisfies
\begin{align}
\DeltaS^{\cutC}_{n,L}(B_R;h,\lambda)
&=
\frac{1}{L^3}
\frac{n}{1-n}
\frac{\mathcal N^{\cutC}_{n,h}}
{\mathcal D^{\cutC}_{n,0}}
+
\cdots,
\lab{eq:Renyi-linearized}
\\
\mathcal N^{\cutC}_{n,h}
&=
\Tr\left[
\left(\rho_{B_R,0}^{\cutC}\right)^{n-1}
\delta\rho_{B_R,h,\lambda}^{\cutC}
\right],
\nonumber\\
\mathcal D^{\cutC}_{n,0}
&=
\Tr\left[
\left(\rho_{B_R,0}^{\cutC}\right)^n
\right].
\nonumber
\end{align}
unless the displayed linear coefficient vanishes.
\end{centralidentity}

The exponent is therefore a measurable part of the finite-volume analysis.  The expected one-particle result is \(p=3\), and the leading coefficient is
\begin{align}
\Fone^{\cutC}_{n}(R;h,\lambda)
=
\lim_{L\to\infty}
L^3\DeltaS^{\cutC}_{n,L}(B_R;h,\lambda),
\lab{eq:Fone-def}
\end{align}
when the limit exists.  The finite-volume estimator is
\begin{align}
\Fone^{\cutC}_{n,L}(R;h,\lambda)
=
L^3\DeltaS^{\cutC}_{n,L}(B_R;h,\lambda).
\lab{eq:FoneL-def}
\end{align}
The coefficient has dimensions of volume when \(L\) and \(R\) are quoted in physical units.

The primary scaling target is therefore the simultaneous test
\begin{align}
\DeltaS^{\cutC}_{n,L}(B_R;h) &\to 0,
\nonumber\\
L^3\DeltaS^{\cutC}_{n,L}(B_R;h) &\to \Fone^{\cutC}_{n}(R;h),
\lab{eq:primary-scaling-target}
\end{align}
at fixed physical \(R\), fixed integer \(n>1\), and fixed \(\cutC\).  A lattice calculation should test both the vanishing of \(\DeltaS\) itself and the stability of the rescaled coefficient.

Several effects could invalidate or obscure the leading \(p=3\) behavior in a finite data set.  The linear coefficient in \cref{eq:Renyi-linearized} could vanish for a symmetry or prescription reason; massless modes or near-threshold multi-particle states could generate unusually slow finite-volume corrections; cut-localized gauge structures could produce a different leading power at fixed cutoff; and choosing a region that scales with \(L\) would define a different observable.  For this reason the exponent should be fitted directly, as in \cref{eq:free-exponent-fit}, before interpreting \(L^3\DeltaS\) as the asymptotic response.

\subsection{Expected \texorpdfstring{$R$}{R}-dependence}

The \(R\)-dependence of \(\Fone^{\cutC}_{n}(R;h)\) is the hadronic output.  A smooth delocalized occupancy contribution gives an \(R^3\) profile.  Correlations concentrated within a length \(\xi\) of the entangling surface can give an \(R^2\)-type contribution.  A useful diagnostic parametrization is
\begin{align}
\Fone^{\cutC}_{n}(R;h)
&=
A^{\rm bulk}_{n}(R;h)\,
\frac{4\pi R^3}{3}
\nonumber\\
&\quad+
A^{\Sigma}_{n}(R;h)\,
(4\pi R^2)\xi
+\cdots .
\lab{eq:F-bulk-surface}
\end{align}
The first term is the smooth occupancy response.  The second represents cut-localized QCD correlations.  Equation~\eqref{eq:F-bulk-surface} is not a controlled expansion unless additional scale separation is established; the coefficients should not be interpreted as model-independent bulk and surface densities.  The separation of these structures is a data question, not an input assumption.

For the initial lattice calculation, fixed integer \(n>1\), especially \(n=2\), is the cleanest target.  The small-fraction limit of the single-quasiparticle occupancy result~\cite{CastroAlvaredo2018PRL,CastroAlvaredo2018JHEP,CastroAlvaredo2019JMP} gives
\begin{align}
\DeltaS^{\rm occ}_{n,L}
\sim
\frac{n}{n-1}\frac{\Vol_R}{L^3},
\qquad n>1.
\lab{eq:fixed-n-occ}
\end{align}
Its quasiparticle interpretation requires both the region and its complement to be large on the relevant localization scale: in the cited massive-free or large-momentum settings, either a short correlation length or a short de Broglie wavelength suffices. For a composite hadron the intrinsic size is an additional resolution scale. Appendix~\ref{app:renyi-occupancy} states this hierarchy and distinguishes the ideal reference from its finite-scale corrections~\cite{Zhang2021JSTAT}.

The corresponding von Neumann limit contains an additional logarithm,
\begin{align}
\DeltaS^{\rm occ}_{\rm EE}
\sim
\frac{\Vol_R}{L^3}
\ln\frac{L^3}{\Vol_R},
\lab{eq:vn-occ}
\end{align}
which is why the fixed-\(n=2\) observable is the natural first numerical target.

\section{Lattice formulation and analysis}
\lab{sec:lattice}

\subsection{Region and radius assignment}

A default lattice ball centered at \(\bm x_0\) is
\begin{align}
B_R(\bm x_0)
=
\{\bm x:\,d_L(\bm x,\bm x_0)<R\},
\lab{eq:lattice-ball}
\end{align}
where \(d_L\) is the minimum-image distance on the periodic box.  The working window is
\begin{align}
a\ll R\ll L/2.
\lab{eq:working-window}
\end{align}
Cut links or faces may be assigned by a midpoint-crossing rule.  Center averaging and cubic-rotation averaging reduce finite-\(a\) geometry effects.

The analysis radius should be stated.  A useful default is the volume-equivalent effective radius
\begin{align}
N_{\regA}(R,\bm x_0)
&=
\#\{\bm x:\bm x\in B_R(\bm x_0)\},
\nonumber\\
\Reff(R,\bm x_0)
&=
\left(\frac{3a^3N_{\regA}(R,\bm x_0)}{4\pi}\right)^{1/3}.
\lab{eq:Reff-def}
\end{align}
One may use \(\Rfit=\Reff\) after center averaging, while also reporting the input geometric radius \(R\).

\begin{figure*}[t]
\centering
\ifusetikzfigures
\resizebox{0.98\textwidth}{!}{%
\begin{tikzpicture}[x=1.02cm,y=0.88cm,>=Latex,font=\scriptsize]
  \def\xL{-2.90}
  \def\xR{ 2.90}
  \def\xa{-1.08}
  \def\xb{ 1.08}
  \def\eps{0.18}
  \def\tT{ 2.25}
  \def\tB{-2.25}
  \def\ysrc{ 1.60}
  \def\ysnk{-1.60}
  \def\Dx{6.80}
  \pgfmathsetmacro{\Dxhalf}{0.5*\Dx}
  \pgfmathsetmacro{\DxcA}{\Dx-1.85}

  \tikzset{
    sheettitle/.style={anchor=south,fill=white,inner sep=1.2pt},
    smallwhite/.style={fill=white,fill opacity=0.88,text opacity=1,inner sep=1.2pt},
    srcsink/.style={align=left,anchor=west,smallwhite}
  }

  \path[use as bounding box] (\xL-0.75,-2.55) rectangle (\Dx+\xR+0.55,\tT+0.52);

  \begin{scope}[shift={(0,0)}]
    \fill[gray!6] (\xL,0) rectangle (\xR,\tT);
    \fill[gray!3] (\xL,0) rectangle (\xR,\tB);
    \draw[->] (\xL-0.42,\tB+0.25) -- (\xL+0.82,\tB+0.25) node[right] {$x$};
    \draw[->] (\xL-0.42,\tB+0.25) -- (\xL-0.42,\tB+1.22) node[above] {$\tau$};
    \draw[very thick,black!55] (\xL,0) -- (\xa,0);
    \draw[very thick,black!55] (\xb,0) -- (\xR,0);
    \draw[very thick] (\xa,\eps) -- (\xb,\eps);
    \draw[very thick] (\xa,-\eps) -- (\xb,-\eps);
    \draw[densely dashed,black!35] (\xa,\tB) -- (\xa,\tT);
    \draw[densely dashed,black!35] (\xb,\tB) -- (\xb,\tT);
    \filldraw[black] (\xa,0) circle (0.8pt);
    \filldraw[black] (\xb,0) circle (0.8pt);
    \filldraw[black] (0,\ysrc) circle (1.25pt);
    \node[srcsink] at (0.18,\ysrc+0.03)
      {$\Sh_{0,\lambda}^{(1)}$\\[-1pt]$\tau_f>0$};
    \filldraw[black] (0,\ysnk) circle (1.25pt);
    \node[srcsink] at (0.18,\ysnk-0.03)
      {$\Sh_{0,\lambda}^{\dagger(1)}$\\[-1pt]$\tau_i<0$};
    \node[sheettitle] at (0,\tT+0.14) {sheet $1$};
    \node[smallwhite] at (-0.15,0.39) {$\Phi_{\regA,+}^{(1)}$};
    \node[smallwhite] at (-0.15,-0.48) {$\Phi_{\regA,-}^{(1)}$};
    \node[smallwhite] at (0,0.92) {$\regA$};
    \node[black!60,smallwhite] at (-2.05,-0.78) {$\regB$};
    \node[black!60,smallwhite] at ( 2.05,-0.78) {$\regB$};
  \end{scope}

  \begin{scope}[shift={(\Dx,0)}]
    \fill[gray!6] (\xL,0) rectangle (\xR,\tT);
    \fill[gray!3] (\xL,0) rectangle (\xR,\tB);
    \draw[very thick,black!55] (\xL,0) -- (\xa,0);
    \draw[very thick,black!55] (\xb,0) -- (\xR,0);
    \draw[very thick] (\xa,\eps) -- (\xb,\eps);
    \draw[very thick] (\xa,-\eps) -- (\xb,-\eps);
    \draw[densely dashed,black!35] (\xa,\tB) -- (\xa,\tT);
    \draw[densely dashed,black!35] (\xb,\tB) -- (\xb,\tT);
    \filldraw[black] (\xa,0) circle (0.8pt);
    \filldraw[black] (\xb,0) circle (0.8pt);
    \filldraw[black] (0,\ysrc) circle (1.25pt);
    \node[srcsink] at (0.18,\ysrc+0.03)
      {$\Sh_{0,\lambda}^{(2)}$\\[-1pt]$\tau_f>0$};
    \filldraw[black] (0,\ysnk) circle (1.25pt);
    \node[srcsink] at (0.18,\ysnk-0.03)
      {$\Sh_{0,\lambda}^{\dagger(2)}$\\[-1pt]$\tau_i<0$};
    \node[sheettitle] at (0,\tT+0.14) {sheet $2$};
    \node[smallwhite] at (0,0.39) {$\Phi_{\regA,+}^{(2)}$};
    \node[smallwhite] at (0,-0.48) {$\Phi_{\regA,-}^{(2)}$};
    \node[smallwhite] at (0,0.92) {$\regA$};
    \node[black!60,smallwhite] at (-2.05,-0.78) {$\regB$};
    \node[black!60,smallwhite] at ( 2.05,-0.78) {$\regB$};
  \end{scope}

  \draw[->,very thick]
    (0,-\eps)
    .. controls (1.85,-0.82) and (\DxcA,-0.82) ..
    (\Dx,\eps)
    node[pos=0.52,below=4pt,align=center,smallwhite]
    {$\Phi_{\regA,-}^{(1)}=\Phi_{\regA,+}^{(2)}$};

  \draw[->,very thick]
    (\Dx,-\eps)
    .. controls (\DxcA,0.82) and (1.85,0.82) ..
    (0,\eps)
    node[pos=0.52,above=4pt,align=center,smallwhite]
    {$\Phi_{\regA,+}^{(1)}=\Phi_{\regA,-}^{(2)}$};

  \node[
    align=center,
    fill=white,
    fill opacity=0.92,
    text opacity=1,
    draw=black!20,
    rounded corners=2pt,
    inner sep=2.2pt
  ] at (\Dxhalf,0.02)
    {$\Tr\!\left(\rho_{B_R,h,\lambda}^{\,2}\right)$};
\end{tikzpicture}}
\else
\includegraphics[width=0.95\textwidth]{source_sink_replica_gluing.pdf}
\fi
\caption{\textbf{Source-sink replica geometry for \(n=2\).}
Each sheet carries the same rest-frame source and sink.  The complement \(\regB\) is glued across the cut within each sheet, while the two open \(\regA\) edges are glued cyclically between sheets.  The two source-sink insertions are replica copies used to form \(\Tr\rho_{B_R,h}^{2}\), not a physical two-hadron state.}
\label{fig:source-sink-replica}
\end{figure*}

\subsection{Full-QCD replicated measure}

In full QCD, the replicated sea-quark determinant is part of the ensemble measure:
\begin{align}
Z^{\cutC}_{n,L}[0;B_R]
&=
\int\!\prod_{k=1}^{n}\mathcal D U^{(k)}\,
\e^{-S^{\cutC}_{g,n}[U;B_R]}
\nonumber\\
&\quad\times
\prod_{f=1}^{N_f}
\det D^{\cutC}_{f,n,L}[U;B_R].
\lab{eq:fullQCD-replica-measure}
\end{align}
A full-QCD calculation therefore requires either direct sampling of the replicated measure or a controlled determinant-ratio/reweighting strategy.  Quenched and partially quenched calculations are useful pilots, but they define approximations to the full-QCD observable.  Determinant-ratio reweighting may suffer an overlap problem that worsens with the area of the cut, the number of sheets, and the four-volume; direct generation of replicated ensembles is therefore the realistic long-term route for precision full-QCD calculations.

For Wilson-type fermions, the replicated Dirac matrix is obtained by replacing temporal hopping terms crossing the entanglement cut with sheet-permuted hopping terms.  Staggered, domain-wall, overlap, and mixed-action implementations require the corresponding locality, hermiticity, and positivity checks on the replicated graph.

\subsection{Operational construction for \texorpdfstring{$n=2$}{n=2}}
\lab{sec:n2-operational}

For \(n=2\) the replicated lattice graph contains two copies of the ordinary lattice, labeled \(k=1,2\), with a sheet permutation only on temporal links or fermion hoppings that cross the Euclidean cut at \(\tau=0\).  At spatial sites in the complement \(\regB\), a hop from \(0^-\) to \(0^+\) remains on the same sheet.  At spatial sites in \(\regA=B_R\), the hop swaps sheets,
\begin{align}
(1,0^-,\bm x) &\longrightarrow (2,0^+,\bm x),
\nonumber\\
(2,0^-,\bm x) &\longrightarrow (1,0^+,\bm x),
\qquad \bm x\in\regA .
\lab{eq:n2-sheet-swap}
\end{align}
The inverse map is used for the backward hop.  Gauge plaquettes, improvement terms, and fermion stencils that touch the cut are built on this two-sheet graph.

The corresponding source-sink replica geometry is shown in \cref{fig:source-sink-replica}.  Each sheet carries the same rest-frame source and sink, while the two open \(\regA\) edges are glued cyclically between sheets.

The numerator for \(n=2\) is
\begin{align}
\left\langle C_{h,\lambda}^{(1)}C_{h,\lambda}^{(2)}\right\rangle^{\cutC}_{2,B_R,L},
\lab{eq:n2-numerator}
\end{align}
where each sheet has the same spatially summed source at \(\tau_i<0\) and sink at \(\tau_f>0\).  The denominator is \(\langle C_{h,\lambda}\rangle_{1,L}^2\), measured with the same action, masses, source, sink, spin projection, and momentum projection on the ordinary one-sheet lattice.  If the numerator and denominator are measured on statistically independent ensembles their covariance is zero; if a reweighting, shared-noise, or matched-ensemble strategy is used, the covariance between numerator and denominator must be retained in the error analysis.

Two limiting cases provide useful analytic checks on the sheet map and normalization.  If \(\regA=\emptyset\), no sheet permutation is present and the replicated path integral factorizes into ordinary one-sheet factors.  If \(\regA\) is the full spatial volume, the subsystem is the entire projected pure state.  In either case,
\begin{align}
\DeltaS^{\cutC}_{n,L}=0
\qquad
(\regA=\emptyset\ \text{or}\ \regA=\text{full spatial volume})
\lab{eq:empty-full-check}
\end{align}
The empty-region relation is an exact ensemble factorization at finite temporal extent and source/sink separation. The full-region relation uses the projected pure-state and exterior-vacuum limits. Finite samples can differ from either target; temporal contamination is an additional issue for the full region, not an explanation of a failure of exact empty factorization.

\subsection{Source-sink separation}
\lab{sec:source-sink-control}

At finite source and sink times, the estimator depends separately on the distance from the source to the cut and from the cut to the sink.  A schematic form is
\begin{align}
&\DeltaS^{\cutC}_{n,L}(R;\tau_f,\tau_i)
\nonumber\\
&\quad=
\DeltaS^{\cutC}_{n,L}(R)
\nonumber\\
&\quad+
A_+(R,L)e^{-\Delta_+\tau_f}
+
A_-(R,L)e^{-\Delta_-|\tau_i|}
\nonumber\\
&\quad+
A_{+-}(R,L)e^{-\Delta_+\tau_f}
 e^{-\Delta_-|\tau_i|}
\nonumber\\
&\quad+
A_{\rm rep}(R,L)
 e^{-\delta_{+}^{(n,\cutC)}\tau_f}
\nonumber\\
&\quad\times
 e^{-\delta_{-}^{(n,\cutC)}|\tau_i|}
\nonumber\\
&\quad+
\cdots .
\lab{eq:excited-state-short}
\end{align}
Here \(\Delta_\pm\) denote ordinary gaps in the source-to-cut and cut-to-sink channels, while \(\delta^{(n,\cutC)}_\pm\) denote gaps of the replicated transfer problem.  A symmetric choice \(\tau_f=-\tau_i\) is convenient, but the double-sided projection in \cref{eq:double-sided-projection} requires both \(\tau_f\) and \(|\tau_i|\) to be large.  Using only \(T_{\rm sep}=\tau_f-\tau_i\) can hide an excited-state contamination caused by keeping one side close to the cut.  Since the target signal is expected to scale as \(1/L^3\), source-sink systematics must be controlled before interpreting the volume dependence.

\subsection{Measurement strategy}

A minimal first calculation is:
\begin{enumerate}
\item Work at fixed integer \(n=2\).
\item Use a rest-frame pion or nucleon source and sink with standard momentum projection.
\item Use the same cut prescription \(\cutC\) in the hadron and vacuum factors.
\item Implement the two-sheet cut graph of \cref{sec:n2-operational} and measure the ratio in \cref{eq:DeltaSn-correlator-ratio}.
\item Compare matched physical \(R\) across at least three spatial volumes for a production scaling study.
\item Vary \(\tau_f\) and \(|\tau_i|\) separately, or use symmetric choices supplemented by multistate analyses.
\item Average over ball centers and cubic rotations.
\item Extract \(\DeltaS^{\cutC}_{2,L}(B_R;h)\) and \(L^3\DeltaS^{\cutC}_{2,L}(B_R;h)\) with the full covariance across \(R\), \(L\), spin projection, source time, and sink time.
\item Use no-hadron, no-cut, full-volume, small-volume, lower-dimensional benchmark, quenched, and partially quenched tests to validate the replicated graph, contractions, and finite-volume analysis.
\end{enumerate}

The logarithm should be applied to the ensemble-averaged ratio,
\begin{align}
\mathcal R^{\cutC}_{n,L}(R;h,\lambda)
&=
\frac{
\left\langle
\prod_{k=1}^{n}C_{h,\lambda}^{(k)}
\right\rangle^{\cutC}_{n,B_R,L}
}{
\langle C_{h,\lambda}\rangle_{1,L}^{\,n}
},\nonumber\\
\DeltaS^{\cutC}_{n,L}
&=
\frac{1}{1-n}\ln \mathcal R^{\cutC}_{n,L}.
\lab{eq:log-ratio-practical}
\end{align}
After exact projection and exact ensemble averaging, the ratio in \cref{eq:log-ratio-practical} is the ratio of positive replica traces and is therefore real and positive.  Configuration-level estimators, stochastic estimates, baryon correlator products, and bootstrap samples need not be positive or real at finite statistics.  Negative or complex samples are statistical diagnostics and should be handled in the correlated error analysis, not by taking configuration-by-configuration logarithms.

\subsection{Volume fits}

At fixed physical \(R\), a practical scaling form is
\begin{align}
&\DeltaS^{\cutC}_{n,L,a,\tau_f,\tau_i}(B_R;h)
\nonumber\\
&\quad=
\frac{1}{L^3}
\Bigg[
\Fone^{\cutC}_{n}(R;h)
+
c_R^{\cutC}(R;h)
\left(\frac{R}{L}\right)^{\omega}
\nonumber\\
&\qquad
+
c_\pi^{\cutC}(R;h)e^{-m_\pi L}
+
c_a^{\cutC}(R;h)
\left(\frac{a}{R}\right)^p
\nonumber\\
&\qquad
+
c_+^{\cutC}(R;h)e^{-\Delta_+\tau_f}
+
c_-^{\cutC}(R;h)e^{-\Delta_-|\tau_i|}
+
\cdots
\Bigg].
\lab{eq:scaling-form-main}
\end{align}
A complementary exponent test is
\begin{align}
\DeltaS^{\cutC}_{n,L_i}(B_R;h)
=
A_n^{\cutC}(R;h)\,L_i^{-p(R)}
+\cdots ,
\lab{eq:free-exponent-fit}
\end{align}
with expected value \(p(R)=3\).

\section{Exploratory quenched lattice implementation}
\lab{sec:numerical}

We implement the source--sink measurement on five finite balls and empty/full controls in quenched SU(3), with $\beta=5.7$, degenerate Wilson valence quarks at $\kappa=0.12$, and two $16^3\times32$ sheets. The charged pseudoscalar meson has local effective mass $am_{\rm eff}\simeq2.07$. The Supplemental Material provides the complete nine-placement measurements and the supporting cut, contraction, and uncertainty definitions~\cite{Mamo:SupplementalMaterial}.\footnote{The Wilson-action scale $r_0/a=2.990(24)$~\cite{EdwardsHellerKlassen1998}, with $r_0=0.5\,\mathrm{fm}$, gives nominal values $a\simeq0.167\,\mathrm{fm}$, $L\simeq2.68\,\mathrm{fm}$ and $M_{\rm eff}\simeq2.44\,\mathrm{GeV}$. The five volume-equivalent radii span $0.20$--$0.83\,\mathrm{fm}$. The mass is a nominal conversion of the rest-channel effective mass; its kinetic counterpart requires a dispersion-relation measurement.}

\begin{figure*}[t]
\centering
\includegraphics[width=0.91\textwidth]{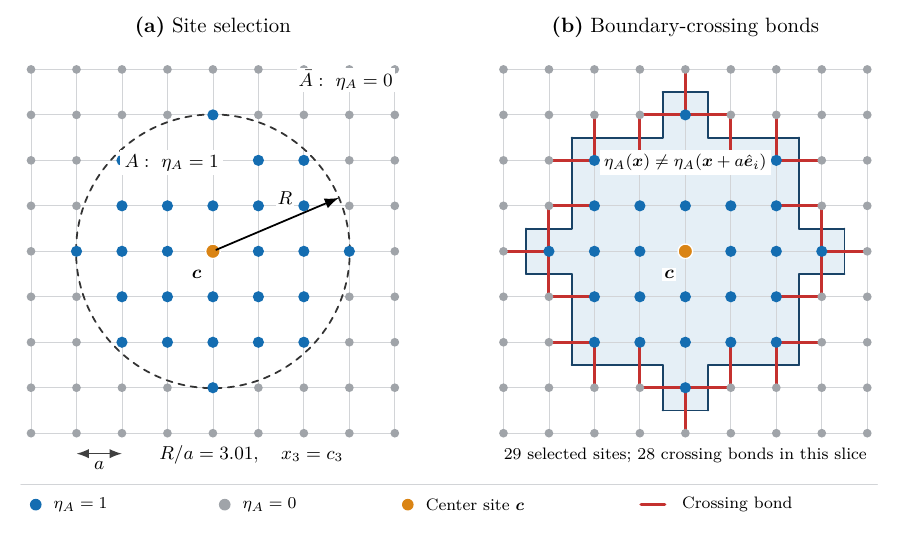}
\caption{Site-based implementation of the ball, shown in the two-dimensional central slice $x_3=c_3$ of the $R/a=3.01$ mask. (a) The dashed circle is the selection guide $|\bm x-\bm c|_{\rm per}=R$; sites satisfying the strict inequality are included. The orange marker is the center site $\bm c=(8,8,8)$ in lattice units. (b) Red nearest-neighbor bonds join sites with unequal membership $\eta_A(\bm x)=\mathbf1_{\bm x\in A}$. The shaded site-centered squares and their stepped outline visualize the same mask; they introduce no new boundary prescription. This slice contains 29 sites and 28 in-plane crossing bonds. The full three-dimensional mask is illustrated in SM Fig.~S2. In the two-sheet cut graph, a boundary-crossing spatial bond supports a unique eight-link seam face in $\mathcal P_8(A)$.}
\label{fig:lattice-ball-2d}
\end{figure*}

\begin{figure*}[t]
\centering
\includegraphics[width=0.87\textwidth]{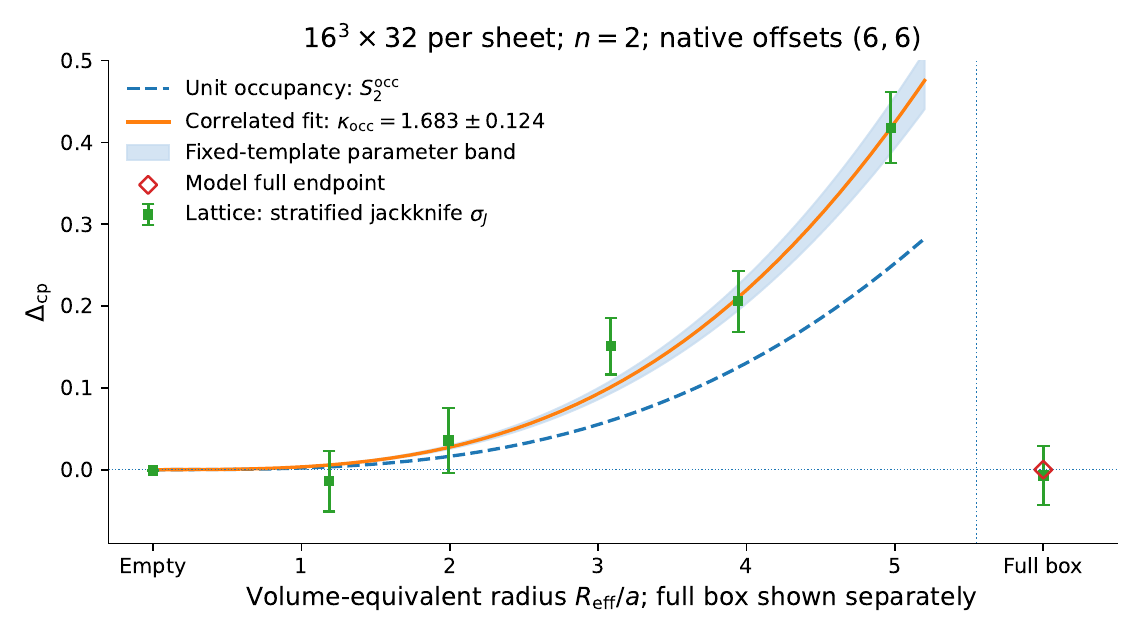}
\caption{Charged-pseudoscalar meson replica response for all five balls and empty/full controls at native offsets $(6,6)$, corresponding to link-centered distances $(\ell_-/a,\ell_+/a)=(5.5,6.5)$. The bars are stratified jackknife standard errors $\sigma_J$ under the within-history independence assumption. The unit-occupancy reference and the fitted normalization of Sec.~\ref{pair:lattice} are shown together. Both use the exact mask fractions; the continuous guide uses their volume-equivalent radius. The full box is a separate endpoint. The band propagates the fitted amplitude with the covariance and template held fixed.}
\lab{fig:radius-response}
\end{figure*}

\subsection{Central-plaquette prescription and estimator}
\lab{sec:cp-prescription}

We use the central-plaquette prescription $\mathcal C_{\rm cp}$ as an explicit Euclidean cut realization. Let $\mathcal P_4(A)$ denote the closed four-link faces on the two-sheet graph and $\mathcal P_8(A)$ the distinct eight-link faces at spatial bonds crossing the mask boundary. Counting each eight-link face once gives
\begin{align}
 S_{g,2}^{\rm cp}[U;A]
 &=\beta\sum_{p\in\mathcal P_4(A)}
 \left(1-\frac13\operatorname{ReTr}U_p\right)\nonumber\\
 &\quad+\frac\beta2\sum_{p\in\mathcal P_8(A)}
 \left(1-\frac13\operatorname{ReTr}U_p\right).
 \lab{eq:cp-action}
\end{align}
The central coefficient follows the $\beta/n$ prescription of Ref.~\cite{ChenDaiPang:1503.01766} at $n=2$. Temporal seam edges exchange sheets inside $A$ and preserve the sheet outside. Fermion seam phases are $-1$ on both complement closures, $+1$ on the region transition $0\to1$, and $-1$ on $1\to0$; reverse hops use the conjugate phase of the actual incoming edge. Each connected temporal trace cycle thus has one antiperiodic closure~\cite{FujimuraNishiokaShimamori:2309.11889}. The SU(3) action contains only the color-link products.

For the charged-pseudoscalar interpolator, the complete noise-pair contraction and measured response are
\begin{align}
 N_A&=A_{11}A_{22}+A_{12}A_{21}-2\operatorname{Re}B,\nonumber\\
 \Delta_{\rm cp}(A)&=-\ln\overline N_A+2\ln\overline C_{\rm ord}.
 \lab{eq:cp-estimator}
\end{align}
The contraction sheet labels are $1,2$. Here $A_{ij}$ is the squared wall-to-sink propagator norm from source sheet $j$ to sink sheet $i$, summed over sink sites, spin, color and diluted source columns. With $W_{ij}$ the corresponding twelve-column wall-solution matrix, $B=\Tr(W_{11}W_{21}^\dagger W_{22}W_{12}^\dagger)$. Products are formed within each independent wall pair, followed by noise and field averaging and then the logarithm. The ordinary mean averages both sheets of the matching empty-region fields and is shared by all regions. 

The balls satisfy the strict minimum-image rule $|\bm x-\bm c|_{\rm per}<R$, with $\bm c=(8,8,8)$ and $R/a=1.01,2.01,3.01,4.01,5.01$. Figure~\ref{fig:lattice-ball-2d} illustrates the site selection and boundary-crossing bonds in a central spatial slice; the corresponding three-dimensional view is in SM Fig.~S2. Each of the seven geometries has one hot-start and one cold-start history. An operational discard of 1000 combined heatbath and overrelaxation sweeps precedes observations at sweeps 1100, 1200, 1300 and 1400. Each observation uses sixteen independent $Z_4$ wall pairs with complete twelve-component spin--color dilution on each source sheet. The independently varied native offsets $d_-/a,d_+/a$ take values $4,6,8$. Their source and sink slices are $N_t-d_-/a$ and $d_+/a$, with the seam on the link $N_t-1\to0$. Link-centered distances are $\ell_-=d_--a/2$ and $\ell_+=d_++a/2$, so $T_{\rm sep}=d_-+d_+=\ell_-+\ell_+$. Figures retain the native labels; SM Sec.~S2 gives their reflection map. The collective study uses 56 gauge observations, with the complete placement dependence displayed in SM Fig.~S1.

\subsection{Finite-region response and occupancy comparison}
\lab{sec:radius-results}

Figure~\ref{fig:radius-response} and Table~\ref{tab:radius-source-errors} present the reference-placement responses obtained from 896 wall-pair records at native offsets $(6,6)$. The five balls contain 7, 33, 123, 257 and 515 of the 4096 spatial sites, with 30, 78, 174, 294 and 486 distinct boundary faces. Their volume-equivalent radii are approximately $1.187a$, $1.990a$, $3.085a$, $3.944a$ and $4.972a$.

\begin{table}[htbp]
\centering\small
\setlength{\tabcolsep}{5pt}
\caption{Reference-placement responses and the exact-mask occupancy reference. The primary uncertainty $\sigma_J$ is a stratified delete-one-field jackknife standard error, conditional on independent observations within each short history. The conditional source-noise comparison and covariance definition are given in SM Sec.~S3.}
\begin{tabular}{lrrr}\toprule
Region & $\Delta_{\rm cp}$ & $\sigma_J$ & $S_2^{\rm occ}$\\\midrule
Empty & $-0.00100$ & $0.00073$ & $0.00000$\\
$R/a=1.01$ & $-0.01384$ & $0.03725$ & $0.00342$\\
$R/a=2.01$ & $0.03577$ & $0.03936$ & $0.01611$\\
$R/a=3.01$ & $0.15089$ & $0.03452$ & $0.06002$\\
$R/a=4.01$ & $0.20610$ & $0.03724$ & $0.12513$\\
$R/a=5.01$ & $0.41779$ & $0.04337$ & $0.24827$\\
Full box & $-0.00726$ & $0.03594$ & $0.00000$\\
\bottomrule\end{tabular}
\label{tab:radius-source-errors}
\end{table}

The single-quasiparticle occupancy result~\cite{CastroAlvaredo2018PRL,CastroAlvaredo2018JHEP,CastroAlvaredo2019JMP}, represented in Appendix~\ref{app:renyi-occupancy}, supplies a parameter-free reference for the meson treated as one unresolved quasiparticle at these exact site fractions:
\[
 f_A=\frac{n_A}{4096},\qquad
 S_2^{\rm occ}(A)=-\ln\!\left[f_A^2+(1-f_A)^2\right].
\]
Table~\ref{tab:radius-source-errors} gives its values. The three largest responses are $2.51$, $1.65$ and $1.68$ times this reference. Its quasiparticle regime requires large region and complement sizes on the localization scale; a composite-particle description also resolves the hadron size. The finite-cutoff excess motivates finite-region, composite and cut-dependent studies, with quantum-chain corrections providing precedent~\cite{Zhang2021JSTAT}.

The primary covariance deletes one wall-averaged observation at a time within each of the fourteen histories, retaining equal hot/cold weights and all source/sink coordinates. Every deletion recomputes the regional means and shared ordinary denominator before taking logarithms. Summing the fourteen within-history jackknife contributions gives the full 63-coordinate covariance. This retains the correlation between the empty numerator and its normalization, and between every pair of regions. SM Sec.~S3 specifies this covariance and the separate fixed-field source-noise estimate.

The five finite-ball means increase with radius at every measured placement. At all eight non-midpoint placements, the three largest-ball means lie between $0.118$ and $0.433$ and exceed both controls; the full-control means lie between $-0.029$ and $0.002$. The half-period placement $(8,8)$, with $T_{\rm sep}=N_ta/2$, is included in SM Fig.~S1 and analyzed with the temporal controls in SM Sec.~S4. For regions measured at a common placement,
\begin{equation}
 \Delta_{\rm cp}(A_2)-\Delta_{\rm cp}(A_1)
 =-\ln\!\left(\frac{\overline N_{A_2}}{\overline N_{A_1}}\right),
 \lab{eq:radius-contrast}
\end{equation}
so the ordinary denominator cancels. At $(6,6)$ the contrasts between the three largest balls and the full control are
\[
 \begin{gathered}
 0.15816\pm0.02237,\quad 0.21336\pm0.02637,\\
 0.42505\pm0.03449,
 \end{gathered}
\]
with jackknife standard errors. These denominator-free contrasts show a positive large-region enhancement on the recorded sampling scale. Their shared full-region numerator remains in the covariance.

The smallest-ball mean is close to zero on its jackknife error scale. The five ball means rise through $R/a=5.01$, which occupies 12.6\% of the box. Intermediate larger regions would locate the turnover toward the full endpoint.

\paragraph{Scope.} The calculation samples a fixed-cutoff, finite-time, quenched charged-pseudoscalar meson response in $\mathcal C_{\rm cp}$. Jackknife errors estimate variation under independence of the four observations within each history. Observable autocorrelations, equilibration, state projection, sea effects and cutoff dependence require longer or matched studies. Source noise is already present in the field means; its separately reported conditional error is not added again. Fitted-parameter errors condition on the estimated covariance, and fit-window variation is reported separately. SM Secs.~S1, S3 and S4 give the complete placement scan, uncertainty definitions, and temporal controls. A ball/complement comparison and a regional-state construction test the relation of the implemented cut to the entropy target.

\section{Occupancy normalization and quark--antiquark interpretation}
\label{pair:lattice}

\begin{figure*}[t]
 \centering
 \includegraphics[width=0.78\textwidth]{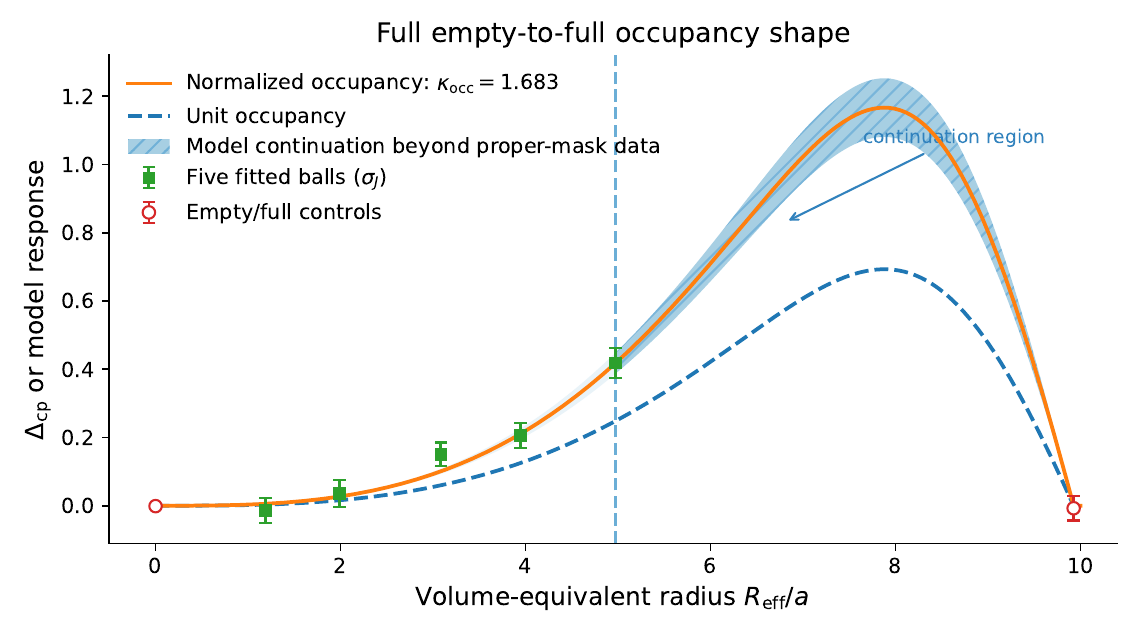}
 \caption{Continuous empty-to-full occupancy comparison at native offsets $(6,6)$. The five-ball fit gives $\kappa_{\rm occ}=1.683\pm0.124$ with the stratified jackknife covariance. All five proper masks and empty/full controls are shown; the controls are excluded from the fit. The shaded range is a model continuation beyond the largest measured proper mask, $f_A=0.125732$. The parameter band holds the occupancy shape and covariance fixed. Both model and data use the volume-equivalent coordinate $R_{\rm eff}/a$.}
 \label{hr:figcontinuous}
\end{figure*}

We summarize the enhancement by fitting all five proper balls in Table~\ref{tab:radius-source-errors} to
\begin{equation}
 \Delta_{{\rm cp},i}=\kappa_{\rm occ}S_2^{\rm occ}(f_i),
 \qquad f_i=N_{A,i}/4096.
 \label{pair:fitmodel}
\end{equation}
The replica index is $n=2$; $\kappa_{\rm occ}$ is an adjustable amplitude. The complete $5\times5$ jackknife covariance gives
\begin{equation}
 \boxed{\kappa_{\rm occ}=1.683\pm0.124,}
 \qquad q^2=10.61\quad(4\ \mathrm{dof}),
 \label{pair:fitresult}
\end{equation}
where $q^2=(y-\kappa s)^TC_J^{-1}(y-\kappa s)$ and $s_i=S_2^{\rm occ}(f_i)$. The error conditions on this estimated covariance and template. Its sampling assumptions are stated in Sec.~\ref{sec:numerical}; SM Sec.~S3 gives the covariance and fit definitions. The quadratic residual is concentrated near the middle radius and is used as a fit diagnostic.

Table~\ref{tab:occupancy-fit-summary} collects the radius-window and reference choices alongside the finite-mask models. The common ordinary normalization correlates the absolute responses; ball--full contrasts cancel it and use the full numerator as their shared reference. The rows are correlated analysis choices. Across the eight non-midpoint placements, fits using each placement's own covariance give $1.602$--$1.746$ for all five balls; SM Table~S3 gives the complete placement comparison.

\begin{table}[t]
\centering\small
\setlength{\tabcolsep}{1.9pt}
\caption{Fit comparisons at native offsets $(6,6)$ using the stratified jackknife covariance. The occupancy rows vary the data vector; the quark--antiquark and boundary models use all five balls. The final columns give the residual degrees of freedom and the numerical value of $q^2/\mathrm{dof}$. Errors on $\kappa_{\rm occ}$ condition on the covariance and template. The rows are correlated analysis choices; SM Secs.~S3 and S5 give the covariance and nonlinear-model profile ranges.}
\label{tab:occupancy-fit-summary}
\begin{tabular}{lcrr}
\toprule
Sample or model & Fitted parameter & dof & $q^2/\mathrm{dof}$\\
\midrule
Five balls (primary) & $\kappa_{\rm occ}=1.683\pm0.124$ & 4 & 2.65\\
Without seven-site mask & $\kappa_{\rm occ}=1.556\pm0.138$ & 3 & 2.06\\
Three largest balls & $\kappa_{\rm occ}=1.489\pm0.151$ & 2 & 2.48\\
Five balls + full & $\kappa_{\rm occ}=1.734\pm0.114$ & 5 & 2.34\\
Ball--full contrasts & $\kappa_{\rm occ}=1.728\pm0.116$ & 4 & 2.91\\
\midrule
Gaussian $q\bar q$ & $\langle r\rangle/a=3.706$ & 4 & 2.28\\
Exponential $q\bar q$ & $\langle r\rangle/a=3.871$ & 4 & 2.34\\
Boundary template & $\delta/a=1.001$ & 4 & 1.60\\
\bottomrule
\end{tabular}
\end{table}

\paragraph{Meson and quark--antiquark interpretation.}
An unresolved meson, treated as a single quasiparticle, has purity $p(f)=f^2+(1-f)^2$ and entropy $S_2^{\rm occ}$. To model its internal spatial structure, we use two distinguishable modes representing a valence quark and antiquark. When these modes are independently occupied and pure, their regional purity is $p(f)^2$ and their entropy is $2S_2^{\rm occ}$. A spatially correlated quark--antiquark pair interpolates between these reference limits through the chance that the boundary separates the constituents. This pure spatial toy model isolates the geometry; color, spin, and gluonic structure require the QCD matching outlined in Appendix~\ref{pair:qcd}. Appendix~\ref{hr:appendix} derives the exact finite-fraction pair purity and the leading-dilution coefficient
\begin{equation}
 \kappa_{\rm pair}(R)=2-\frac{\langle C_A\rangle_h}{V_A},
 \qquad 1\leq\kappa_{\rm pair}\leq2,
 \label{pair:main-overlap}
\end{equation}
where $C_A$ is the overlap of translated regions. Its large-ball correction to $L^3\Delta S_2$ is $\mathcal A_\Sigma\langle r\rangle_h/2$. The fitted constant in Eq.~\eqref{pair:fitresult} is a finite-time, finite-fraction summary; Eq.~\eqref{pair:main-overlap} has a specified dilute limit.

\paragraph{Gaussian and exponential quark--antiquark fits.}
The corresponding rows of Table~\ref{tab:occupancy-fit-summary} use normalized relative wavefunctions on the periodic spatial lattice,
\begin{align}
 \phi_G(\bm r;w)&=Z_G(w)^{-1/2}e^{-r_{\min}^2/(4w^2)},\nonumber\\
 \phi_E(\bm r;w)&=Z_E(w)^{-1/2}e^{-r_{\min}/(2w)},
 \label{pair:fit-wavefunctions}
\end{align}
where $r_{\min}=|\bm r|_{\rm per}$, $w$ is the width, and $\sum_{\bm r}|\phi_X(\bm r;w)|^2=1$ for $X=G,E$. The quark and antiquark positions enter through $\Psi_X(\bm x,\bm y)=N_s^{-3/2}\phi_X(\bm x-\bm y;w)$, with $N_s=16$. On each measured mask, the exact finite-fraction fitting function is
\begin{align}
 S_{2,X}^{\rm pair}(A;w)&=-\ln P_{2,X}(A;w),\nonumber\\
 P_{2,X}&=(1-2f_A+p_{11,X})^2+p_{11,X}^2
          +2\Tr K_{10,X}^2,
 \label{pair:fit-purity}
\end{align}
where $p_{11,X}=\sum_{\bm x,\bm y\in A}|\Psi_X(\bm x,\bm y)|^2$ and $K_{10,X}$ is the split-sector kernel in Eq.~\eqref{pair:kernel}, with the complement integral replaced by its periodic site sum. The factor two uses the real, even wavefunctions in Eq.~\eqref{pair:fit-wavefunctions}. SM Sec.~S5 specifies the finite-mask matrices.

For each shape, the single width $w_X^{\rm fit}$ minimizes the same correlated residual $q^2$, with $\kappa s$ replaced by the five values of $S_{2,X}^{\rm pair}(A;w)$. There is no independently fitted amplitude. Table~\ref{tab:occupancy-fit-summary} reports the resulting mean quark--antiquark separation,
\begin{equation}
 \frac{\langle r\rangle_X}{a}
 =\sum_{\bm r}\frac{r_{\min}}{a}
       |\phi_X(\bm r;w_X^{\rm fit})|^2.
 \label{pair:fit-mean-separation}
\end{equation}
The fitted widths are $w_G/a=2.3303$ and $w_E/a=1.3733$. The area and dilute expressions in Eqs.~\eqref{pair:area} and \eqref{pair:expansion} interpret the same separation at large $R$; the fits use the unexpanded purity in Eq.~\eqref{pair:fit-purity}.

Figure~\ref{hr:figfit} displays the four larger-mask ratios on a linear scale; its caption gives the seven-site result, which remains in every five-ball fit and in the complete ratio display of SM Fig.~S6. At the largest mask, the fitted Gaussian gives $S_2^{\rm pair}/S_2^{\rm occ}=1.671$, whereas its overlap coefficient is $1.519$. This illustrates why finite-fraction and dilute coefficients are distinguished. The separations characterize these meson toy wavefunctions; a QCD size comparison requires a specified gauge-invariant spatial correlator and state preparation.

A boundary-sensitive template provides a complementary description,
\begin{equation}
 \Delta_\delta(A)=\left(1+\frac{\delta}{R_{\rm eff}}\right)^3 S_2^{\rm occ}(f_A).
 \label{pair:boundary-template}
\end{equation}
Table~\ref{tab:occupancy-fit-summary} gives its fit on the same five masks. At small fraction and $\delta/R\ll1$, its area coefficient is $B_\Sigma=2\delta$. Equating this coefficient with the pair result $B_\Sigma=\langle r\rangle/2$ defines the leading-area matching condition $\delta=\langle r\rangle/4$. It relates the two parametrizations analytically; their higher-order geometric terms and finite-fraction predictions differ. This template is used over the measured radii; its continuation to the empty limit requires a separate prescription. Internal correlations, gluonic dressing and the regulated cut can all contribute to a boundary response. Related flux-tube tomography illustrates the geometric information obtained by severing an extended excitation~\cite{Amorosso2026Tomography}.

Figure~\ref{hr:figcontinuous} retains the occupancy continuation over the full fraction range using
\begin{equation}
 R_{\rm eff}/a=16\left(\frac{3f_A}{4\pi}\right)^{1/3}.
 \label{pair:radius}
\end{equation}
Both occupancy templates return to zero at $f_A=1$; their maxima at $f_A=1/2$ are $\ln2$ and $\kappa_{\rm occ}\ln2\simeq1.167$. The measurements cover only the five proper masks and endpoints shown.

The area derivation further motivates
\begin{equation}
 \Delta_{{\rm cp},i}=2f_i\kappa_V+
 \frac{4\pi(R_{{\rm eff},i}/a)^2}{4096}\frac{B_\Sigma}{a}.
 \label{pair:vsfit}
\end{equation}
Its two fitted coefficients have correlation $-0.980$ and $q^2=5.34$ for three residual degrees of freedom (Fig.~\ref{hr:figvs}). The supporting covariance, finite-mask calculations and profile ranges are given in SM Secs.~S3 and S5. Matched volumes, valence masses, spacings and shapes can separate the geometric contributions. Appendix~\ref{pair:qcd} outlines how QCD replica matrix elements determine their hadronic and cut-dependent content.

\section{Conclusions}
\lab{sec:conclusion}

We have defined a finite-volume lattice-QCD observable for the spatial R\'enyi response of a momentum-projected hadron.  At fixed \(\cutC\), integer \(n>1\), region \(B_R\), and spin projection \(\lambda\), the vacuum-subtracted entropy is computed from
\begin{align}
\DeltaS^{\cutC}_{n,L}(B_R;h,\lambda)
=
\frac{1}{1-n}
\ln
\frac{
\left\langle
\prod_{k=1}^{n}C_{h,\lambda}^{(k)}
\right\rangle^{\cutC}_{n,B_R,L}
}{
\langle C_{h,\lambda}\rangle_{1,L}^{\,n}
}.
\lab{eq:conclusion-main}
\end{align}
The numerator is a hadron source-sink correlator on the replicated cut graph.  The denominator is the corresponding product of ordinary source-sink correlators.  The formula is the direct lattice representation of the density-matrix replica trace, with all absolute partition functions canceled.  The observable remains prescription-defined: continuum extrapolation, when attempted, must hold the gauge-theory cut prescription and radius assignment fixed, and comparisons among prescriptions provide a useful systematic check.

For a box-normalized rest-frame hadron, the leading fixed-\(R\) response is expected to scale as \(1/L^3\).  The main physics output is therefore
\begin{align}
\Fone^{\cutC}_{n}(R;h)
=
\lim_{L\to\infty}
L^3\DeltaS^{\cutC}_{n,L}(B_R;h),
\end{align}
when the limit exists.  Its \(R\)-dependence can distinguish a smooth occupancy-type response from additional cut-localized QCD correlations.  A fixed \(n=2\) calculation provides the cleanest first numerical target.

The four-dimensional R\'enyi exponent is to be determined by a matched-volume lattice study.

The lattice response defined here can also serve as an anchor for phenomenological descriptions of hadron entanglement.  Effective pictures used in DIS, small-\(x\) evolution, rapidity-space entropy, string fragmentation, or infinite-momentum-frame wave packets should be viewed as approximations to, or limits of, a QCD reduced-state observable.  A measured \(R\)- and \(n\)-dependence of \(\Fone_n^{\cutC}(R;h)\) would provide nonperturbative information that such descriptions must reproduce after the appropriate state preparation, boost, and matching limits are specified.  Establishing that matching is separate from the Euclidean finite-volume definition, but the present construction supplies the lattice quantity to which those downstream frameworks can be compared.

The full-QCD implementation is conceptually direct: generate or reweight the replicated cut ensemble, compute the source-sink correlator product on that graph, and form the ratio in \cref{eq:conclusion-main}.  Quenched and partially quenched calculations develop the measurement strategy. The full-QCD observable includes both the replicated sea-quark determinant and valence contractions on the cut geometry.  Because determinant-ratio reweighting may face a severe overlap problem, precision full-QCD calculations should be viewed as replicated-ensemble calculations rather than ordinary two-point-function measurements.

The quenched calculation supplies a concrete finite-region realization on five balls and empty/full controls. The five ball means increase with radius at all nine source/sink placements. At the reference placement $d_-=d_+=6a$, the three largest responses are approximately $2.51$, $1.65$ and $1.68$ times the ideal occupancy reference, and their ball--full enhancements persist under the jackknife analysis with the ordinary denominator canceled. The collective full-covariance fit gives $\kappa_{\rm occ}=1.683\pm0.124$, with quadratic residual $q^2=10.61$ for four residual degrees of freedom. The coefficient describes the selected radius range; restricted-radius and reference choices are compared in Sec.~\ref{pair:lattice}.

The quark--antiquark toy model gives this normalization a mesonic interpretation. In the fixed-region dilution limit, the overlap of translated regions interpolates between unit occupancy for an unresolved meson and double occupancy for its separately resolved quark and antiquark. For a large nonwrapping ball, its leading correction is $\mathcal A_\Sigma\langle r\rangle_h/2$ in $L^3\Delta S_2$. The exact finite-mask pair fits and a boundary-length template provide complementary descriptions of the enhancement. Valence-mass and lattice-spacing studies, compared with an independently defined spatial correlator, can test the hadronic and regulator-dependent origins of the inferred length. Shape and complement measurements test the regional geometry. Appendix~\ref{hr:appendix} supplies the analytic derivation; the Supplemental Material supplies the finite-mask calculations, essential fit definitions, and temporal and source-noise controls. The scope stated in Sec.~\ref{sec:numerical} guides the next measurements: longer histories, controlled state preparation, matched volumes and cutoffs, and ball/complement tests.

\begin{acknowledgments}
I thank Christian Weiss and Peter Schweitzer for discussions. I also thank Rocco Amorosso, Sergey Syritsyn, and Raju Venugopalan for discussions. I thank Jefferson Lab for hospitality during the completion of this work. This work was supported by DOE Grant No. DE-FG02-04ER41309, NSF Grant No. 2412625, and DOE Award No. DE-SC002364 under the umbrella of the Quark-Gluon Tomography (QGT) Topical Collaboration.
\end{acknowledgments}

\section*{Data Availability}
The numerical results reported in this article are supported by the complete source/sink-placement means, reference-placement covariance, uncertainty definitions, and finite-mask model specifications in the Supplemental Material~\cite{Mamo:SupplementalMaterial}.

\appendix

\section{Kernel derivation}
\lab{app:kernel-derivation}

The source-sink-prepared hadron kernel is
\begin{align}
\rho_{h,\lambda,L}[\Phi_+,\Phi_-]
&=
\frac{1}{\mathcal N_h}
\int\mathcal D\Phi\;e^{-S_E[\Phi]}
\nonumber\\
&\quad\times
\Sh_{0,\lambda}[\Phi]
\Sh_{0,\lambda}^{\dagger}[\Phi]
\bigg|_{\substack{\Phi(0^-)=\Phi_-\\\Phi(0^+)=\Phi_+}}.
\lab{eq:rhoh-kernel-app}
\end{align}
Tracing out the complement gives
\begin{align}
&\rho_{\regA,h,\lambda,L}
[\Phi_{\regA,+},\Phi_{\regA,-}]
\nonumber\\
&\quad=
\int\mathcal D\Phi_{\regB}\;
\rho_{h,\lambda,L}
\big[
(\Phi_{\regA,+},\Phi_{\regB}),
(\Phi_{\regA,-},\Phi_{\regB})
\big].
\lab{eq:traceB-app}
\end{align}
For integer \(n\), cyclic gluing along \(\regA\) gives
\begin{align}
\Tr \rho_{B_R,h,\lambda,L}^{\,n}
&=
\frac{Z^{\cutC}_{n,L}[h_\lambda;B_R]}
{Z_{1,L}[h_\lambda]^n},
\nonumber\\
\Tr \rho_{B_R,0,L}^{\,n}
&=
\frac{Z^{\cutC}_{n,L}[0;B_R]}
{Z_{1,L}[0]^n}.
\lab{eq:kernel-traces-app}
\end{align}
Taking the ratio and using
\begin{align}
Z^{\cutC}_{n,L}[h_\lambda;B_R]
&=
Z^{\cutC}_{n,L}[0;B_R]
\left\langle
\prod_{k=1}^{n}
C_{h,\lambda}^{(k)}
\right\rangle^{\cutC}_{n,B_R,L},
\\
Z_{1,L}[h_\lambda]
&=
Z_{1,L}[0]\langle C_{h,\lambda}\rangle_{1,L}
\end{align}
gives \cref{eq:DeltaSn-correlator-ratio}.

\section{Gauge-link and fermion sheet maps}
\lab{app:sheet-maps}

One explicit sheet map is as follows.  Let \(k,l\in\mathbb Z_n\) and
\begin{align}
\eta_{\regA}(\bm x)
=
\begin{cases}
0, & \bm x\in\regB,\\
1, & \bm x\in\regA.
\end{cases}
\lab{eq:etaA-def-app}
\end{align}
A temporal gauge link crossing the cut from \(0^-\) to \(0^+\) is represented as
\begin{align}
U_{0,\cutC}^{(k)}(0^-,\bm x):
(k,0^-,\bm x)
\longrightarrow
(k+\eta_{\regA}(\bm x)\;{\rm mod}\;n,0^+,\bm x).
\lab{eq:gauge-link-sheet-map-app}
\end{align}
The conjugate link uses the inverse sheet map.  Plaquettes and improvement terms crossing the cut are built from these directed links on the replicated graph.

For a nearest-neighbor fermion operator, away from the cut,
\begin{align}
\big(D^{\cutC}_{n,L}[B_R]\big)_{(k,x),(l,y)}
=
D[U^{(k)}]_{xy}\,\delta_{kl}.
\lab{eq:replicated-dirac-away-app}
\end{align}
For a temporal hop crossing the cut at spatial point \(\bm x\), let \(H^{(k)}_{+}(0^-,\bm x)\) denote the usual forward temporal hopping block of the chosen Dirac operator on sheet \(k\), including the gauge link, spin projector, and hopping coefficient.  Then
\begin{align}
&\big(D^{\cutC}_{n,L}\big)_{(k,0^-,\bm x),(l,0^+,\bm x)}
\nonumber\\
&\qquad=
H^{(k)}_{+}(0^-,\bm x)
\begin{cases}
\delta_{l,k}, & \bm x\in\regB,\\
\delta_{l,k+1\;({\rm mod}\; n)}, & \bm x\in\regA.
\end{cases}
\lab{eq:replicated-dirac-forward-app}
\end{align}
The backward hop uses the corresponding backward hopping block and the inverse cyclic map on \(\regA\), while remaining on the same sheet on \(\regB\).  The spin projectors, improvement terms, and temporal boundary conditions are those of the chosen lattice action.

The limiting cases in \cref{eq:empty-full-check} also follow directly from this map.  For \(\regA=\emptyset\), \(\eta_{\regA}=0\) everywhere and the replicated graph factorizes.  For \(\regA\) equal to the full spatial volume, the cyclic sheet map acts on the entire time cut and computes the trace of the full normalized pure density matrix.

\section{Finite-cutoff R\'enyi expansion and occupancy check}
\lab{app:renyi-occupancy}

At fixed integer \(n>1\), assume the finite-cutoff reduced density matrices lie in a domain where the R\'enyi functional is differentiable and \(\Tr\rho_0^n\neq0\).  Let
\begin{align}
\rho_L
=
\rho_0+\epsilon_L\,\delta\rho,
\qquad
\epsilon_L\sim L^{-3},
\qquad
\Tr\delta\rho=0.
\end{align}
Then
\begin{align}
\Tr\rho_L^n
&=
\Tr\rho_0^n
+
n\epsilon_L
\Tr\left(\rho_0^{n-1}\delta\rho\right)
+
O(\epsilon_L^2),
\nonumber\\
S_n(\rho_L)-S_n(\rho_0)
&=
\epsilon_L
\frac{n}{1-n}
\frac{
\Tr\left(\rho_0^{n-1}\delta\rho\right)
}{
\Tr\rho_0^n
}
+
O(\epsilon_L^2).
\lab{eq:app-linearized-Renyi}
\end{align}
This is the finite-cutoff expansion used in \cref{sec:scaling}.

For the hadronic comparison, a meson whose internal structure is unresolved is treated as one quasiparticle. The single-quasiparticle excess entropy of Refs.~\cite{CastroAlvaredo2018PRL,CastroAlvaredo2018JHEP,CastroAlvaredo2019JMP} has the occupancy representation
\begin{align}
\rho^{\rm occ}_{B_R}
&=
f(R)|1\rangle\langle1|
+
[1-f(R)]|0\rangle\langle0|,
\nonumber\\
f(R)
&\simeq \frac{\Vol_R}{L^3}
=
\frac{4\pi R^3}{3L^3}.
\end{align}
For fixed integer \(n>1\), this two-level model gives exactly the single-particle form~\cite{CastroAlvaredo2018PRL}
\begin{align}
S^{\rm occ}_n(R,L)
&=
\frac{1}{1-n}
\ln\left[f(R)^n+(1-f(R))^n\right]
\nonumber\\
&=
\frac{n}{n-1}f(R)+O(f^2).
\lab{eq:occ-Sn-app}
\end{align}
The quasiparticle limit requires large region and complement scales,
\[
 \min(\xi,\lambda_{\rm dB})\ll\min(\ell_A,\ell_{\bar A}),
\]
with correlation length $\xi$ or de Broglie wavelength $\lambda_{\rm dB}=2\pi/|\bm p|$ supplying localization~\cite{CastroAlvaredo2018PRL,CastroAlvaredo2019JMP}. At rest the correlation-length condition applies. The cited large-region limit keeps $f$ fixed; the ideal model also supplies the fixed-$R$, small-$f$ benchmark below. An effective composite-particle description additionally resolves the hadron size. Quantum-chain corrections are discussed in Ref.~\cite{Zhang2021JSTAT}, with the occupancy comparison summarized in SM Sec.~S1.

Thus
\begin{align}
L^3S^{\rm occ}_n(R,L)
\longrightarrow
\frac{n}{n-1}\frac{4\pi R^3}{3}.
\lab{eq:occ-L3-limit-app}
\end{align}
For \(n=2\),
\begin{align}
L^3S^{\rm occ}_2(R,L)
\longrightarrow
\frac{8\pi R^3}{3}.
\end{align}
A contribution localized near the entangling surface over a correlation length \(\xi\) instead scales as
\begin{align}
\DeltaS^{\Sigma}_{n,L}
\sim
\frac{4\pi R^2\xi}{L^3}.
\end{align}
Both mechanisms vanish as \(1/L^3\), but they give different \(R\)-dependence in \(L^3\DeltaS_{n,L}\).

\section{Correlated finite-volume fitting}
\lab{app:correlated-fits}

Let
\begin{align}
y_{ij}
=
\DeltaS^{\cutC}_{n,L_j}(B_{R_i};h),
\qquad
F_i
=
\Fone^{\cutC}_{n}(R_i;h).
\end{align}
A simple fixed-exponent fit is
\begin{align}
y_{ij}
=
\frac{1}{L_j^3}
\left[
F_i
+
G_i\left(\frac{R_i}{L_j}\right)^\omega
+
H_i e^{-m_\pi L_j}
\right].
\lab{eq:app-fit-fixed}
\end{align}
With limited volumes, one should use the simplest stable version of this form.  A free-exponent test is
\begin{align}
y_{ij}
=
A_i L_j^{-p}
+
\cdots ,
\lab{eq:app-fit-p}
\end{align}
with expected value \(p=3\) for a one-particle state in three spatial dimensions. The four-dimensional fit should also be repeated with \(p\) fixed to \(3\) to extract \(\Fone_n(R;h)\).

For each bootstrap or jackknife sample, repeat the full fit and compute
\begin{align}
\Fone^{\cutC}_{n,L_j}(R_i;h)
=
L_j^3 y_{ij},
\qquad
\Fone^{\cutC}_{n}(R_i;h)
=
F_i.
\end{align}
The final uncertainty should include statistical covariance, fit-window dependence, covariance regularization, separate source-to-cut and cut-to-sink extrapolations, lattice spacing, and cut discretization.



\section{A quark--antiquark model for meson occupancy and boundary response}
\label{hr:appendix}
\providecommand{\HRMainInterpretationSection}{\ref{pair:lattice}}

The single-meson occupancy reference admits a simple spatial quark--antiquark extension: an entangling cut can leave the two valence constituents on opposite sides. An exact two-body partial trace relates this possibility to the overlap of translated regions. It gives an analytic interpolation between an unresolved meson and a resolved quark--antiquark pair, and an area-times-internal-separation correction for a large ball. The two distinguishable spatial modes represent the valence constituents in a toy model; the treatment of color, spin, and gluonic correlations belongs to the QCD matching in Sec.~\ref{pair:qcd}. Flux-tube entanglement tomography offers a related illustration of a boundary resolving an extended excitation~\cite{Amorosso2026Tomography}. These results supply the analytic basis for the occupancy-normalization discussion in Sec.~\HRMainInterpretationSection{}. The finite-mask implementation and fit definitions are supplied in SM Secs.~S3 and S5~\cite{Mamo:SupplementalMaterial}.

\begin{figure*}[t]
 \centering
 \includegraphics[width=0.74\textwidth]{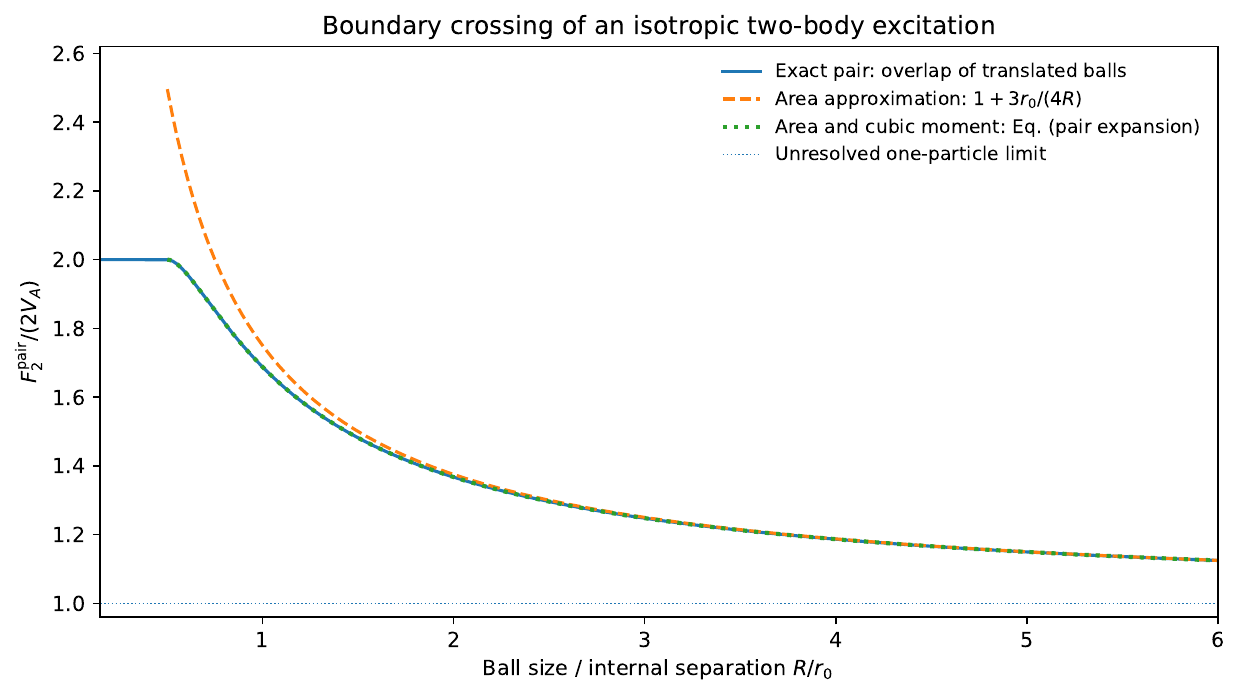}
 \caption{Analytic interpolation for the isotropic quark--antiquark meson toy model with fixed separation $r_0$. The exact piecewise overlap gives Eq.~\eqref{pair:interpolation}; the area term alone gives $1+3r_0/(4R)$. Including the cubic moment reproduces the exact result for $R\geq r_0/2$. The plotted ratio is the leading-dilution coefficient $F_2^{\rm pair}/(2V_A)$, with resolved-quark--antiquark limit two and unresolved-meson limit one.}
 \label{hr:figpair}
\end{figure*}

\subsection{From single-meson occupancy to quark--antiquark purity}
\label{pair:purity}

Write $V=L^3$, $f_A=V_A/V$, and $p(f)=f^2+(1-f)^2$. In the effective single-meson quasiparticle description, the two regional Schmidt weights are $f_A$ and $1-f_A$, so
\begin{equation}
 P_2^{\rm occ}(A)=p(f_A),\qquad
 S_2^{\rm occ}(f_A)=-\ln p(f_A).
 \label{pair:occ}
\end{equation}
This is the single-quasiparticle excess-entropy reference of Refs.~\cite{CastroAlvaredo2018PRL,CastroAlvaredo2018JHEP,CastroAlvaredo2019JMP}, with the localization regime stated in Appendix~C. Two distinguishable, independently occupied pure modes instead have a tensor-product regional state, giving $P_2=p(f_A)^2$ and $S_2=2S_2^{\rm occ}$. These provide the unit and double-occupancy limits.

To resolve the meson's internal spatial structure, consider a pure quark--antiquark toy state with normalized wave function $\Psi(\mathbf x,\mathbf y)$, where $\mathbf x$ and $\mathbf y$ label the quark and antiquark positions, fixed constituent numbers, explicit spatial tensor factors, and a Fock-vacuum reference. After tracing $\bar A$, the regional state separates into orthogonal $(1,1),(1,0),(0,1),(0,0)$ occupation sectors, with probabilities $p_{ij}$. The both-inside block is a pure subnormalized wave function, while the both-outside block is the regional vacuum. A split block has kernel
\begin{equation}
 K_{10}(\mathbf x,\mathbf x')=
 \int_{\bar A}d^3y\,
 \Psi(\mathbf x,\mathbf y)\Psi^*(\mathbf x',\mathbf y),
 \quad \mathbf x,\mathbf x'\in A.
 \label{pair:kernel}
\end{equation}
Define $\gamma_{10}=\operatorname{Tr}(K_{10}/p_{10})^2$ and similarly $\gamma_{01}$, with zero-weight terms understood by continuity. Orthogonality of the sectors gives the exact purity
\begin{equation}
 P_2(A)=p_{00}^2+p_{11}^2+
 p_{10}^2\gamma_{10}+p_{01}^2\gamma_{01}.
 \label{pair:sector}
\end{equation}
The split-sector purities retain the internal coherence left by the partial trace.

For a homogeneous state
\begin{equation}
 \Psi(\mathbf x,\mathbf y)=V^{-1/2}\varphi(\mathbf x-\mathbf y),
 \qquad \int_Vd^3r\,|\varphi(\mathbf r)|^2=1,
 \label{pair:wavefunction}
\end{equation}
both one-body marginals are uniform. Let $s_A=p_{10}+p_{01}$ be the probability of a split pair. Then $p_{10}=p_{01}=s_A/2$, $p_{11}=f_A-s_A/2$, and $p_{00}=1-f_A-s_A/2$. Therefore
\begin{align}
 P_2(A)&=p(f_A)-s_A+
 \frac{s_A^2}{4}(2+\gamma_{10}+\gamma_{01}),
 \label{pair:exactpurity}\\
 S_2^{\rm pair}(A)&=S_2^{\rm occ}(f_A)+
 \frac{s_A}{p(f_A)}+O(s_A^2).
 \label{pair:entropy}
\end{align}
The first correction is positive and follows from the density-matrix calculation. An unresolved pointlike meson has $s_A\to0$ and the unit-occupancy response. Independent uniform quark and antiquark modes have $s_A=2f_A(1-f_A)$ and $\gamma_{10}=\gamma_{01}=1$, recovering the exact double-occupancy result. A localized quark--antiquark pair approaches the unresolved-meson limit when the region is large relative to its internal separation.

\begin{figure*}[t]
 \centering
 \includegraphics[width=0.74\textwidth]{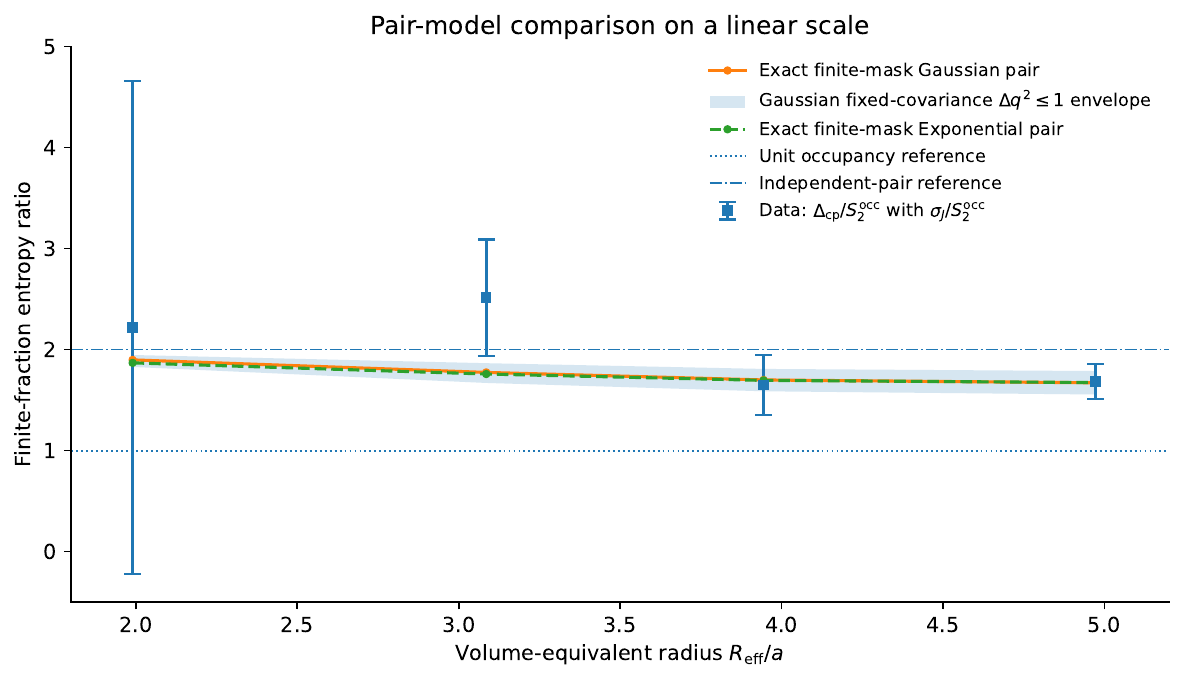}
 \caption{Finite-fraction entropy ratios for the four larger proper masks on a linear axis, with complete jackknife error bars and exact finite-mask quark--antiquark model predictions. The Gaussian band spans the fixed-covariance profile $q^2\leq q^2_{\min}+1$; connecting markers guide the eye. The seven-site mask at input radius $R/a=1.01$ gives $-4.05\pm10.90$ and remains in Table~\ref{tab:radius-source-errors}, Fig.~\ref{fig:radius-response}, and every five-ball fit. SM Fig.~S6 displays all five ratios with their full errors. Lines at one and two denote the unit-occupancy and independent-mode reference values.}
 \label{hr:figfit}
\end{figure*}

\begin{figure*}[t]
 \centering
 \includegraphics[width=0.74\textwidth]{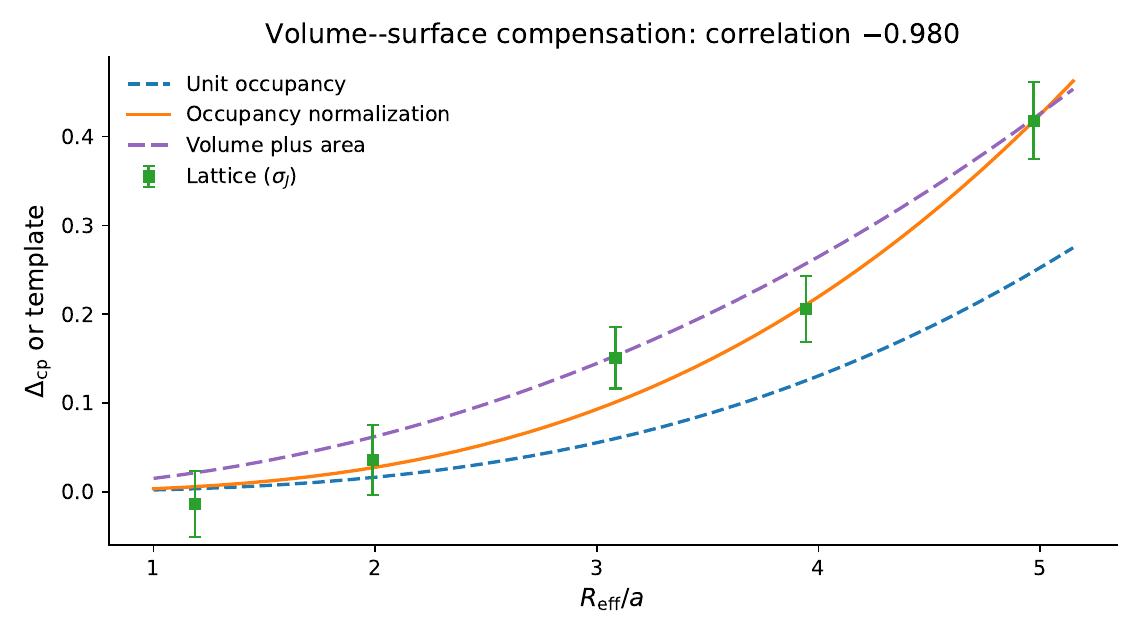}
 \caption{The same data and jackknife covariance compared with the occupancy-amplitude and volume-plus-surface templates. The display is restricted to the measured radius range. The bulk--surface correlation $-0.980$ identifies the parameter combination constrained by this sample. Fit coefficients and uncertainty conventions are given in SM Secs.~S3 and S5. The plotted bars are $\sigma_J$ under the stated independence assumption.}
 \label{hr:figvs}
\end{figure*}

\subsection{Boundary crossing, the area term, and the analytic interpolation}
\label{pair:geometry}

Let $w_h(\mathbf r)=|\varphi(\mathbf r)|^2$ and let $\chi_A$ be the region indicator. The split probability is
\begin{align}
 s_A&=\frac1V\int d^3r\,w_h(\mathbf r)\int d^3X\,
 [\chi_++\chi_--2\chi_+\chi_-],\nonumber\\
 \chi_\pm&=\chi_A(\mathbf X\pm\mathbf r/2).
 \label{pair:split}
\end{align}
The center integral is the volume of the symmetric difference of two translated copies of $A$. Defining their overlap by
\begin{equation}
 C_A(\mathbf r)=\int d^3x\,\chi_A(\mathbf x)\chi_A(\mathbf x+\mathbf r),
 \label{pair:overlap}
\end{equation}
one obtains
\begin{equation}
 s_A=\frac2V\big[V_A-\langle C_A(\mathbf r)\rangle_h\big].
 \label{pair:splitoverlap}
\end{equation}
At fixed $A$ and finite internal size, $f_A,s_A=O(V^{-1})$. Expanding Eq.~\eqref{pair:exactpurity} before the logarithm gives
\begin{align}
 F_2^{\rm pair}(A)&\equiv\lim_{V\to\infty}V S_2^{\rm pair}(A)
 =4V_A-2\langle C_A\rangle_h,\nonumber\\
 \kappa_{\rm pair}(A)&\equiv\frac{F_2^{\rm pair}(A)}{2V_A}
 =2-\frac{\langle C_A\rangle_h}{V_A}.
 \label{pair:kappa}
\end{align}
Since $0\leq C_A\leq V_A$, the leading-dilution coefficient obeys
\begin{equation}
 \boxed{1\leq\kappa_{\rm pair}\leq2.}
 \label{pair:bound}
\end{equation}
Coincident translated regions give one; disjoint translated regions give two. Thus the interpolation measures whether the cut treats the quark--antiquark pair as one localized meson or resolves its two spatial constituents. Equivalently, $\kappa_{\rm pair}=\lim_{V\to\infty}S_2^{\rm pair}/S_2^{\rm occ}$ at fixed region. This bound applies to the specified pure fixed-pair model with its Fock-vacuum reference.

The area factor has a local origin. A displacement $\mathbf r$ changes membership over a boundary thickness $|\mathbf r\cdot\hat{\mathbf n}|$:
\begin{equation}
 |A\triangle(A+\mathbf r)|
 =\int_{\partial A}dS\,|\mathbf r\cdot\hat{\mathbf n}|+\cdots.
 \label{pair:localboundary}
\end{equation}
For an isotropic internal density, $\langle|\mathbf r\cdot\hat{\mathbf n}|\rangle=\langle r\rangle/2$, and hence $s_A=\mathcal A_\Sigma\langle r\rangle/(2V)+\cdots$. The surface area counts possible center positions along the cut, and the internal separation supplies the thickness of the layer in which the pair is split.

For a nonwrapping ball, the entire overlap is elementary. Integrating the two spherical caps gives
\begin{align}
 C_R(r)&=2\pi\int_0^{R-r/2}dz\,[R^2-(z+r/2)^2]\nonumber\\
 &=\frac{4\pi R^3}{3}-\pi R^2r+\frac{\pi r^3}{12},
 \quad 0\leq r\leq2R,
 \label{pair:sphereoverlap}
\end{align}
with $C_R(r)=0$ for $r>2R$. When $w_h$ is supported at $r<2R$, Eqs.~\eqref{pair:kappa} and \eqref{pair:sphereoverlap} yield
\begin{align}
 \boxed{F_2^{\rm pair}(B_R)
 =2V_A+\frac{\mathcal A_\Sigma}{2}\langle r\rangle_h
 -\frac\pi6\langle r^3\rangle_h,}
 \label{pair:area}\\
 \kappa_{\rm pair}(R)
 =1+\frac{3\langle r\rangle_h}{4R}
 -\frac{\langle r^3\rangle_h}{16R^3}.
 \label{pair:expansion}
\end{align}
For a distribution with longer tails, averaging the piecewise overlap gives the exact leading-dilution result. The first term beyond area in Eq.~\eqref{pair:area} is independent of $R$; the spherical overlap produces no term proportional to $R$ at this order.

For fixed isotropic separation $r_0$, the interpolation is explicit:
\begin{equation}
 \kappa_{\rm pair}(R;r_0)=
 \begin{cases}
 2,&0<R\leq r_0/2,\\[2pt]
 1+\dfrac{3r_0}{4R}-\dfrac{r_0^3}{16R^3},&R\geq r_0/2.
 \end{cases}
 \label{pair:interpolation}
\end{equation}
Figure~\ref{hr:figpair} displays this continuous crossover. At $R\ll r_0$, the ball can contain either constituent separately. At $R\gg r_0$, almost all centers put the complete pair inside or outside, and the boundary correction decreases as $r_0/R$.

\subsection{How QCD determines the coefficients}
\label{pair:qcd}

The pair calculation gives a microscopic example of
\begin{equation}
 F_2^{\mathcal C}(B_R;h)
 =2V_A\kappa_V^{\mathcal C}
 +\mathcal A_\Sigma B_\Sigma^{\mathcal C}+\cdots,
 \label{pair:organization}
\end{equation}
with $\kappa_V=1$ and $B_\Sigma=\langle r\rangle_h/2$ in the unresolved-pair regime. More generally, the leading pure-cluster model gives twice the mean volume of the union $\bigcup_j(A-\mathbf r_j)$. Its multiparton inclusion--exclusion terms recover unit occupancy for an unresolved bound cluster. This identifies both internal correlations and their geometry as the dynamical information needed beyond the single-particle reference.

In the QCD replica construction, a vacuum-normalized twist-defect matrix element supplies that information~\cite{hr:Cardy2008,hr:Hung2014,hr:Bianchi2016}. Matching can organize its response through ordinary hadron matrix elements, multiparton kernels, and operators intrinsic to the cut. Selected local quark/gluon towers give PDF moments; suitable nonlocal and off-forward preparations connect to GPDs and GTMDs~\cite{Diehl:2003ny,hr:Meissner2008}. A single ball-integrated insertion between identical momentum eigenstates selects its forward matrix element. Transfer-dependent structure enters through off-forward preparation, spatial localization, or multiple insertions with compensating transfers. Null-limit replica-defect expansions provide a route to light-ray structures with the replica number and operator conventions specified~\cite{hr:Balakrishnan2019}.

Determining the QCD coefficients calls for operator-inserted replica measurements with the state, cut, and renormalization conventions matched. The finite-mask calculation in SM Sec.~S5 supplies the explicit toy-model realization used here. The immediate analytic result is the geometric crossover: unit occupancy for an unresolved meson, double occupancy for a resolved quark--antiquark pair in the dilute model, and a calculable area correction between them. The lattice fits quantify an effective response on which this QCD matching program can be built.

\FloatBarrier

%

\onecolumngrid
\clearpage
\begingroup
\fontsize{11}{13.3}\selectfont
\setlength{\parindent}{1.1em}
\setlength{\parskip}{3pt}
\makeatletter
\@removefromreset{equation}{section}
\let\@hangfrom@section\ArxivNormalHangFrom
\let\@sectioncntformat\ArxivNormalSectionCntFormat
\let\ArxivOriginalLabel\label
\def\label#1{\def\ArxivLabelArg{#1}\def\ArxivLastBib{LastBibItem}%
\ifx\ArxivLabelArg\ArxivLastBib\ArxivOriginalLabel{SMLastBibItem}%
\else\ArxivOriginalLabel{#1}\fi}
\makeatother
\setcounter{section}{0}
\setcounter{subsection}{0}
\setcounter{equation}{0}
\setcounter{figure}{0}
\setcounter{table}{0}
\setcounter{footnote}{0}
\renewcommand{\thesection}{S\arabic{section}}
\renewcommand{\thesubsection}{\thesection.\arabic{subsection}}
\renewcommand{\theequation}{S\arabic{equation}}
\renewcommand{\thefigure}{S\arabic{figure}}
\renewcommand{\thetable}{S\arabic{table}}
\renewcommand{\theHsection}{supp.\arabic{section}}
\renewcommand{\theHsubsection}{supp.\arabic{section}.\arabic{subsection}}
\renewcommand{\theHequation}{supp.\arabic{equation}}
\renewcommand{\theHfigure}{supp.\arabic{figure}}
\renewcommand{\theHtable}{supp.\arabic{table}}
\providecommand{\Dcp}{\Delta_{\rm cp}}
\providecommand{\Socc}{S_2^{\rm occ}}
\providecommand{\CJ}{C_J}
\renewcommand{\refname}{Supplemental references}
\pdfbookmark[0]{Supplemental Material}{supplemental-material}

\begin{center}
{\large\bfseries Supplemental Material for}\\[4pt]
{\Large\bfseries Momentum-projected hadron entanglement\\from lattice-QCD replica correlators}\\[8pt]
Kiminad A.~Mamo\\[-1pt]
{\small Department of Physics, University of Connecticut, Storrs, CT 06269-3046, USA\\\texttt{ska25005@uconn.edu}}
\end{center}

The figures and definitions below complete the five-radius charged-pseudoscalar meson study: source/sink dependence, the cut and contraction, uncertainties, temporal controls, and the finite-mask pair-model fits.

\section{Measured regions and source--sink dependence}
\label{sm:results}

For a regulated regional state construction with prescription $\mathcal C$, the vacuum-subtracted second-R\'enyi response is
\begin{equation}
 \Delta S_{2,L}^{\mathcal C}(A;h)
 =-\ln\frac{\Tr[(\rho_{A,h}^{\mathcal C})^2]}{\Tr[(\rho_{A,0}^{\mathcal C})^2]}
 =-\ln\frac{\langle C_h^{(1)}C_h^{(2)}\rangle_{2,A}^{\mathcal C}}
 {\langle C_h\rangle_1^2},
 \label{sm:target}
\end{equation}
after source-side, sink-side and exterior-vacuum projection. Here $h$ is a unit-normalized finite-volume state and both reduced density matrices have unit trace. At the measured finite times, the central-plaquette calculation reports
\begin{equation}
 \Dcp(A;m_-,m_+)=-\ln\overline N_A+2\ln\overline C_{\rm ord},
 \label{sm:response}
\end{equation}
with the numerator, ordinary mean and order of averages specified in Sec.~\ref{sm:cut}. The cut convention is held fixed throughout.

The calculation uses quenched SU(3), Wilson gauge coupling $\beta=5.7$, degenerate Wilson valence parameter $\kappa=0.12$, and two $16^3\times32$ sheets. The charged-pseudoscalar meson effective mass is $am_{\rm eff}\simeq2.07$. Each of the seven geometries has a hot-start and a cold-start history. After an operational discard of 1000 combined heatbath and overrelaxation sweeps, four fields per history are measured at sweeps 1100, 1200, 1300 and 1400. Each field and source distance uses sixteen independent $Z_4$ wall pairs with twelve-component spin--color dilution. The sea-quark determinant is omitted. These parameters specify a finite-time, fixed-cutoff calculation; the temporal and sampling controls are described below.

The strict periodic masks obey $|\bm x-(8,8,8)|_{\rm per}<R$. At input radii $R/a=1.01,2.01,3.01,4.01,5.01$, their site counts are $n_A=7,33,123,257,515$, and their boundary-face counts are $30,78,174,294,486$. The reference uses the exact fraction $f_A=n_A/4096$:
\begin{equation}
 \Socc(f_A)=-\ln[f_A^2+(1-f_A)^2].
 \label{sm:occupancy}
\end{equation}
For an unresolved meson treated as one quasiparticle, this is the two-level single-occupancy reference~\cite{smref:CastroAlvaredo2018PRL,smref:CastroAlvaredo2018JHEP,smref:CastroAlvaredo2019JMP}. Its quasiparticle interpretation requires the region and complement to be large on the relevant correlation-length or wavelength scale; the intrinsic size additionally enters an unresolved composite-hadron description. Finite-scale corrections are illustrated in Ref.~\cite{smref:Zhang2021JSTAT}. The volume coordinate $R_{\rm eff}/a=(3n_A/4\pi)^{1/3}$ is used when continuous curves are compared with the discrete masks. Empty and full controls have $f_A=0$ and $1$.

\begin{figure}[p]
\centering
\includegraphics[width=.327\textwidth]{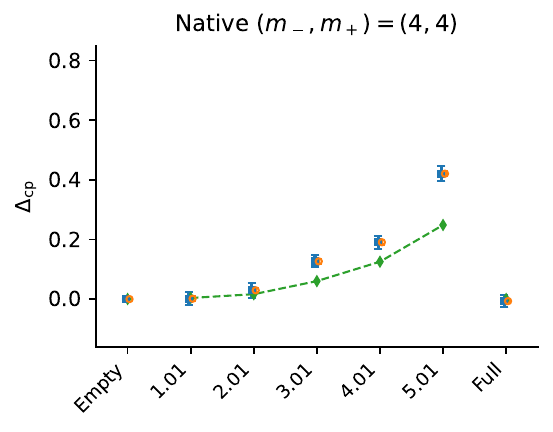}\hfill
\includegraphics[width=.327\textwidth]{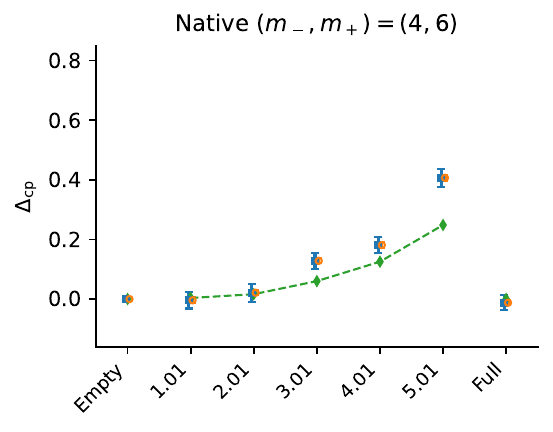}\hfill
\includegraphics[width=.327\textwidth]{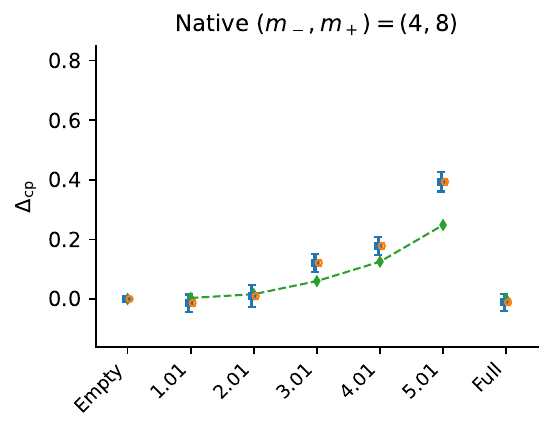}\\[3pt]
\includegraphics[width=.327\textwidth]{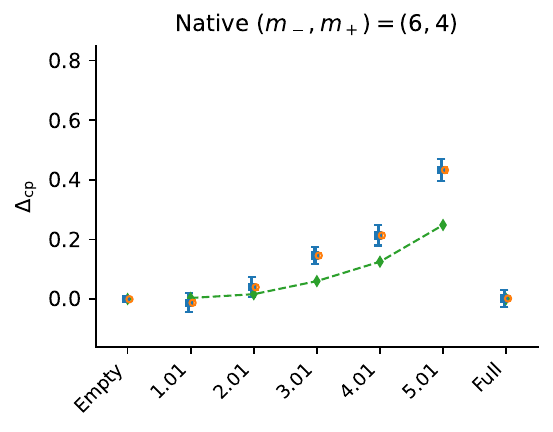}\hfill
\includegraphics[width=.327\textwidth]{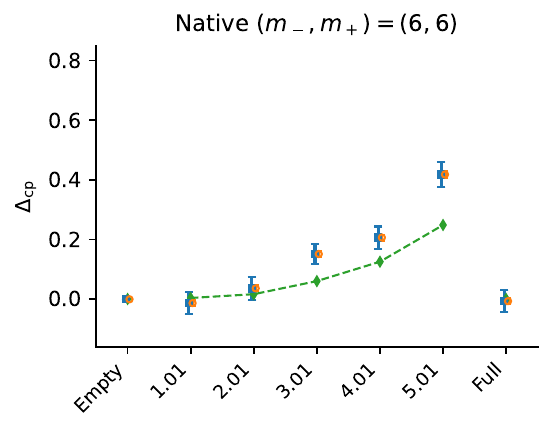}\hfill
\includegraphics[width=.327\textwidth]{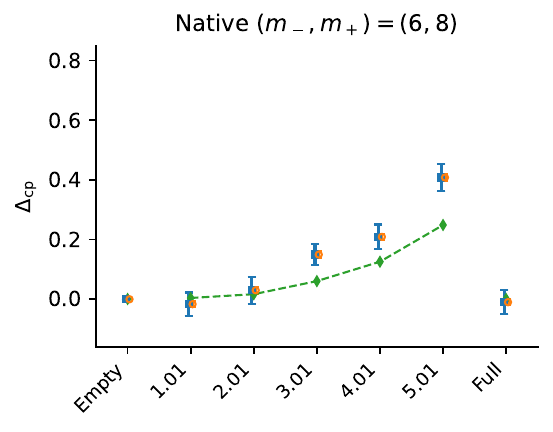}\\[3pt]
\includegraphics[width=.327\textwidth]{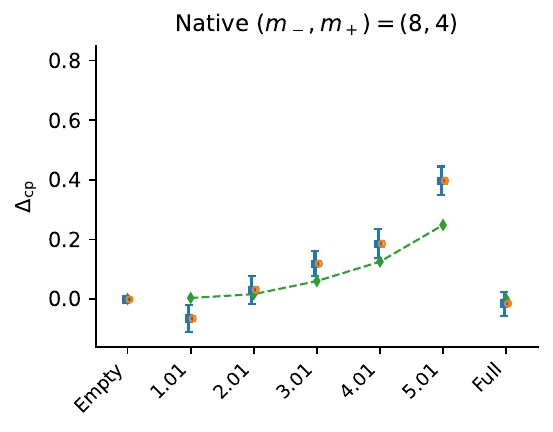}\hfill
\includegraphics[width=.327\textwidth]{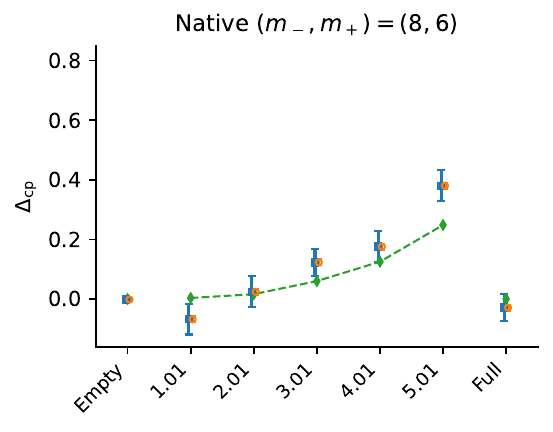}\hfill
\includegraphics[width=.327\textwidth]{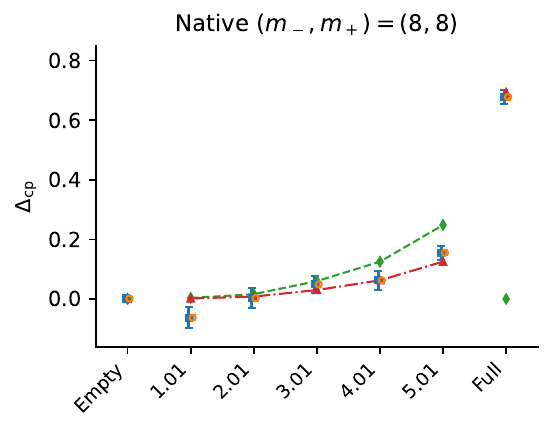}
\caption{Complete five-ball study with empty/full controls at all nine native source/sink offsets. Filled squares carry the primary stratified-jackknife errors; open circles carry conditional source-noise errors. Dashed diamonds give the unit-occupancy reference. The last panel also shows the equal-mixture occupancy guide of Sec.~\ref{sm:temporal}, with full endpoint $\ln2$. Finite-ball labels are input $R/a$; the full box is separate. The same fields and common-source sink measurements correlate the panels. Native offsets and link-centered distances are related in Sec.~\ref{sm:cut}.}
\label{sm:fig-grid}
\end{figure}

The five finite-ball means increase with radius in every panel of Fig.~\ref{sm:fig-grid}. Table~\ref{sm:tab-grid} lists the 63 regional means across the nine source/sink placements. The half-period point $(8,8)$ is included alongside the other placements and analyzed in Sec.~\ref{sm:temporal}. All responses use their directly measured normalization.

\begin{table}[htbp]
\centering\small
\setlength{\tabcolsep}{4.0pt}
\caption{Complete seven-region means. Native offsets $m_\pm$ are integers. Column headings for proper balls are input radii $R/a$. Figure~\ref{sm:fig-grid} gives the corresponding errors; Sec.~\ref{sm:uncertainties} specifies their meaning.}
\label{sm:tab-grid}
\begin{tabular}{crrrrrrr}
\toprule
$(m_-,m_+)$ & Empty & 1.01 & 2.01 & 3.01 & 4.01 & 5.01 & Full\\\midrule
$(4,4)$ & $-0.00037$ & $0.00096$ & $0.02907$ & $0.12631$ & $0.19041$ & $0.42048$ & $-0.00706$\\
$(4,6)$ & $-0.00037$ & $-0.00440$ & $0.02103$ & $0.12809$ & $0.18087$ & $0.40655$ & $-0.01245$\\
$(4,8)$ & $-0.00005$ & $-0.01359$ & $0.00935$ & $0.12128$ & $0.17820$ & $0.39312$ & $-0.01112$\\
$(6,4)$ & $-0.00073$ & $-0.01269$ & $0.03959$ & $0.14561$ & $0.21318$ & $0.43256$ & $0.00171$\\
$(6,6)$ & $-0.00100$ & $-0.01384$ & $0.03577$ & $0.15089$ & $0.20610$ & $0.41779$ & $-0.00726$\\
$(6,8)$ & $-0.00092$ & $-0.01735$ & $0.02933$ & $0.14903$ & $0.20848$ & $0.40765$ & $-0.01064$\\
$(8,4)$ & $-0.00156$ & $-0.06589$ & $0.03080$ & $0.11835$ & $0.18527$ & $0.39686$ & $-0.01565$\\
$(8,6)$ & $-0.00170$ & $-0.06813$ & $0.02413$ & $0.12268$ & $0.17562$ & $0.38000$ & $-0.02866$\\
$(8,8)$ & $0.00153$ & $-0.06278$ & $0.00263$ & $0.04921$ & $0.06278$ & $0.15522$ & $0.67755$\\
\bottomrule
\end{tabular}
\end{table}
\FloatBarrier

\section{Cut prescription and charged-pseudoscalar contraction}
\label{sm:cut}

Figure~\ref{sm:fig-ball3d} shows the spatial mask and its boundary-crossing bonds before the temporal gluing in Eq.~\eqref{sm:seam}. The center is the lattice site $\bm c=(8,8,8)$. The central-plaquette prescription below refers to seam faces associated with the boundary bonds, rather than a distinguished spatial cube at the ball center.

\begin{figure}[!htbp]
\centering
\includegraphics[width=0.94\textwidth]{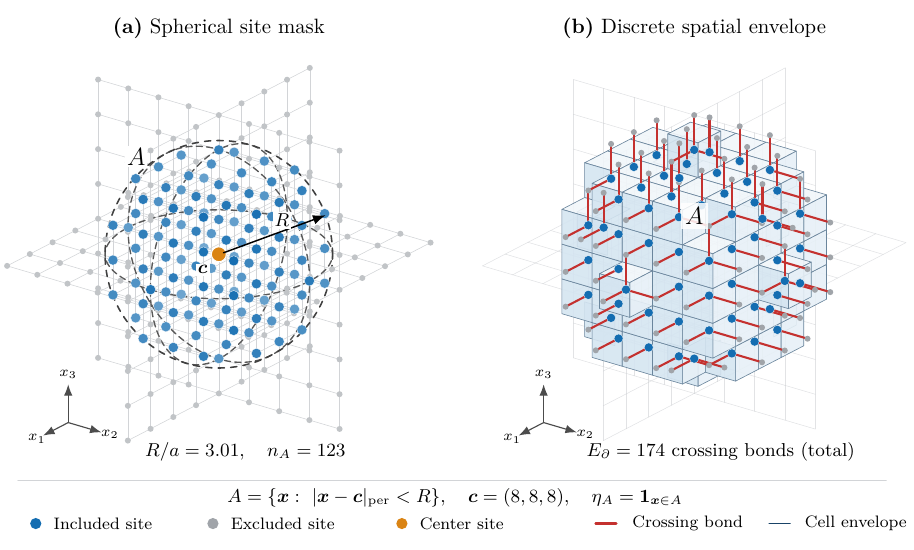}
\caption{Three-dimensional illustration of the strict periodic site mask at $R/a=3.01$. (a) Included sites obey $|\bm x-\bm c|_{\rm per}<R$, with the center site marked in orange and dashed spherical guides indicating the input radius. (b) Site-centered cubes visualize the same 123-site set; their exposed faces form a display envelope halfway along boundary-crossing bonds. Red bonds connect sites with different $\eta_A$. For readability, the right view displays the foreground bonds and surface sites; the complete mask has $E_\partial=174$ boundary-crossing bonds. These correspond to 174 unique eight-link seam faces at $n=2$. The site-cell envelope is only a spatial visualization; Eq.~\eqref{sm:action} defines the cut action. Both panels show a local portion of the lattice, with periodic copies omitted.}
\label{sm:fig-ball3d}
\end{figure}

Write $v=(k,t,\bm x)$, with sheet $k=0,1$ and $0\leq t<32$. Spatial edges preserve the sheet. The temporal seam connects
\begin{equation}
 (k,31,\bm x)\longmapsto(k\mathbin{\mathrm{xor}}\eta_A(\bm x),0,\bm x),
 \qquad \eta_A(\bm x)=\mathbf1_{\bm x\in A}.
 \label{sm:seam}
\end{equation}
Every positive edge carries an SU(3) link and the reverse carries its Hermitian conjugate. At a spatial bond whose endpoint masks agree, the seam walk $+i,+0,-i,-0$ gives one four-link face per sheet. For unequal endpoint masks, two walks close an eight-link face; starting on the other sheet traces that same face cyclically. Counting each distinct eight-link face once, the normalized cut measure is proportional to $\exp(-S_{g,2}^{\rm cp})$, with
\begin{equation}
 S_{g,2}^{\rm cp}
 =\beta\sum_{p\in\mathcal P_4}\left(1-\tfrac13\operatorname{ReTr}U_p\right)
 +\frac\beta2\sum_{p\in\mathcal P_8}\left(1-\tfrac13\operatorname{ReTr}U_p\right).
 \label{sm:action}
\end{equation}
The coefficient is the $n=2$ central-face prescription of Ref.~\cite{smref:ChenDaiPang:1503.01766}. Each geometry has its own sampled cut measure. If $E_\partial$ counts boundary-crossing spatial bonds, $|\mathcal P_8|=E_\partial$ and $|\mathcal P_4|=12(16^3\,32)-2E_\partial$.

Fermion spin phases are separate from the color links. Each complement closure has phase $-1$; region transitions have $+1$ for $0\to1$ and $-1$ for $1\to0$. Each connected temporal cycle thus has one antiperiodic closure, consistent with the fermionic replica convention in Ref.~\cite{smref:FujimuraNishiokaShimamori:2309.11889}. For an oriented edge $e:v\to w$ in direction $\mu$,
\begin{equation}
 D_{vw}=-\kappa(1-\gamma_\mu)z_eU_e,
 \qquad D_{wv}=-\kappa(1+\gamma_\mu)z_e^*U_e^\dagger,
 \label{sm:wilson}
\end{equation}
with identity diagonal. The reverse hop uses the phase of the actual incoming edge. These spin phases do not enter Eq.~\eqref{sm:action}.

The source and sink slices are $32-m_-$ and $m_+$. The seam is on a link, so the distances from its midpoint obey
\begin{equation}
 \ell_-/a=m_--\tfrac12,\qquad \ell_+/a=m_++\tfrac12,
 \qquad T_{\rm sep}/a=m_-+m_+.
 \label{sm:times}
\end{equation}
The link-centered reflection sends $(m_-,m_+)\mapsto(m_++1,m_--1)$; in particular $(6,6)$ pairs with $(7,5)$. A reflected correlator also transforms links, spin phases and source operators. Simply exchanging the two native labels is a different comparison.

For the spatially summed charged-pseudoscalar bilinear, relabel the contraction sheets by $1,2$. Let $S_{ij}$ be the propagator from source sheet $j$ to sink sheet $i$, including slice coordinates and spin--color indices. The complete contraction is
\begin{align}
 A_{ij}&=\Tr(S_{ij}S_{ij}^\dagger),\qquad
 B=\Tr(S_{11}S_{21}^\dagger S_{22}S_{12}^\dagger),\nonumber\\
 N_A&=A_{11}A_{22}+A_{12}A_{21}-2\operatorname{Re}B.
 \label{sm:wick}
\end{align}
In explicit indices, $B=\sum_{\alpha\beta pq}(S_{11})_{\alpha p}(S_{21})_{\beta p}^*(S_{22})_{\beta q}(S_{12})_{\alpha q}^*$, with $\alpha,\beta$ in the two sink spaces and $p,q$ in the two source spaces. Degenerate flavors give conjugate exchange terms. For a wall-pair realization, the $S_{ij}$ are replaced by their twelve-column diluted wall-solution matrices. The two source sheets use independent spatial $Z_4$ noise, while factors from the same source sheet share its noise within a contraction.

Products are formed within each complete wall pair. At a fixed placement, the pooled means are
\begin{equation}
 \overline N_A=\frac18\sum_{u\in A}\frac1{16}\sum_{r=1}^{16}N_{A,u,r},
 \qquad
 \overline C_{\rm ord}=\frac18\sum_{u\in\emptyset}\frac1{16}\sum_{r=1}^{16}
 \frac{C_{1,u,r}+C_{2,u,r}}2.
 \label{sm:averages}
\end{equation}
The logarithm in Eq.~\eqref{sm:response} follows these averages. The same ordinary mean is used for every region at the matching placement.

The empty measure and Dirac operator factor into two ordinary-sheet systems. Its shared numerator/denominator fields strongly cancel fluctuations. The full geometry is one $2N_t$ temporal circle and also probes state preparation. The replica-vacuum factor is implicit in the normalized cut expectation. A reduced-state interpretation of $\Ccp$ requires matched boundary variables and trace conventions; a proper ball/complement comparison is a proposed test.

\section{Uncertainty definitions and correlated normalization}
\label{sm:uncertainties}

Figure~\ref{sm:fig-errors} compares the field uncertainty with the conditional stochastic-source uncertainty on the same sample. Table~\ref{sm:tab-inputs} gives the corresponding reference-placement means, errors, and mask geometry. Source noise is already present in the wall-averaged field observations, so these errors are kept separate.

\begin{figure}[!htbp]
\centering\includegraphics[width=.90\textwidth]{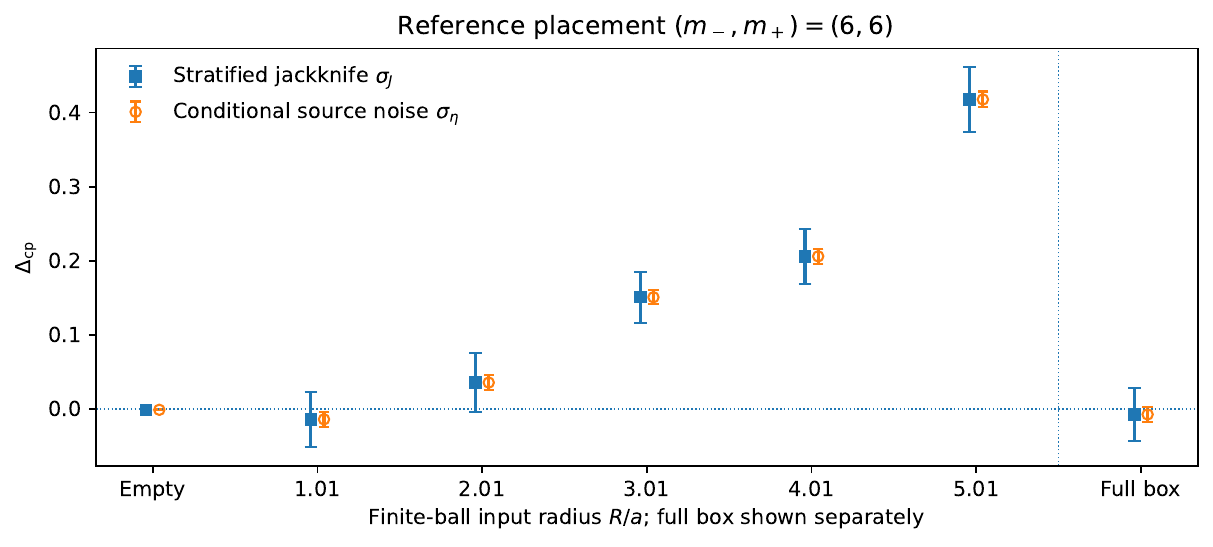}
\caption{Reference-placement errors for the same five balls and empty/full controls. Filled squares carry the stratified-jackknife field error $\sigma_J$; open circles carry the conditional source-noise error $\sigma_\eta$. Horizontal offsets separate the two evaluations of each mean. The ordinary denominator is shared across all regions in both analyses.}
\label{sm:fig-errors}
\end{figure}

The primary estimate deletes one of the four wall-averaged observations within each of the fourteen histories. Each history keeps its original weight; its remaining three observations are reaveraged, and all affected numerators, ordinary means and logarithms are recomputed together. If $y_{h,-u}$ is the full regional/placement vector after this deletion and $\overline y_{h,-}=\tfrac14\sum_u y_{h,-u}$, then
\begin{equation}
 C_J=\sum_{h=1}^{14}\frac34\sum_{u=1}^{4}
 (y_{h,-u}-\overline y_{h,-})(y_{h,-u}-\overline y_{h,-})^T.
 \label{sm:jackknife}
\end{equation}
The plotted central values are the complete-sample means. This variance estimate assumes independence of the four observations within each history; longer histories are required to establish autocorrelation and equilibration control~\cite{smref:hr:Wolff2004}. The sixteen-pair conditional calculation instead resamples complete noise-pair vectors within each fixed field, with $10^5$ replicates. All sinks from one source share the same draw; empty numerator and ordinary components remain joint, and a single ordinary-denominator draw is shared by every region. It holds the gauge fields fixed.

\begin{table}[htbp]
\centering\small
\caption{Reference-placement inputs. The five proper-mask rows, in increasing radius order, define the occupancy and pair fits. Empty/full controls are included for the regional comparisons.}
\label{sm:tab-inputs}
\begin{tabular}{lrrrrr}\toprule
Region & $n_A$ & $R_{\rm eff}/a$ & $\Dcp$ & $\sigma_J$ & $\sigma_\eta$\\\midrule
Empty & 0 & 0 & $-0.00099694$ & 0.00072702 & 0.00043968\\
$R/a=1.01$ & 7 & 1.186688 & $-0.01384392$ & 0.03725279 & 0.00974898\\
$R/a=2.01$ & 33 & 1.989795 & $0.03576809$ & 0.03936195 & 0.00987774\\
$R/a=3.01$ & 123 & 3.085121 & $0.15089336$ & 0.03452483 & 0.00971641\\
$R/a=4.01$ & 257 & 3.944102 & $0.20609932$ & 0.03724100 & 0.00989519\\
$R/a=5.01$ & 515 & 4.972478 & $0.41778881$ & 0.04337253 & 0.01075163\\
Full & 4096 & 9.925608 & $-0.00726280$ & 0.03594231 & 0.00968311\\
\bottomrule\end{tabular}
\end{table}

For the five proper masks, the full covariance can be compactly specified as
\begin{align}
 C_J^{(5)}&=10^{-3}\big[\operatorname{diag}(d_i)+v\mathbf1\mathbf1^T\big],\nonumber\\
 v&=0.991745482,\nonumber\\
 (d_1,\ldots,d_5)&=(0.396024580,\ 0.557617819,\ 0.200218156,\ 0.395146709,\ 0.889430897).
 \label{sm:covnumbers}
\end{align}
Here $\mathbf1$ is the five-component all-ones vector. The full-control variance is $0.001291849587$, and its covariance with each proper mask is $0.000991745482$. The empty-control variance is $5.285551435\times10^{-7}$ and its covariance with each nonempty region is $1.807205902\times10^{-5}$. These entries specify the complete seven-region reference covariance to the printed precision. The common ordinary normalization supplies the positive off-diagonal component. In a difference $d_i=\Delta_i-\Delta_{\rm full}$ it cancels exactly, while the full numerator becomes the shared reference; its covariance is formed as $C_d=HC_JH^T$, where $H$ takes the five differences.

With $s_i=\Socc(f_i)$ and $y_i=\Dcp(A_i)$, the fit and fixed-covariance parameter error are
\begin{equation}
 \widehat\kappa_{\rm occ}=\frac{s^TC^{-1}y}{s^TC^{-1}s},\qquad
 \sigma_\kappa^2=\frac1{s^TC^{-1}s},\qquad
 q^2=(y-\widehat\kappa_{\rm occ}s)^TC^{-1}(y-\widehat\kappa_{\rm occ}s).
 \label{sm:fit}
\end{equation}
They give $\kappa_{\rm occ}=1.68296\pm0.12393$ and $q^2=10.6104$ for four residual degrees of freedom. $q^2$ is a quadratic diagnostic with no covariance-calibrated tail probability~\cite{smref:ChristErankiKelly2024Pvalues}. The parameter error holds the estimated covariance and model fixed. Radius-window and reference changes are correlated analysis choices. Dropping the seven-site mask gives $1.55647\pm0.13769$; using the three largest balls gives $1.48870\pm0.15098$; adding the full control gives $1.73385\pm0.11388$. Table~\ref{sm:tab-placement-fits} extends the five-ball fit to all nine source/sink placements, using each placement's own covariance.

\begin{table}[htbp]
\centering\small
\caption{Five-ball occupancy fits at each native placement, using that placement's own jackknife covariance; every row has four residual degrees of freedom. The panels in Fig.~\ref{sm:fig-grid} share fields and sources. The half-period row is included separately as a finite-time comparison.}
\label{sm:tab-placement-fits}
\begin{tabular}{crr}\toprule
$(m_-,m_+)$ & $\kappa_{\rm occ}$ & $q^2/\mathrm{dof}$\\\midrule
$(4,4)$ & $1.64211\pm0.07613$ & $2.7262$\\
$(4,6)$ & $1.60193\pm0.08378$ & $2.1172$\\
$(4,8)$ & $1.60643\pm0.09817$ & $1.1373$\\
$(6,4)$ & $1.74593\pm0.11112$ & $2.7752$\\
$(6,6)$ & $1.68296\pm0.12393$ & $2.6526$\\
$(6,8)$ & $1.68002\pm0.14104$ & $2.1351$\\
$(8,4)$ & $1.69457\pm0.12702$ & $2.6894$\\
$(8,6)$ & $1.61436\pm0.13897$ & $2.8583$\\
$(8,8)$ & $0.67373\pm0.07593$ & $1.3024$\\
\bottomrule\end{tabular}
\end{table}
The reference-placement ball--full fit gives $1.72790\pm0.11625$, with $q^2=11.6295$ for four degrees of freedom. Its five differences are $-0.00658$, $0.04303$, $0.15816$, $0.21336$, $0.42505$, with jackknife errors $0.02638$, $0.02929$, $0.02237$, $0.02637$, $0.03449$. Neither control is imposed as an offset correction on the primary five-ball fit.
\FloatBarrier

\section{Temporal and source-preparation controls}
\label{sm:temporal}

Figure~\ref{sm:fig-temporal} resolves the full-circle crossover with a finer sink scan. This diagnostic uses four fields per endpoint, from both histories at sweeps 1100 and 1400, sixteen wall pairs, native $m_-=6$ and $m_+=2,\ldots,12$. Its ordinary mean is formed from these four empty fields. It is distinct from the eight-field-per-region sample of Fig.~\ref{sm:fig-grid}.

\begin{figure}[!htbp]
\centering\includegraphics[width=.94\textwidth]{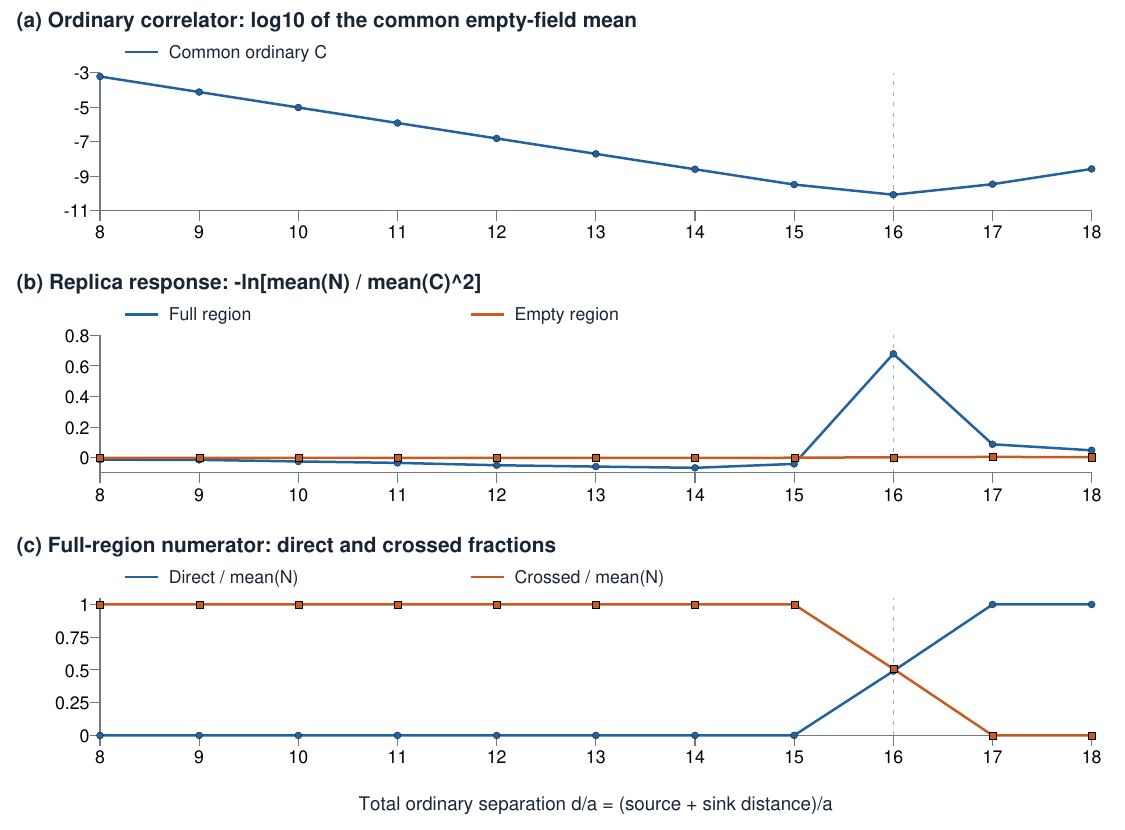}
\caption{Endpoint temporal scan: the ordinary correlator, empty/full responses, and direct/crossed fractions of the full numerator. The correlator minimum and pairing crossover occur at total separation $T_{\rm sep}/a=16=N_t/2$. Lines connect measured means over the entire sink grid. This four-field sample uses its own ordinary normalization. The complete nine-placement radius study, including its half-period point, is shown in Fig.~\ref{sm:fig-grid}.}
\label{sm:fig-temporal}
\end{figure}

The full response at total separation $16a$ is $0.679655\pm0.010528$, where the error is conditional source noise on this fixed-field sample. Its direct and crossed fractions are $0.491481$ and $0.508519$; the interference fraction has magnitude below $1.24\times10^{-9}$ throughout the scan. The measured non-midpoint tails remain visible. A simple propagation model with forward/backward amplitudes $F=Ze^{-a m_{\rm eff}\tau}$ and $B=Ze^{-a m_{\rm eff}(N_t-\tau)}$, $\tau=T_{\rm sep}/a$, gives
\begin{equation}
 \Delta_{\rm image}(\tau)
 =-\ln\frac{F^2+B^2}{(F+B)^2}
 =\ln\!\left[1+\operatorname{sech}\big(am_{\rm eff}(N_t-2\tau)\big)\right].
 \label{sm:image}
\end{equation}
The approximation $\overline N_{\rm full}\simeq F^2+B^2$ explains the scale $\ln2$ and the interchange of pairings. A matched-$N_t$ study at fixed source/cut/sink distances tests this interpretation independently of the endpoint normalization.

The ordinary means give the local estimator
\begin{equation}
 am_{\rm eff}^{(2)}(t)=\frac12\operatorname{arcosh}
 \frac{\overline C(t-2a)+\overline C(t+2a)}{2\overline C(t)}.
 \label{sm:mass}
\end{equation}
At central separations $10a,12a,14a$ with source offsets $4,6,8$, respectively, the values are $2.07398$, $2.06802$, $2.06658$. The single-state image model gives $B/F\leq2.57\times10^{-4}$ at the eight non-midpoint placements. Excitation gaps and overlaps separately govern projection onto the selected hadron. The effective mass comes from zero-momentum propagation; a kinetic mass requires a meson dispersion measurement.

Within the ideal occupancy model, an equal incoherent mixture of a one-particle state and vacuum changes the regional occupancy probability to $f_A/2$. Its reference curve is
\begin{equation}
 S_{2,\mathrm{mix}}^{\rm occ}(f_A)
 =-\ln[(f_A/2)^2+(1-f_A/2)^2],
 \label{sm:mixture}
\end{equation}
with empty value zero and full value $\ln2$. This explicitly assumed temporal model supplies the second guide in the $(8,8)$ panel. In the eight-field study, the three largest-ball responses at $(8,8)$ are $0.04921,0.06278,0.15522$. Relative to the same-source baseline $[\Delta_A(8,4)+\Delta_A(8,6)]/2$, their decreases are $0.07131\pm0.03653$, $0.11766\pm0.04088$, $0.23321\pm0.04031$ with the joint jackknife. At native source offset $6$, the largest-ball contrast $(6,8)-(6,4)$ is $-0.02492\pm0.01344$. These measured placement changes motivate independently varied source/sink distances and larger temporal extents; the current radius fit is a finite-time comparison.
\FloatBarrier

\section{Finite-mask quark--antiquark models and meson interpretation}
\label{sm:models}

Figure~\ref{sm:fig-pair} tests the quark--antiquark meson toy model on all five periodic masks, using one fitted width. The occupancy-amplitude and boundary-length fits provide comparison templates with the same input covariance.

Represent the meson's valence quark and antiquark by two distinguishable, fixed-number spatial modes, with a Fock-vacuum reference. Color, spin, and gluonic degrees of freedom are outside this spatial toy calculation. Let
\begin{align}
 \Psi(\bm x,\bm y)&=4096^{-1/2}\phi(\bm x-\bm y),\qquad
 \sum_{\bm r}|\phi(\bm r)|^2=1,\nonumber\\
 \phi_G(\bm r)&=Z_G^{-1/2}e^{-r_{\min}^2/(4w^2)},\qquad
 \phi_E(\bm r)=Z_E^{-1/2}e^{-r_{\min}/(2w)}.
 \label{sm:wavefunctions}
\end{align}
The sums cover the $16^3$ periodic lattice. The minimum-image separation $r_{\min}$ and width $w$ are lengths; numerical evaluation uses $r_{\min}/a$ and $w/a$. For $x,x'\in A$, define $Q_{xx'}=\phi(x-x')$ and $G_{xx'}=\sum_{y\in V}\phi(x-y)\phi^*(x'-y)$. The full finite-mask purity follows from
\begin{align}
 K_{10}&=(G_A-QQ^\dagger)/4096,\qquad
 p_{11}=\Tr(QQ^\dagger)/4096,\nonumber\\
 p_{10}=p_{01}&=f_A-p_{11},\qquad p_{00}=1-2f_A+p_{11},\nonumber\\
 P_2(A)&=p_{00}^2+p_{11}^2+2\Tr K_{10}^2,\qquad
 S_2^{\rm pair}(A)=-\ln P_2(A).
 \label{sm:pairpurity}
\end{align}
The equal split-sector contributions use the real, even wavefunctions in Eq.~\eqref{sm:wavefunctions}. Their normalized purities, $\gamma_{10}=\Tr K_{10}^2/p_{10}^2$ and $\gamma_{01}$, retain the coherence left by the partial trace. Evaluating the periodic autocorrelation or summing the complement columns gives the same kernel. Coincident quark--antiquark pairs reproduce unresolved-meson occupancy, and independent uniform quark and antiquark modes reproduce $2\Socc$.

\begin{figure}[!htbp]
\centering\includegraphics[width=.90\textwidth]{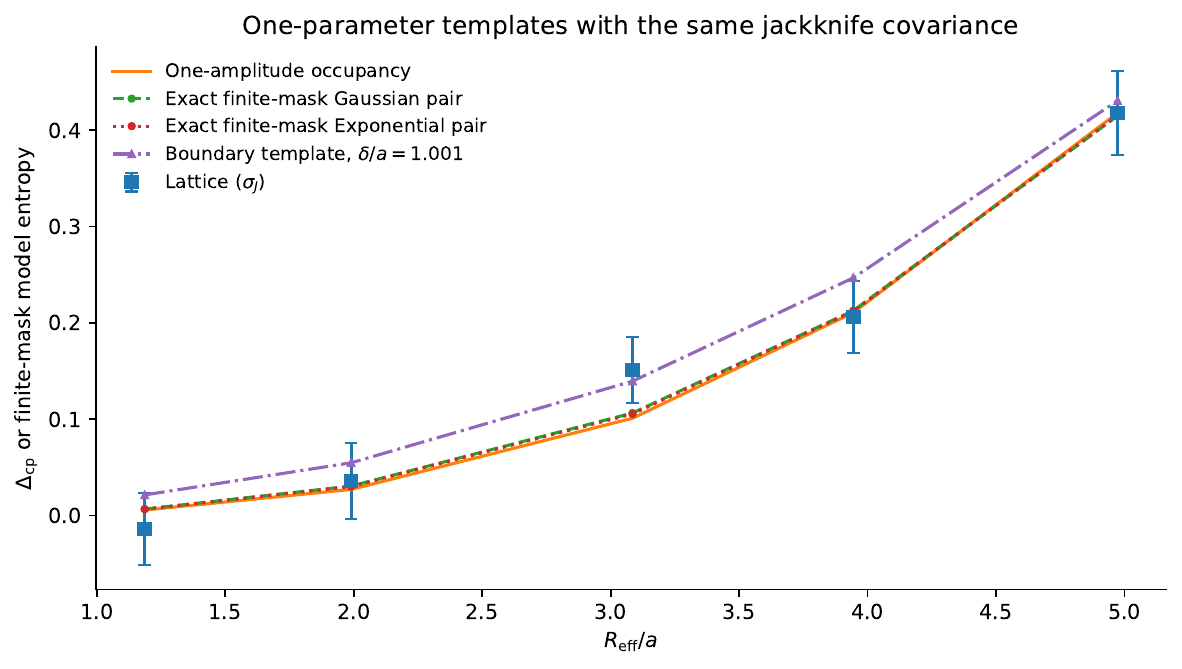}
\caption{Exact finite-mask Gaussian and exponential quark--antiquark entropies in the meson toy model, the occupancy-amplitude fit, and the boundary-length template on the five measured masks. All fits use the same $C_J^{(5)}$ and all five means in Table~\ref{sm:tab-inputs}. Model segments connect discrete-mask predictions. The bars are the primary jackknife errors.}
\label{sm:fig-pair}
\end{figure}

For each $X=G,E$, the fitted width $w_X^{\rm fit}$ minimizes the quadratic residual in Eq.~\eqref{sm:fit}, with $\kappa s$ replaced by the five exact entropies from Eq.~\eqref{sm:pairpurity}. The normalization $Z_X(w)$ is fixed by the site sum, so the model has one free width and no additional amplitude. The mean quark--antiquark separation reported in the article is then
\begin{equation}
 \frac{\langle r\rangle_X}{a}
 =\sum_{\bm r}\frac{r_{\min}}{a}|\phi_X(\bm r;w_X^{\rm fit})|^2.
 \label{sm:mean-separation}
\end{equation}
Table~\ref{sm:tab-pairs} specifies the fitted widths, the resulting mean separations, and the profiles used for the model envelope in the article. The profile criterion holds the covariance fixed; it is not assigned a calibrated coverage probability.

\begin{table}[htbp]
\centering\small
\caption{Exact periodic quark--antiquark fits for the meson toy model. Lengths are in units of $a$; the profile interval is the mean-separation range with $q^2\leq q^2_{\min}+1$. Each fit has one size parameter and four residual degrees of freedom.}
\label{sm:tab-pairs}
\begin{tabular}{lrrrr}\toprule
Wavefunction & $w/a$ & $\langle r\rangle/a$ & Profile interval & $q^2/\mathrm{dof}$\\\midrule
Gaussian & 2.3303 & 3.7060 & $[2.9527,4.6475]$ & $2.2808$\\
Exponential & 1.3733 & 3.8705 & $[3.0326,4.9007]$ & $2.3435$\\
\bottomrule\end{tabular}
\end{table}

For the fitted Gaussian, the five entropies are $0.00674,0.03058,0.10645,0.21260,0.41484$. Its split-sector purities decrease from $0.924$ to $0.272$ over these masks. At the largest mask $S_2^{\rm pair}/\Socc=1.67097$, while the overlap coefficient $2-p_{11}/f_A=1.51917$. The first is a finite-fraction entropy ratio; the second is the geometric quantity that enters the fixed-region dilute coefficient. The $[1,2]$ bound of the latter is therefore kept distinct from the finite-fraction ratios in Fig.~\ref{sm:fig-ratios}.

\begin{figure}[!htbp]
\centering\includegraphics[width=.91\textwidth]{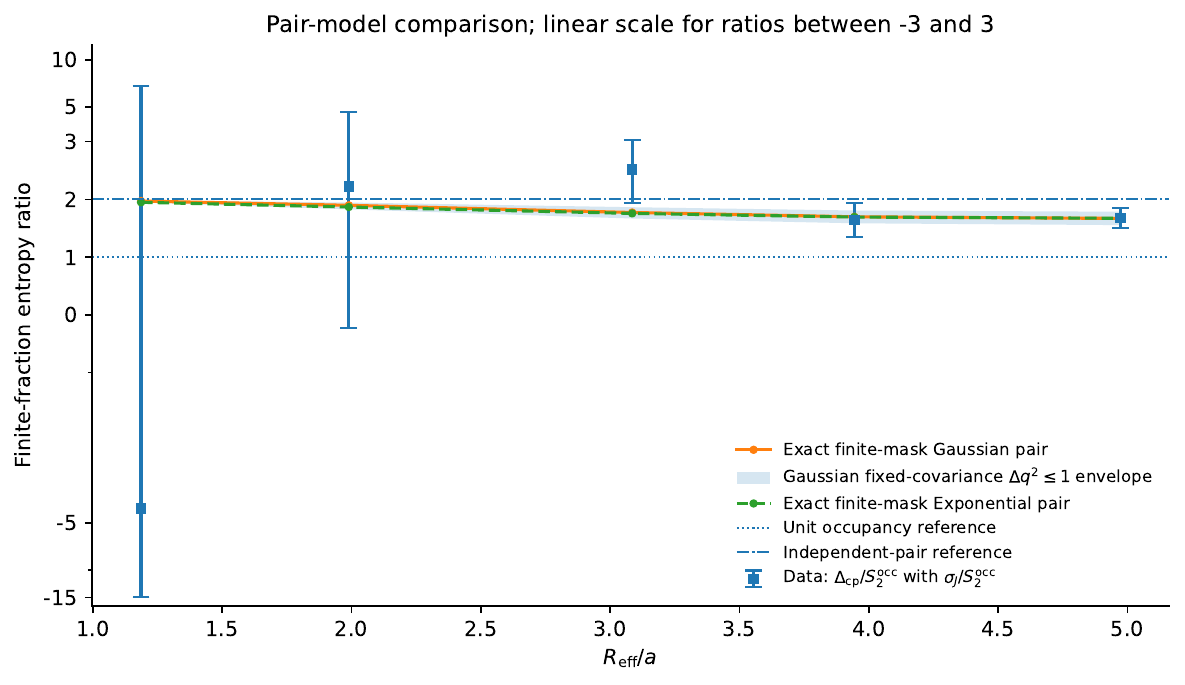}
\caption{Complete five-mask entropy-ratio comparison. The vertical scale is linear between $-3$ and $3$ and logarithmic outside, retaining the seven-site value $-4.05\pm10.90$. Its large ratio error follows from the small occupancy denominator. The Gaussian band uses the fixed-covariance profile in Table~\ref{sm:tab-pairs}. The unit and independent-pair lines are reference limits. All five points enter the fits, including the small mask displayed only in the caption of the article's linear-axis figure.}
\label{sm:fig-ratios}
\end{figure}

A separate finite-radius template is
\begin{equation}
 \Delta_\delta(A)=\left(1+\frac{\delta}{R_{\rm eff}}\right)^3\Socc(f_A).
 \label{sm:boundary}
\end{equation}
It gives $\delta/a=1.00141$, profile interval $[0.85241,1.14408]$, and $q^2=6.3859$ for four residual degrees of freedom. At small fraction its leading area coefficient is $B_\Sigma=2\delta$. Matching that coefficient to the pair result $B_\Sigma=\langle r\rangle/2$ defines $\delta=\langle r\rangle/4$. This is an analytic matching condition; complete finite-radius fits include different higher-order terms. The boundary template is evaluated on the measured masks: at fixed positive $\delta$ its continuum $R\to0$ limit is $8\pi\delta^3/(3V)$, so it is not used to define an empty-endpoint extrapolation.

For the two-term form $\Delta_i=2f_i\kappa_V+g_iB_\Sigma/a$, with $g_i=4\pi(R_{{\rm eff},i}/a)^2/4096$, the fit gives
\begin{equation}
 \kappa_V=0.237\pm0.621,\quad B_\Sigma/a=4.760\pm2.032,\quad
 \operatorname{Corr}(\kappa_V,B_\Sigma/a)=-0.9803,
 \label{sm:volsurface}
\end{equation}
with $q^2=5.3402$ for three residual degrees of freedom. The covariance of the two fitted parameters is $(X^TC^{-1}X)^{-1}$, with columns $2f$ and $g$. Their strong compensation limits a separate volume/surface determination from these radii. Raw boundary-bond counts approach $a^{-2}\int_\Sigma dS(|n_x|+|n_y|+|n_z|)$; the spherical angular average is $3/2$, so they are distinct from smooth area unless a matching convention is specified.

With the nominal scale $r_0/a=2.990(24)$~\cite{smref:EdwardsHellerKlassen1998} and $r_0=0.5\,\mathrm{fm}$, the mean toy separations are $0.620$ and $0.647\,\mathrm{fm}$. The equal-mass Gaussian meson toy model has quark/antiquark-to-center rms distance $2.0095a=0.336\,\mathrm{fm}$. A QCD spatial-amplitude comparison requires a specified gauge or link path, normalization and state projection, followed by a matched-profile test of Eq.~\eqref{sm:pairpurity}.

Valence-mass, spacing, and shape studies address the origin of the fitted length. Quenching permits reuse of the cut gauge measures for new valence contractions. A ball/complement comparison tests complementary power traces with matched regional factors and exterior-vacuum projection. These measurements and the spatial-amplitude test remain prospective.

\FloatBarrier
\begingroup\footnotesize\setlength{\bibsep}{1pt}
%

\endgroup

\endgroup

\begin{thebibliography}{57}%
\makeatletter
\providecommand \@ifxundefined [1]{%
 \@ifx{#1\undefined}
}%
\providecommand \@ifnum [1]{%
 \ifnum #1\expandafter \@firstoftwo
 \else \expandafter \@secondoftwo
 \fi
}%
\providecommand \@ifx [1]{%
 \ifx #1\expandafter \@firstoftwo
 \else \expandafter \@secondoftwo
 \fi
}%
\providecommand \natexlab [1]{#1}%
\providecommand \enquote  [1]{``#1''}%
\providecommand \bibnamefont  [1]{#1}%
\providecommand \bibfnamefont [1]{#1}%
\providecommand \citenamefont [1]{#1}%
\providecommand \href@noop [0]{\@secondoftwo}%
\providecommand \href [0]{\begingroup \@sanitize@url \@href}%
\providecommand \@href[1]{\@@startlink{#1}\@@href}%
\providecommand \@@href[1]{\endgroup#1\@@endlink}%
\providecommand \@sanitize@url [0]{\catcode `\\12\catcode `\$12\catcode
  `\&12\catcode `\#12\catcode `\^12\catcode `\_12\catcode `\%12\relax}%
\providecommand \@@startlink[1]{}%
\providecommand \@@endlink[0]{}%
\providecommand \url  [0]{\begingroup\@sanitize@url \@url }%
\providecommand \@url [1]{\endgroup\@href {#1}{\urlprefix }}%
\providecommand \urlprefix  [0]{URL }%
\providecommand \Eprint [0]{\href }%
\providecommand \doibase [0]{https://doi.org/}%
\providecommand \selectlanguage [0]{\@gobble}%
\providecommand \bibinfo  [0]{\@secondoftwo}%
\providecommand \bibfield  [0]{\@secondoftwo}%
\providecommand \translation [1]{[#1]}%
\providecommand \BibitemOpen [0]{}%
\providecommand \bibitemStop [0]{}%
\providecommand \bibitemNoStop [0]{.\EOS\space}%
\providecommand \EOS [0]{\spacefactor3000\relax}%
\providecommand \BibitemShut  [1]{\csname bibitem#1\endcsname}%
\let\auto@bib@innerbib\@empty
\bibitem [{\citenamefont {Ji}(1997)}]{Ji:1996ek}%
  \BibitemOpen
  \bibfield  {author} {\bibinfo {author} {\bibfnamefont {X.}~\bibnamefont
  {Ji}},\ }\href {https://doi.org/10.1103/PhysRevD.55.7114} {\bibfield
  {journal} {\bibinfo  {journal} {Phys. Rev. D}\ }\textbf {\bibinfo {volume}
  {55}},\ \bibinfo {pages} {7114} (\bibinfo {year} {1997})},\ \Eprint
  {https://arxiv.org/abs/hep-ph/9609381} {arXiv:hep-ph/9609381 [hep-ph]}
  \BibitemShut {NoStop}%
\bibitem [{\citenamefont {Polyakov}(2003)}]{Polyakov:2002yz}%
  \BibitemOpen
  \bibfield  {author} {\bibinfo {author} {\bibfnamefont {M.~V.}\ \bibnamefont
  {Polyakov}},\ }\href {https://doi.org/10.1016/S0370-2693(03)00036-4}
  {\bibfield  {journal} {\bibinfo  {journal} {Phys. Lett. B}\ }\textbf
  {\bibinfo {volume} {555}},\ \bibinfo {pages} {57} (\bibinfo {year} {2003})},\
  \Eprint {https://arxiv.org/abs/hep-ph/0210165} {arXiv:hep-ph/0210165}
  \BibitemShut {NoStop}%
\bibitem [{\citenamefont {Polyakov}\ and\ \citenamefont
  {Schweitzer}(2019)}]{Polyakov:2018zvc}%
  \BibitemOpen
  \bibfield  {author} {\bibinfo {author} {\bibfnamefont {M.~V.}\ \bibnamefont
  {Polyakov}}\ and\ \bibinfo {author} {\bibfnamefont {P.}~\bibnamefont
  {Schweitzer}},\ }\href {https://doi.org/10.1140/epja/i2019-12885-0}
  {\bibfield  {journal} {\bibinfo  {journal} {Eur. Phys. J. A}\ }\textbf
  {\bibinfo {volume} {55}},\ \bibinfo {pages} {191} (\bibinfo {year} {2019})},\
  \Eprint {https://arxiv.org/abs/1805.06596} {arXiv:1805.06596 [hep-ph]}
  \BibitemShut {NoStop}%
\bibitem [{\citenamefont {Hackett}\ \emph {et~al.}(2024)\citenamefont
  {Hackett}, \citenamefont {Pefkou},\ and\ \citenamefont
  {Shanahan}}]{Hackett:2023rif}%
  \BibitemOpen
  \bibfield  {author} {\bibinfo {author} {\bibfnamefont {D.~C.}\ \bibnamefont
  {Hackett}}, \bibinfo {author} {\bibfnamefont {D.~A.}\ \bibnamefont
  {Pefkou}},\ and\ \bibinfo {author} {\bibfnamefont {P.~E.}\ \bibnamefont
  {Shanahan}},\ }\href {https://doi.org/10.1103/PhysRevLett.132.251904}
  {\bibfield  {journal} {\bibinfo  {journal} {Phys. Rev. Lett.}\ }\textbf
  {\bibinfo {volume} {132}},\ \bibinfo {pages} {251904} (\bibinfo {year}
  {2024})},\ \Eprint {https://arxiv.org/abs/2310.08484} {arXiv:2310.08484
  [hep-lat]} \BibitemShut {NoStop}%
\bibitem [{\citenamefont {Kharzeev}\ and\ \citenamefont
  {Levin}(2017)}]{Kharzeev:2017qzs}%
  \BibitemOpen
  \bibfield  {author} {\bibinfo {author} {\bibfnamefont {D.~E.}\ \bibnamefont
  {Kharzeev}}\ and\ \bibinfo {author} {\bibfnamefont {E.~M.}\ \bibnamefont
  {Levin}},\ }\href {https://doi.org/10.1103/PhysRevD.95.114008} {\bibfield
  {journal} {\bibinfo  {journal} {Phys. Rev. D}\ }\textbf {\bibinfo {volume}
  {95}},\ \bibinfo {pages} {114008} (\bibinfo {year} {2017})},\ \Eprint
  {https://arxiv.org/abs/1702.03489} {arXiv:1702.03489 [hep-ph]} \BibitemShut
  {NoStop}%
\bibitem [{\citenamefont {G{\"u}rsoy}\ \emph {et~al.}(2024)\citenamefont
  {G{\"u}rsoy}, \citenamefont {Kharzeev},\ and\ \citenamefont
  {Pedraza}}]{Gursoy:2023hge}%
  \BibitemOpen
  \bibfield  {author} {\bibinfo {author} {\bibfnamefont {U.}~\bibnamefont
  {G{\"u}rsoy}}, \bibinfo {author} {\bibfnamefont {D.~E.}\ \bibnamefont
  {Kharzeev}},\ and\ \bibinfo {author} {\bibfnamefont {J.~F.}\ \bibnamefont
  {Pedraza}},\ }\href {https://doi.org/10.1103/PhysRevD.110.074008} {\bibfield
  {journal} {\bibinfo  {journal} {Phys. Rev. D}\ }\textbf {\bibinfo {volume}
  {110}},\ \bibinfo {pages} {074008} (\bibinfo {year} {2024})},\ \Eprint
  {https://arxiv.org/abs/2306.16145} {arXiv:2306.16145 [hep-th]} \BibitemShut
  {NoStop}%
\bibitem [{\citenamefont {Goykhman}(2015)}]{Goykhman:2015sga}%
  \BibitemOpen
  \bibfield  {author} {\bibinfo {author} {\bibfnamefont {M.}~\bibnamefont
  {Goykhman}},\ }\href {https://doi.org/10.1103/PhysRevD.92.025048} {\bibfield
  {journal} {\bibinfo  {journal} {Phys. Rev. D}\ }\textbf {\bibinfo {volume}
  {92}},\ \bibinfo {pages} {025048} (\bibinfo {year} {2015})},\ \Eprint
  {https://arxiv.org/abs/1501.07590} {arXiv:1501.07590 [hep-th]} \BibitemShut
  {NoStop}%
\bibitem [{\citenamefont {Liu}\ \emph {et~al.}(2023)\citenamefont {Liu},
  \citenamefont {Nowak},\ and\ \citenamefont {Zahed}}]{Liu:2022qqf}%
  \BibitemOpen
  \bibfield  {author} {\bibinfo {author} {\bibfnamefont {Y.}~\bibnamefont
  {Liu}}, \bibinfo {author} {\bibfnamefont {M.~A.}\ \bibnamefont {Nowak}},\
  and\ \bibinfo {author} {\bibfnamefont {I.}~\bibnamefont {Zahed}},\ }\href
  {https://doi.org/10.1103/PhysRevD.107.054010} {\bibfield  {journal} {\bibinfo
   {journal} {Phys. Rev. D}\ }\textbf {\bibinfo {volume} {107}},\ \bibinfo
  {pages} {054010} (\bibinfo {year} {2023})},\ \Eprint
  {https://arxiv.org/abs/2205.06724} {arXiv:2205.06724 [hep-ph]} \BibitemShut
  {NoStop}%
\bibitem [{\citenamefont {Mamo}(2025)}]{Mamo:2025}%
  \BibitemOpen
  \bibfield  {author} {\bibinfo {author} {\bibfnamefont {K.~A.}\ \bibnamefont
  {Mamo}},\ }\href {https://doi.org/10.1103/PhysRevD.112.L111506} {\bibfield
  {journal} {\bibinfo  {journal} {Phys. Rev. D}\ }\textbf {\bibinfo {volume}
  {112}},\ \bibinfo {pages} {L111506} (\bibinfo {year} {2025})},\ \Eprint
  {https://arxiv.org/abs/2507.00176} {arXiv:2507.00176 [hep-ph]} \BibitemShut
  {NoStop}%
\bibitem [{\citenamefont {Buividovich}\ and\ \citenamefont
  {Polikarpov}(2008{\natexlab{a}})}]{Buividovich:0806.3376}%
  \BibitemOpen
  \bibfield  {author} {\bibinfo {author} {\bibfnamefont {P.~V.}\ \bibnamefont
  {Buividovich}}\ and\ \bibinfo {author} {\bibfnamefont {M.~I.}\ \bibnamefont
  {Polikarpov}},\ }\href {https://doi.org/10.1016/j.physletb.2008.10.032}
  {\bibfield  {journal} {\bibinfo  {journal} {Phys. Lett. B}\ }\textbf
  {\bibinfo {volume} {670}},\ \bibinfo {pages} {141} (\bibinfo {year}
  {2008}{\natexlab{a}})},\ \Eprint {https://arxiv.org/abs/0806.3376}
  {arXiv:0806.3376 [hep-th]} \BibitemShut {NoStop}%
\bibitem [{\citenamefont {Buividovich}\ and\ \citenamefont
  {Polikarpov}(2008{\natexlab{b}})}]{Buividovich:08:2}%
  \BibitemOpen
  \bibfield  {author} {\bibinfo {author} {\bibfnamefont {P.~V.}\ \bibnamefont
  {Buividovich}}\ and\ \bibinfo {author} {\bibfnamefont {M.~I.}\ \bibnamefont
  {Polikarpov}},\ }\href {https://doi.org/10.1016/j.nuclphysb.2008.05.002}
  {\bibfield  {journal} {\bibinfo  {journal} {Nucl. Phys. B}\ }\textbf
  {\bibinfo {volume} {802}},\ \bibinfo {pages} {458} (\bibinfo {year}
  {2008}{\natexlab{b}})},\ \Eprint {https://arxiv.org/abs/0802.4247}
  {arXiv:0802.4247 [hep-lat]} \BibitemShut {NoStop}%
\bibitem [{\citenamefont {Velytsky}(2008)}]{Velytsky:2008sv}%
  \BibitemOpen
  \bibfield  {author} {\bibinfo {author} {\bibfnamefont {A.}~\bibnamefont
  {Velytsky}},\ }\href {https://doi.org/10.22323/1.066.0256} {\bibfield
  {journal} {\bibinfo  {journal} {PoS}\ }\textbf {\bibinfo {volume}
  {LATTICE2008}},\ \bibinfo {pages} {256} (\bibinfo {year} {2008})},\ \Eprint
  {https://arxiv.org/abs/0809.4502} {arXiv:0809.4502 [hep-lat]} \BibitemShut
  {NoStop}%
\bibitem [{\citenamefont {Nakagawa}\ \emph {et~al.}(2009)\citenamefont
  {Nakagawa}, \citenamefont {Nakamura}, \citenamefont {Motoki},\ and\
  \citenamefont {Zakharov}}]{Nakagawa:2009jk}%
  \BibitemOpen
  \bibfield  {author} {\bibinfo {author} {\bibfnamefont {Y.}~\bibnamefont
  {Nakagawa}}, \bibinfo {author} {\bibfnamefont {A.}~\bibnamefont {Nakamura}},
  \bibinfo {author} {\bibfnamefont {S.}~\bibnamefont {Motoki}},\ and\ \bibinfo
  {author} {\bibfnamefont {V.~I.}\ \bibnamefont {Zakharov}},\ }\href
  {https://doi.org/10.22323/1.091.0188} {\bibfield  {journal} {\bibinfo
  {journal} {PoS}\ }\textbf {\bibinfo {volume} {LAT2009}},\ \bibinfo {pages}
  {188} (\bibinfo {year} {2009})},\ \Eprint {https://arxiv.org/abs/0911.2596}
  {arXiv:0911.2596 [hep-lat]} \BibitemShut {NoStop}%
\bibitem [{\citenamefont {Nakagawa}\ \emph {et~al.}(2010)\citenamefont
  {Nakagawa}, \citenamefont {Nakamura}, \citenamefont {Motoki},\ and\
  \citenamefont {Zakharov}}]{Nakagawa:2010kjk}%
  \BibitemOpen
  \bibfield  {author} {\bibinfo {author} {\bibfnamefont {Y.}~\bibnamefont
  {Nakagawa}}, \bibinfo {author} {\bibfnamefont {A.}~\bibnamefont {Nakamura}},
  \bibinfo {author} {\bibfnamefont {S.}~\bibnamefont {Motoki}},\ and\ \bibinfo
  {author} {\bibfnamefont {V.~I.}\ \bibnamefont {Zakharov}},\ }\href@noop {}
  {\bibfield  {journal} {\bibinfo  {journal} {PoS}\ }\textbf {\bibinfo {volume}
  {LATTICE2010}},\ \bibinfo {pages} {281} (\bibinfo {year} {2010})},\ \Eprint
  {https://arxiv.org/abs/1104.1011} {arXiv:1104.1011 [hep-lat]} \BibitemShut
  {NoStop}%
\bibitem [{\citenamefont {Itou}\ \emph {et~al.}(2016)\citenamefont {Itou},
  \citenamefont {Nagata}, \citenamefont {Nakagawa}, \citenamefont {Nakamura},\
  and\ \citenamefont {Zakharov}}]{Itou:2015cyu}%
  \BibitemOpen
  \bibfield  {author} {\bibinfo {author} {\bibfnamefont {E.}~\bibnamefont
  {Itou}}, \bibinfo {author} {\bibfnamefont {T.}~\bibnamefont {Nagata}},
  \bibinfo {author} {\bibfnamefont {Y.}~\bibnamefont {Nakagawa}}, \bibinfo
  {author} {\bibfnamefont {A.}~\bibnamefont {Nakamura}},\ and\ \bibinfo
  {author} {\bibfnamefont {V.~I.}\ \bibnamefont {Zakharov}},\ }\href
  {https://doi.org/10.1093/ptep/ptw050} {\bibfield  {journal} {\bibinfo
  {journal} {Prog. Theor. Exp. Phys.}\ }\textbf {\bibinfo {volume} {2016}},\
  \bibinfo {pages} {061B01} (\bibinfo {year} {2016})},\ \Eprint
  {https://arxiv.org/abs/1512.01334} {arXiv:1512.01334 [hep-th]} \BibitemShut
  {NoStop}%
\bibitem [{\citenamefont {Rabenstein}\ \emph {et~al.}(2019)\citenamefont
  {Rabenstein}, \citenamefont {Bodendorfer}, \citenamefont {Buividovich},\ and\
  \citenamefont {Sch{\"a}fer}}]{Rabenstein:2018bri}%
  \BibitemOpen
  \bibfield  {author} {\bibinfo {author} {\bibfnamefont {A.}~\bibnamefont
  {Rabenstein}}, \bibinfo {author} {\bibfnamefont {N.}~\bibnamefont
  {Bodendorfer}}, \bibinfo {author} {\bibfnamefont {P.}~\bibnamefont
  {Buividovich}},\ and\ \bibinfo {author} {\bibfnamefont {A.}~\bibnamefont
  {Sch{\"a}fer}},\ }\href {https://doi.org/10.1103/PhysRevD.100.034504}
  {\bibfield  {journal} {\bibinfo  {journal} {Phys. Rev. D}\ }\textbf {\bibinfo
  {volume} {100}},\ \bibinfo {pages} {034504} (\bibinfo {year} {2019})},\
  \Eprint {https://arxiv.org/abs/1812.04279} {arXiv:1812.04279 [hep-lat]}
  \BibitemShut {NoStop}%
\bibitem [{\citenamefont {Rindlisbacher}\ \emph {et~al.}(2022)\citenamefont
  {Rindlisbacher}, \citenamefont {Jokela}, \citenamefont {P{\"o}nni},
  \citenamefont {Rummukainen},\ and\ \citenamefont
  {Salami}}]{Rindlisbacher:2022}%
  \BibitemOpen
  \bibfield  {author} {\bibinfo {author} {\bibfnamefont {T.}~\bibnamefont
  {Rindlisbacher}}, \bibinfo {author} {\bibfnamefont {N.}~\bibnamefont
  {Jokela}}, \bibinfo {author} {\bibfnamefont {A.}~\bibnamefont {P{\"o}nni}},
  \bibinfo {author} {\bibfnamefont {K.}~\bibnamefont {Rummukainen}},\ and\
  \bibinfo {author} {\bibfnamefont {A.}~\bibnamefont {Salami}},\ }\href
  {https://doi.org/10.22323/1.430.0031} {\bibfield  {journal} {\bibinfo
  {journal} {PoS}\ }\textbf {\bibinfo {volume} {LATTICE2022}},\ \bibinfo
  {pages} {031} (\bibinfo {year} {2022})},\ \Eprint
  {https://arxiv.org/abs/2211.00425} {arXiv:2211.00425 [hep-lat]} \BibitemShut
  {NoStop}%
\bibitem [{\citenamefont {Jokela}\ \emph {et~al.}(2023)\citenamefont {Jokela},
  \citenamefont {P{\"o}nni}, \citenamefont {Rindlisbacher}, \citenamefont
  {Rummukainen},\ and\ \citenamefont {Salami}}]{Rindlisbacher:2023}%
  \BibitemOpen
  \bibfield  {author} {\bibinfo {author} {\bibfnamefont {N.}~\bibnamefont
  {Jokela}}, \bibinfo {author} {\bibfnamefont {A.}~\bibnamefont {P{\"o}nni}},
  \bibinfo {author} {\bibfnamefont {T.}~\bibnamefont {Rindlisbacher}}, \bibinfo
  {author} {\bibfnamefont {K.}~\bibnamefont {Rummukainen}},\ and\ \bibinfo
  {author} {\bibfnamefont {A.}~\bibnamefont {Salami}},\ }\href
  {https://doi.org/10.1007/JHEP12(2023)137} {\bibfield  {journal} {\bibinfo
  {journal} {JHEP}\ }\textbf {\bibinfo {volume} {2023}}\bibfield  {number}
  {\bibinfo  {number} { (12)},\ \bibinfo {pages} {137}},\ }\Eprint
  {https://arxiv.org/abs/2304.08949} {arXiv:2304.08949 [hep-th]} \BibitemShut
  {NoStop}%
\bibitem [{\citenamefont {Bombelli}\ \emph {et~al.}(1986)\citenamefont
  {Bombelli}, \citenamefont {Koul}, \citenamefont {Lee},\ and\ \citenamefont
  {Sorkin}}]{Bombelli:1986rw}%
  \BibitemOpen
  \bibfield  {author} {\bibinfo {author} {\bibfnamefont {L.}~\bibnamefont
  {Bombelli}}, \bibinfo {author} {\bibfnamefont {R.~K.}\ \bibnamefont {Koul}},
  \bibinfo {author} {\bibfnamefont {J.}~\bibnamefont {Lee}},\ and\ \bibinfo
  {author} {\bibfnamefont {R.~D.}\ \bibnamefont {Sorkin}},\ }\href
  {https://doi.org/10.1103/PhysRevD.34.373} {\bibfield  {journal} {\bibinfo
  {journal} {Phys. Rev. D}\ }\textbf {\bibinfo {volume} {34}},\ \bibinfo
  {pages} {373} (\bibinfo {year} {1986})}\BibitemShut {NoStop}%
\bibitem [{\citenamefont {Srednicki}(1993)}]{Srednicki:1993im}%
  \BibitemOpen
  \bibfield  {author} {\bibinfo {author} {\bibfnamefont {M.}~\bibnamefont
  {Srednicki}},\ }\href {https://doi.org/10.1103/PhysRevLett.71.666} {\bibfield
   {journal} {\bibinfo  {journal} {Phys. Rev. Lett.}\ }\textbf {\bibinfo
  {volume} {71}},\ \bibinfo {pages} {666} (\bibinfo {year} {1993})}\BibitemShut
  {NoStop}%
\bibitem [{\citenamefont {Callan}\ and\ \citenamefont
  {Wilczek}(1994)}]{Callan:1994py}%
  \BibitemOpen
  \bibfield  {author} {\bibinfo {author} {\bibfnamefont {C.~G.}\ \bibnamefont
  {Callan}}\ and\ \bibinfo {author} {\bibfnamefont {F.}~\bibnamefont
  {Wilczek}},\ }\href {https://doi.org/10.1016/0370-2693(94)91007-3} {\bibfield
   {journal} {\bibinfo  {journal} {Phys. Lett. B}\ }\textbf {\bibinfo {volume}
  {333}},\ \bibinfo {pages} {55} (\bibinfo {year} {1994})},\ \Eprint
  {https://arxiv.org/abs/hep-th/9401072} {arXiv:hep-th/9401072 [hep-th]}
  \BibitemShut {NoStop}%
\bibitem [{\citenamefont {Calabrese}\ and\ \citenamefont
  {Cardy}(2004)}]{Calabrese:2004eu}%
  \BibitemOpen
  \bibfield  {author} {\bibinfo {author} {\bibfnamefont {P.}~\bibnamefont
  {Calabrese}}\ and\ \bibinfo {author} {\bibfnamefont {J.}~\bibnamefont
  {Cardy}},\ }\href {https://doi.org/10.1088/1742-5468/2004/06/P06002}
  {\bibfield  {journal} {\bibinfo  {journal} {J. Stat. Mech.}\ }\textbf
  {\bibinfo {volume} {2004}},\ \bibinfo {pages} {P06002} (\bibinfo {year}
  {2004})},\ \Eprint {https://arxiv.org/abs/hep-th/0405152}
  {arXiv:hep-th/0405152 [hep-th]} \BibitemShut {NoStop}%
\bibitem [{\citenamefont {Ryu}\ and\ \citenamefont
  {Takayanagi}(2006{\natexlab{a}})}]{Ryu:2006bv}%
  \BibitemOpen
  \bibfield  {author} {\bibinfo {author} {\bibfnamefont {S.}~\bibnamefont
  {Ryu}}\ and\ \bibinfo {author} {\bibfnamefont {T.}~\bibnamefont
  {Takayanagi}},\ }\href {https://doi.org/10.1103/PhysRevLett.96.181602}
  {\bibfield  {journal} {\bibinfo  {journal} {Phys. Rev. Lett.}\ }\textbf
  {\bibinfo {volume} {96}},\ \bibinfo {pages} {181602} (\bibinfo {year}
  {2006}{\natexlab{a}})},\ \Eprint {https://arxiv.org/abs/hep-th/0603001}
  {arXiv:hep-th/0603001 [hep-th]} \BibitemShut {NoStop}%
\bibitem [{\citenamefont {Ryu}\ and\ \citenamefont
  {Takayanagi}(2006{\natexlab{b}})}]{Ryu:2006ef}%
  \BibitemOpen
  \bibfield  {author} {\bibinfo {author} {\bibfnamefont {S.}~\bibnamefont
  {Ryu}}\ and\ \bibinfo {author} {\bibfnamefont {T.}~\bibnamefont
  {Takayanagi}},\ }\href {https://doi.org/10.1088/1126-6708/2006/08/045}
  {\bibfield  {journal} {\bibinfo  {journal} {JHEP}\ }\textbf {\bibinfo
  {volume} {2006}}\bibfield  {number} {\bibinfo  {number} { (08)},\ \bibinfo
  {pages} {045}},\ }\Eprint {https://arxiv.org/abs/hep-th/0605073}
  {arXiv:hep-th/0605073 [hep-th]} \BibitemShut {NoStop}%
\bibitem [{\citenamefont {Klebanov}\ \emph {et~al.}(2008)\citenamefont
  {Klebanov}, \citenamefont {Kutasov},\ and\ \citenamefont
  {Murugan}}]{Klebanov:2007ws}%
  \BibitemOpen
  \bibfield  {author} {\bibinfo {author} {\bibfnamefont {I.~R.}\ \bibnamefont
  {Klebanov}}, \bibinfo {author} {\bibfnamefont {D.}~\bibnamefont {Kutasov}},\
  and\ \bibinfo {author} {\bibfnamefont {A.}~\bibnamefont {Murugan}},\ }\href
  {https://doi.org/10.1016/j.nuclphysb.2007.12.017} {\bibfield  {journal}
  {\bibinfo  {journal} {Nucl. Phys. B}\ }\textbf {\bibinfo {volume} {796}},\
  \bibinfo {pages} {274} (\bibinfo {year} {2008})},\ \Eprint
  {https://arxiv.org/abs/0709.2140} {arXiv:0709.2140 [hep-th]} \BibitemShut
  {NoStop}%
\bibitem [{\citenamefont {Solodukhin}(2008)}]{Solodukhin:2008dh}%
  \BibitemOpen
  \bibfield  {author} {\bibinfo {author} {\bibfnamefont {S.~N.}\ \bibnamefont
  {Solodukhin}},\ }\href {https://doi.org/10.1016/j.physletb.2008.05.071}
  {\bibfield  {journal} {\bibinfo  {journal} {Phys. Lett. B}\ }\textbf
  {\bibinfo {volume} {665}},\ \bibinfo {pages} {305} (\bibinfo {year}
  {2008})},\ \Eprint {https://arxiv.org/abs/0802.3117} {arXiv:0802.3117}
  \BibitemShut {NoStop}%
\bibitem [{\citenamefont {Casini}\ \emph {et~al.}(2011)\citenamefont {Casini},
  \citenamefont {Huerta},\ and\ \citenamefont {Myers}}]{Casini:2011kv}%
  \BibitemOpen
  \bibfield  {author} {\bibinfo {author} {\bibfnamefont {H.}~\bibnamefont
  {Casini}}, \bibinfo {author} {\bibfnamefont {M.}~\bibnamefont {Huerta}},\
  and\ \bibinfo {author} {\bibfnamefont {R.~C.}\ \bibnamefont {Myers}},\ }\href
  {https://doi.org/10.1007/JHEP05(2011)036} {\bibfield  {journal} {\bibinfo
  {journal} {JHEP}\ }\textbf {\bibinfo {volume} {2011}}\bibfield  {number}
  {\bibinfo  {number} { (05)},\ \bibinfo {pages} {036}},\ }\Eprint
  {https://arxiv.org/abs/1102.0440} {arXiv:1102.0440 [hep-th]} \BibitemShut
  {NoStop}%
\bibitem [{\citenamefont {Casini}\ \emph {et~al.}(2014)\citenamefont {Casini},
  \citenamefont {Huerta},\ and\ \citenamefont
  {Rosabal}}]{CasiniHuertaRosabal:2013}%
  \BibitemOpen
  \bibfield  {author} {\bibinfo {author} {\bibfnamefont {H.}~\bibnamefont
  {Casini}}, \bibinfo {author} {\bibfnamefont {M.}~\bibnamefont {Huerta}},\
  and\ \bibinfo {author} {\bibfnamefont {J.~A.}\ \bibnamefont {Rosabal}},\
  }\href {https://doi.org/10.1103/PhysRevD.89.085012} {\bibfield  {journal}
  {\bibinfo  {journal} {Phys. Rev. D}\ }\textbf {\bibinfo {volume} {89}},\
  \bibinfo {pages} {085012} (\bibinfo {year} {2014})},\ \Eprint
  {https://arxiv.org/abs/1312.1183} {arXiv:1312.1183 [hep-th]} \BibitemShut
  {NoStop}%
\bibitem [{\citenamefont {Donnelly}\ and\ \citenamefont
  {Wall}(2015)}]{DonnellyWall:2014edge}%
  \BibitemOpen
  \bibfield  {author} {\bibinfo {author} {\bibfnamefont {W.}~\bibnamefont
  {Donnelly}}\ and\ \bibinfo {author} {\bibfnamefont {A.~C.}\ \bibnamefont
  {Wall}},\ }\href {https://doi.org/10.1103/PhysRevLett.114.111603} {\bibfield
  {journal} {\bibinfo  {journal} {Phys. Rev. Lett.}\ }\textbf {\bibinfo
  {volume} {114}},\ \bibinfo {pages} {111603} (\bibinfo {year} {2015})},\
  \Eprint {https://arxiv.org/abs/1412.1895} {arXiv:1412.1895 [hep-th]}
  \BibitemShut {NoStop}%
\bibitem [{\citenamefont {Huang}(2015)}]{Huang:2014pfa}%
  \BibitemOpen
  \bibfield  {author} {\bibinfo {author} {\bibfnamefont {K.-W.}\ \bibnamefont
  {Huang}},\ }\href {https://doi.org/10.1103/PhysRevD.92.025010} {\bibfield
  {journal} {\bibinfo  {journal} {Phys. Rev. D}\ }\textbf {\bibinfo {volume}
  {92}},\ \bibinfo {pages} {025010} (\bibinfo {year} {2015})},\ \Eprint
  {https://arxiv.org/abs/1412.2730} {arXiv:1412.2730 [hep-th]} \BibitemShut
  {NoStop}%
\bibitem [{\citenamefont {Agarwal}\ \emph {et~al.}(2017)\citenamefont
  {Agarwal}, \citenamefont {Karabali},\ and\ \citenamefont
  {Nair}}]{Agarwal:2016cir}%
  \BibitemOpen
  \bibfield  {author} {\bibinfo {author} {\bibfnamefont {A.}~\bibnamefont
  {Agarwal}}, \bibinfo {author} {\bibfnamefont {D.}~\bibnamefont {Karabali}},\
  and\ \bibinfo {author} {\bibfnamefont {V.~P.}\ \bibnamefont {Nair}},\ }\href
  {https://doi.org/10.1103/PhysRevD.96.125008} {\bibfield  {journal} {\bibinfo
  {journal} {Phys. Rev. D}\ }\textbf {\bibinfo {volume} {96}},\ \bibinfo
  {pages} {125008} (\bibinfo {year} {2017})},\ \Eprint
  {https://arxiv.org/abs/1701.00014} {arXiv:1701.00014 [hep-th]} \BibitemShut
  {NoStop}%
\bibitem [{\citenamefont {Donnelly}\ \emph {et~al.}(2020)\citenamefont
  {Donnelly}, \citenamefont {Timmerman},\ and\ \citenamefont
  {Vald{\'e}s-Meller}}]{Donnelly:2019zde}%
  \BibitemOpen
  \bibfield  {author} {\bibinfo {author} {\bibfnamefont {W.}~\bibnamefont
  {Donnelly}}, \bibinfo {author} {\bibfnamefont {S.}~\bibnamefont
  {Timmerman}},\ and\ \bibinfo {author} {\bibfnamefont {N.}~\bibnamefont
  {Vald{\'e}s-Meller}},\ }\href {https://doi.org/10.1007/JHEP04(2020)182}
  {\bibfield  {journal} {\bibinfo  {journal} {JHEP}\ }\textbf {\bibinfo
  {volume} {2020}}\bibfield  {number} {\bibinfo  {number} { (04)},\ \bibinfo
  {pages} {182}},\ }\Eprint {https://arxiv.org/abs/1911.09302}
  {arXiv:1911.09302 [hep-th]} \BibitemShut {NoStop}%
\bibitem [{\citenamefont {Anegawa}\ \emph {et~al.}(2022)\citenamefont
  {Anegawa}, \citenamefont {Iizuka},\ and\ \citenamefont
  {Kabat}}]{Anegawa:2021osi}%
  \BibitemOpen
  \bibfield  {author} {\bibinfo {author} {\bibfnamefont {T.}~\bibnamefont
  {Anegawa}}, \bibinfo {author} {\bibfnamefont {N.}~\bibnamefont {Iizuka}},\
  and\ \bibinfo {author} {\bibfnamefont {D.}~\bibnamefont {Kabat}},\ }\href
  {https://doi.org/10.1103/PhysRevD.105.085003} {\bibfield  {journal} {\bibinfo
   {journal} {Phys. Rev. D}\ }\textbf {\bibinfo {volume} {105}},\ \bibinfo
  {pages} {085003} (\bibinfo {year} {2022})},\ \Eprint
  {https://arxiv.org/abs/2111.03886} {arXiv:2111.03886 [hep-th]} \BibitemShut
  {NoStop}%
\bibitem [{\citenamefont {Panizza}\ \emph {et~al.}(2022)\citenamefont
  {Panizza}, \citenamefont {de~Almeida},\ and\ \citenamefont
  {Hauke}}]{Panizza:2022gvd}%
  \BibitemOpen
  \bibfield  {author} {\bibinfo {author} {\bibfnamefont {V.}~\bibnamefont
  {Panizza}}, \bibinfo {author} {\bibfnamefont {R.~C.}\ \bibnamefont
  {de~Almeida}},\ and\ \bibinfo {author} {\bibfnamefont {P.}~\bibnamefont
  {Hauke}},\ }\href {https://doi.org/10.1007/JHEP09(2022)196} {\bibfield
  {journal} {\bibinfo  {journal} {JHEP}\ }\textbf {\bibinfo {volume}
  {2022}}\bibfield  {number} {\bibinfo  {number} { (09)},\ \bibinfo {pages}
  {196}},\ }\Eprint {https://arxiv.org/abs/2207.00605} {arXiv:2207.00605
  [hep-th]} \BibitemShut {NoStop}%
\bibitem [{\citenamefont {Bulgarelli}\ and\ \citenamefont
  {Panero}(2023)}]{Bulgarelli:2023ofi}%
  \BibitemOpen
  \bibfield  {author} {\bibinfo {author} {\bibfnamefont {A.}~\bibnamefont
  {Bulgarelli}}\ and\ \bibinfo {author} {\bibfnamefont {M.}~\bibnamefont
  {Panero}},\ }\href {https://doi.org/10.1007/JHEP06(2023)030} {\bibfield
  {journal} {\bibinfo  {journal} {JHEP}\ }\textbf {\bibinfo {volume}
  {2023}}\bibfield  {number} {\bibinfo  {number} { (06)},\ \bibinfo {pages}
  {030}},\ }\Eprint {https://arxiv.org/abs/2304.03311} {arXiv:2304.03311
  [quant-ph]} \BibitemShut {NoStop}%
\bibitem [{\citenamefont {Takahashi}\ and\ \citenamefont
  {Kanada-En'yo}(2019)}]{Takahashi:2019xjv}%
  \BibitemOpen
  \bibfield  {author} {\bibinfo {author} {\bibfnamefont {T.~T.}\ \bibnamefont
  {Takahashi}}\ and\ \bibinfo {author} {\bibfnamefont {Y.}~\bibnamefont
  {Kanada-En'yo}},\ }\href {https://doi.org/10.1103/PhysRevD.100.114502}
  {\bibfield  {journal} {\bibinfo  {journal} {Phys. Rev. D}\ }\textbf {\bibinfo
  {volume} {100}},\ \bibinfo {pages} {114502} (\bibinfo {year} {2019})},\
  \Eprint {https://arxiv.org/abs/1910.00859} {arXiv:1910.00859 [hep-lat]}
  \BibitemShut {NoStop}%
\bibitem [{\citenamefont {Takahashi}\ and\ \citenamefont
  {Kanada-En'yo}(2021)}]{Takahashi:2020vct}%
  \BibitemOpen
  \bibfield  {author} {\bibinfo {author} {\bibfnamefont {T.~T.}\ \bibnamefont
  {Takahashi}}\ and\ \bibinfo {author} {\bibfnamefont {Y.}~\bibnamefont
  {Kanada-En'yo}},\ }\href {https://doi.org/10.1103/PhysRevD.103.034504}
  {\bibfield  {journal} {\bibinfo  {journal} {Phys. Rev. D}\ }\textbf {\bibinfo
  {volume} {103}},\ \bibinfo {pages} {034504} (\bibinfo {year} {2021})},\
  \Eprint {https://arxiv.org/abs/2011.10950} {arXiv:2011.10950 [hep-lat]}
  \BibitemShut {NoStop}%
\bibitem [{\citenamefont {Amorosso}\ \emph
  {et~al.}(2025{\natexlab{a}})\citenamefont {Amorosso}, \citenamefont
  {Syritsyn},\ and\ \citenamefont {Venugopalan}}]{Amorosso:2024glf}%
  \BibitemOpen
  \bibfield  {author} {\bibinfo {author} {\bibfnamefont {R.}~\bibnamefont
  {Amorosso}}, \bibinfo {author} {\bibfnamefont {S.}~\bibnamefont {Syritsyn}},\
  and\ \bibinfo {author} {\bibfnamefont {R.}~\bibnamefont {Venugopalan}},\
  }\href {https://doi.org/10.1016/j.physletb.2025.139806} {\bibfield  {journal}
  {\bibinfo  {journal} {Phys. Lett. B}\ }\textbf {\bibinfo {volume} {868}},\
  \bibinfo {pages} {139806} (\bibinfo {year} {2025}{\natexlab{a}})},\ \Eprint
  {https://arxiv.org/abs/2411.12818} {arXiv:2411.12818 [hep-lat]} \BibitemShut
  {NoStop}%
\bibitem [{\citenamefont {Amorosso}\ \emph {et~al.}(2024)\citenamefont
  {Amorosso}, \citenamefont {Syritsyn},\ and\ \citenamefont
  {Venugopalan}}]{Amorosso:2024leg}%
  \BibitemOpen
  \bibfield  {author} {\bibinfo {author} {\bibfnamefont {R.}~\bibnamefont
  {Amorosso}}, \bibinfo {author} {\bibfnamefont {S.}~\bibnamefont {Syritsyn}},\
  and\ \bibinfo {author} {\bibfnamefont {R.}~\bibnamefont {Venugopalan}},\
  }\href {https://doi.org/10.1007/JHEP12(2024)177} {\bibfield  {journal}
  {\bibinfo  {journal} {JHEP}\ }\textbf {\bibinfo {volume} {2024}}\bibfield
  {number} {\bibinfo  {number} { (12)},\ \bibinfo {pages} {177}},\ }\Eprint
  {https://arxiv.org/abs/2410.00112} {arXiv:2410.00112 [hep-lat]} \BibitemShut
  {NoStop}%
\bibitem [{\citenamefont {Amorosso}\ \emph
  {et~al.}(2025{\natexlab{b}})\citenamefont {Amorosso}, \citenamefont
  {Syritsyn},\ and\ \citenamefont {Venugopalan}}]{Amorosso:2025tgg}%
  \BibitemOpen
  \bibfield  {author} {\bibinfo {author} {\bibfnamefont {R.}~\bibnamefont
  {Amorosso}}, \bibinfo {author} {\bibfnamefont {S.}~\bibnamefont {Syritsyn}},\
  and\ \bibinfo {author} {\bibfnamefont {R.}~\bibnamefont {Venugopalan}},\
  }\href {https://doi.org/10.22323/1.466.0409} {\bibfield  {journal} {\bibinfo
  {journal} {PoS}\ }\textbf {\bibinfo {volume} {LATTICE2024}},\ \bibinfo
  {pages} {409} (\bibinfo {year} {2025}{\natexlab{b}})},\ \Eprint
  {https://arxiv.org/abs/2502.08737} {arXiv:2502.08737 [hep-lat]} \BibitemShut
  {NoStop}%
\bibitem [{\citenamefont {Grieninger}\ \emph {et~al.}(2026)\citenamefont
  {Grieninger}, \citenamefont {Kharzeev},\ and\ \citenamefont
  {Marroquin}}]{Grieninger:2025rdi}%
  \BibitemOpen
  \bibfield  {author} {\bibinfo {author} {\bibfnamefont {S.}~\bibnamefont
  {Grieninger}}, \bibinfo {author} {\bibfnamefont {D.~E.}\ \bibnamefont
  {Kharzeev}},\ and\ \bibinfo {author} {\bibfnamefont {E.}~\bibnamefont
  {Marroquin}},\ }\href {https://doi.org/10.1103/v8mv-185x} {\bibfield
  {journal} {\bibinfo  {journal} {Phys. Rev. D}\ }\textbf {\bibinfo {volume}
  {113}},\ \bibinfo {pages} {036013} (\bibinfo {year} {2026})},\ \Eprint
  {https://arxiv.org/abs/2510.23919} {arXiv:2510.23919 [hep-ph]} \BibitemShut
  {NoStop}%
\bibitem [{\citenamefont {Horn}\ \emph {et~al.}(2026)\citenamefont {Horn},
  \citenamefont {M{\"u}ller},\ and\ \citenamefont {Yao}}]{Horn:2026bfs}%
  \BibitemOpen
  \bibfield  {author} {\bibinfo {author} {\bibfnamefont {D.}~\bibnamefont
  {Horn}}, \bibinfo {author} {\bibfnamefont {B.}~\bibnamefont {M{\"u}ller}},\
  and\ \bibinfo {author} {\bibfnamefont {X.}~\bibnamefont {Yao}},\ }\href@noop
  {} {\bibinfo {title} {{Gluon Entanglement Entropy inside a Nucleon: A Toy
  Model}}} (\bibinfo {year} {2026}),\ \Eprint
  {https://arxiv.org/abs/2605.11171} {arXiv:2605.11171 [hep-ph]} \BibitemShut
  {NoStop}%
\bibitem [{\citenamefont {Amorosso}\ \emph {et~al.}(2026)\citenamefont
  {Amorosso}, \citenamefont {Syritsyn},\ and\ \citenamefont
  {Venugopalan}}]{Amorosso2026Tomography}%
  \BibitemOpen
  \bibfield  {author} {\bibinfo {author} {\bibfnamefont {R.}~\bibnamefont
  {Amorosso}}, \bibinfo {author} {\bibfnamefont {S.}~\bibnamefont {Syritsyn}},\
  and\ \bibinfo {author} {\bibfnamefont {R.}~\bibnamefont {Venugopalan}},\
  }\href@noop {} {\bibinfo {title} {Entanglement enabled tomography of flux
  tubes in (2+1)d yang--mills theory}} (\bibinfo {year} {2026}),\ \Eprint
  {https://arxiv.org/abs/2601.17199} {arXiv:2601.17199 [hep-th]} \BibitemShut
  {NoStop}%
\bibitem [{Mam()}]{Mamo:SupplementalMaterial}%
  \BibitemOpen
  \href@noop {} {}\bibinfo {note} {See Supplemental Material for the complete
  source/sink-placement results, cut and contraction definitions, uncertainty
  and temporal controls, and finite-mask pair-model calculations}\BibitemShut
  {NoStop}%
\bibitem [{\citenamefont {Castro-Alvaredo}\ \emph
  {et~al.}(2018{\natexlab{a}})\citenamefont {Castro-Alvaredo}, \citenamefont
  {De~Fazio}, \citenamefont {Doyon},\ and\ \citenamefont
  {Sz{\'e}cs{\'e}nyi}}]{CastroAlvaredo2018PRL}%
  \BibitemOpen
  \bibfield  {author} {\bibinfo {author} {\bibfnamefont {O.~A.}\ \bibnamefont
  {Castro-Alvaredo}}, \bibinfo {author} {\bibfnamefont {C.}~\bibnamefont
  {De~Fazio}}, \bibinfo {author} {\bibfnamefont {B.}~\bibnamefont {Doyon}},\
  and\ \bibinfo {author} {\bibfnamefont {I.~M.}\ \bibnamefont
  {Sz{\'e}cs{\'e}nyi}},\ }\href
  {https://doi.org/10.1103/PhysRevLett.121.170602} {\bibfield  {journal}
  {\bibinfo  {journal} {Phys. Rev. Lett.}\ }\textbf {\bibinfo {volume} {121}},\
  \bibinfo {pages} {170602} (\bibinfo {year} {2018}{\natexlab{a}})},\ \Eprint
  {https://arxiv.org/abs/1805.04948} {arXiv:1805.04948 [cond-mat.stat-mech]}
  \BibitemShut {NoStop}%
\bibitem [{\citenamefont {Castro-Alvaredo}\ \emph
  {et~al.}(2018{\natexlab{b}})\citenamefont {Castro-Alvaredo}, \citenamefont
  {De~Fazio}, \citenamefont {Doyon},\ and\ \citenamefont
  {Sz{\'e}cs{\'e}nyi}}]{CastroAlvaredo2018JHEP}%
  \BibitemOpen
  \bibfield  {author} {\bibinfo {author} {\bibfnamefont {O.~A.}\ \bibnamefont
  {Castro-Alvaredo}}, \bibinfo {author} {\bibfnamefont {C.}~\bibnamefont
  {De~Fazio}}, \bibinfo {author} {\bibfnamefont {B.}~\bibnamefont {Doyon}},\
  and\ \bibinfo {author} {\bibfnamefont {I.~M.}\ \bibnamefont
  {Sz{\'e}cs{\'e}nyi}},\ }\href {https://doi.org/10.1007/JHEP10(2018)039}
  {\bibfield  {journal} {\bibinfo  {journal} {JHEP}\ }\textbf {\bibinfo
  {volume} {2018}}\bibfield  {number} {\bibinfo  {number} { (10)},\ \bibinfo
  {pages} {039}},\ }\Eprint {https://arxiv.org/abs/1806.03247}
  {arXiv:1806.03247 [hep-th]} \BibitemShut {NoStop}%
\bibitem [{\citenamefont {Castro-Alvaredo}\ \emph {et~al.}(2019)\citenamefont
  {Castro-Alvaredo}, \citenamefont {De~Fazio}, \citenamefont {Doyon},\ and\
  \citenamefont {Sz{\'e}cs{\'e}nyi}}]{CastroAlvaredo2019JMP}%
  \BibitemOpen
  \bibfield  {author} {\bibinfo {author} {\bibfnamefont {O.~A.}\ \bibnamefont
  {Castro-Alvaredo}}, \bibinfo {author} {\bibfnamefont {C.}~\bibnamefont
  {De~Fazio}}, \bibinfo {author} {\bibfnamefont {B.}~\bibnamefont {Doyon}},\
  and\ \bibinfo {author} {\bibfnamefont {I.~M.}\ \bibnamefont
  {Sz{\'e}cs{\'e}nyi}},\ }\href {https://doi.org/10.1063/1.5098892} {\bibfield
  {journal} {\bibinfo  {journal} {J. Math. Phys.}\ }\textbf {\bibinfo {volume}
  {60}},\ \bibinfo {pages} {082301} (\bibinfo {year} {2019})},\ \Eprint
  {https://arxiv.org/abs/1904.02615} {arXiv:1904.02615 [math-ph]} \BibitemShut
  {NoStop}%
\bibitem [{\citenamefont {Zhang}\ and\ \citenamefont
  {Rajabpour}(2021)}]{Zhang2021JSTAT}%
  \BibitemOpen
  \bibfield  {author} {\bibinfo {author} {\bibfnamefont {J.}~\bibnamefont
  {Zhang}}\ and\ \bibinfo {author} {\bibfnamefont {M.~A.}\ \bibnamefont
  {Rajabpour}},\ }\href {https://doi.org/10.1088/1742-5468/ac1f28} {\bibfield
  {journal} {\bibinfo  {journal} {J. Stat. Mech.}\ }\textbf {\bibinfo {volume}
  {2021}},\ \bibinfo {pages} {093101} (\bibinfo {year} {2021})},\ \Eprint
  {https://arxiv.org/abs/2010.16348} {arXiv:2010.16348 [cond-mat.stat-mech]}
  \BibitemShut {NoStop}%
\bibitem [{\citenamefont {Edwards}\ \emph {et~al.}(1998)\citenamefont
  {Edwards}, \citenamefont {Heller},\ and\ \citenamefont
  {Klassen}}]{EdwardsHellerKlassen1998}%
  \BibitemOpen
  \bibfield  {author} {\bibinfo {author} {\bibfnamefont {R.~G.}\ \bibnamefont
  {Edwards}}, \bibinfo {author} {\bibfnamefont {U.~M.}\ \bibnamefont
  {Heller}},\ and\ \bibinfo {author} {\bibfnamefont {T.~R.}\ \bibnamefont
  {Klassen}},\ }\href {https://doi.org/10.1016/S0550-3213(98)80003-5}
  {\bibfield  {journal} {\bibinfo  {journal} {Nucl. Phys. B}\ }\textbf
  {\bibinfo {volume} {517}},\ \bibinfo {pages} {377} (\bibinfo {year}
  {1998})},\ \Eprint {https://arxiv.org/abs/hep-lat/9711003}
  {arXiv:hep-lat/9711003 [hep-lat]} \BibitemShut {NoStop}%
\bibitem [{\citenamefont {Chen}\ \emph {et~al.}(2020)\citenamefont {Chen},
  \citenamefont {Dai},\ and\ \citenamefont {Pang}}]{ChenDaiPang:1503.01766}%
  \BibitemOpen
  \bibfield  {author} {\bibinfo {author} {\bibfnamefont {J.-W.}\ \bibnamefont
  {Chen}}, \bibinfo {author} {\bibfnamefont {S.-H.}\ \bibnamefont {Dai}},\ and\
  \bibinfo {author} {\bibfnamefont {J.-Y.}\ \bibnamefont {Pang}},\ }\href
  {https://doi.org/10.1016/j.nuclphysb.2019.114892} {\bibfield  {journal}
  {\bibinfo  {journal} {Nucl. Phys. B}\ }\textbf {\bibinfo {volume} {951}},\
  \bibinfo {pages} {114892} (\bibinfo {year} {2020})},\ \Eprint
  {https://arxiv.org/abs/1503.01766} {arXiv:1503.01766 [hep-th]} \BibitemShut
  {NoStop}%
\bibitem [{\citenamefont {Fujimura}\ \emph {et~al.}(2023)\citenamefont
  {Fujimura}, \citenamefont {Nishioka},\ and\ \citenamefont
  {Shimamori}}]{FujimuraNishiokaShimamori:2309.11889}%
  \BibitemOpen
  \bibfield  {author} {\bibinfo {author} {\bibfnamefont {H.}~\bibnamefont
  {Fujimura}}, \bibinfo {author} {\bibfnamefont {T.}~\bibnamefont {Nishioka}},\
  and\ \bibinfo {author} {\bibfnamefont {S.}~\bibnamefont {Shimamori}},\
  }\href@noop {} {\bibinfo {title} {Entanglement r\'enyi entropy and
  boson-fermion duality in massless thirring model}} (\bibinfo {year} {2023}),\
  \Eprint {https://arxiv.org/abs/2309.11889} {arXiv:2309.11889 [hep-th]}
  \BibitemShut {NoStop}%
\bibitem [{\citenamefont {Cardy}\ \emph {et~al.}(2008)\citenamefont {Cardy},
  \citenamefont {Castro-Alvaredo},\ and\ \citenamefont {Doyon}}]{hr:Cardy2008}%
  \BibitemOpen
  \bibfield  {author} {\bibinfo {author} {\bibfnamefont {J.~L.}\ \bibnamefont
  {Cardy}}, \bibinfo {author} {\bibfnamefont {O.~A.}\ \bibnamefont
  {Castro-Alvaredo}},\ and\ \bibinfo {author} {\bibfnamefont {B.}~\bibnamefont
  {Doyon}},\ }\href {https://doi.org/10.1007/s10955-007-9422-x} {\bibfield
  {journal} {\bibinfo  {journal} {J. Stat. Phys.}\ }\textbf {\bibinfo {volume}
  {130}},\ \bibinfo {pages} {129} (\bibinfo {year} {2008})},\ \Eprint
  {https://arxiv.org/abs/0706.3384} {arXiv:0706.3384 [hep-th]} \BibitemShut
  {NoStop}%
\bibitem [{\citenamefont {Hung}\ \emph {et~al.}(2014)\citenamefont {Hung},
  \citenamefont {Myers},\ and\ \citenamefont {Smolkin}}]{hr:Hung2014}%
  \BibitemOpen
  \bibfield  {author} {\bibinfo {author} {\bibfnamefont {L.-Y.}\ \bibnamefont
  {Hung}}, \bibinfo {author} {\bibfnamefont {R.~C.}\ \bibnamefont {Myers}},\
  and\ \bibinfo {author} {\bibfnamefont {M.}~\bibnamefont {Smolkin}},\ }\href
  {https://doi.org/10.1007/JHEP10(2014)178} {\bibfield  {journal} {\bibinfo
  {journal} {JHEP}\ }\textbf {\bibinfo {volume} {2014}}\bibfield  {number}
  {\bibinfo  {number} { (10)},\ \bibinfo {pages} {178}},\ }\Eprint
  {https://arxiv.org/abs/1407.6429} {arXiv:1407.6429 [hep-th]} \BibitemShut
  {NoStop}%
\bibitem [{\citenamefont {Bianchi}\ \emph {et~al.}(2016)\citenamefont
  {Bianchi}, \citenamefont {Meineri}, \citenamefont {Myers},\ and\
  \citenamefont {Smolkin}}]{hr:Bianchi2016}%
  \BibitemOpen
  \bibfield  {author} {\bibinfo {author} {\bibfnamefont {L.}~\bibnamefont
  {Bianchi}}, \bibinfo {author} {\bibfnamefont {M.}~\bibnamefont {Meineri}},
  \bibinfo {author} {\bibfnamefont {R.~C.}\ \bibnamefont {Myers}},\ and\
  \bibinfo {author} {\bibfnamefont {M.}~\bibnamefont {Smolkin}},\ }\href
  {https://doi.org/10.1007/JHEP07(2016)076} {\bibfield  {journal} {\bibinfo
  {journal} {JHEP}\ }\textbf {\bibinfo {volume} {2016}}\bibfield  {number}
  {\bibinfo  {number} { (07)},\ \bibinfo {pages} {076}},\ }\Eprint
  {https://arxiv.org/abs/1511.06713} {arXiv:1511.06713 [hep-th]} \BibitemShut
  {NoStop}%
\bibitem [{\citenamefont {Diehl}(2003)}]{Diehl:2003ny}%
  \BibitemOpen
  \bibfield  {author} {\bibinfo {author} {\bibfnamefont {M.}~\bibnamefont
  {Diehl}},\ }\href {https://doi.org/10.1016/j.physrep.2003.08.002} {\bibfield
  {journal} {\bibinfo  {journal} {Phys. Rept.}\ }\textbf {\bibinfo {volume}
  {388}},\ \bibinfo {pages} {41} (\bibinfo {year} {2003})},\ \Eprint
  {https://arxiv.org/abs/hep-ph/0307382} {arXiv:hep-ph/0307382} \BibitemShut
  {NoStop}%
\bibitem [{\citenamefont {Meissner}\ \emph {et~al.}(2008)\citenamefont
  {Meissner}, \citenamefont {Metz}, \citenamefont {Schlegel},\ and\
  \citenamefont {Goeke}}]{hr:Meissner2008}%
  \BibitemOpen
  \bibfield  {author} {\bibinfo {author} {\bibfnamefont {S.}~\bibnamefont
  {Meissner}}, \bibinfo {author} {\bibfnamefont {A.}~\bibnamefont {Metz}},
  \bibinfo {author} {\bibfnamefont {M.}~\bibnamefont {Schlegel}},\ and\
  \bibinfo {author} {\bibfnamefont {K.}~\bibnamefont {Goeke}},\ }\href
  {https://doi.org/10.1088/1126-6708/2008/08/038} {\bibfield  {journal}
  {\bibinfo  {journal} {JHEP}\ }\textbf {\bibinfo {volume} {2008}}\bibfield
  {number} {\bibinfo  {number} { (08)},\ \bibinfo {pages} {038}},\ }\Eprint
  {https://arxiv.org/abs/0805.3165} {arXiv:0805.3165 [hep-ph]} \BibitemShut
  {NoStop}%
\bibitem [{\citenamefont {Balakrishnan}\ \emph {et~al.}(2019)\citenamefont
  {Balakrishnan}, \citenamefont {Chandrasekaran}, \citenamefont {Faulkner},
  \citenamefont {Levine},\ and\ \citenamefont
  {Shahbazi-Moghaddam}}]{hr:Balakrishnan2019}%
  \BibitemOpen
  \bibfield  {author} {\bibinfo {author} {\bibfnamefont {S.}~\bibnamefont
  {Balakrishnan}}, \bibinfo {author} {\bibfnamefont {V.}~\bibnamefont
  {Chandrasekaran}}, \bibinfo {author} {\bibfnamefont {T.}~\bibnamefont
  {Faulkner}}, \bibinfo {author} {\bibfnamefont {A.}~\bibnamefont {Levine}},\
  and\ \bibinfo {author} {\bibfnamefont {A.}~\bibnamefont
  {Shahbazi-Moghaddam}},\ }\href {https://arxiv.org/abs/1906.08274} {\bibinfo
  {title} {Entropy variations and light ray operators from replica defects}}
  (\bibinfo {year} {2019}),\ \Eprint {https://arxiv.org/abs/1906.08274}
  {arXiv:1906.08274 [hep-th]} \BibitemShut {NoStop}%
\end{thebibliography}

\begin{thebibliography}{9}%
\makeatletter
\providecommand \@ifxundefined [1]{%
 \@ifx{#1\undefined}
}%
\providecommand \@ifnum [1]{%
 \ifnum #1\expandafter \@firstoftwo
 \else \expandafter \@secondoftwo
 \fi
}%
\providecommand \@ifx [1]{%
 \ifx #1\expandafter \@firstoftwo
 \else \expandafter \@secondoftwo
 \fi
}%
\providecommand \natexlab [1]{#1}%
\providecommand \enquote  [1]{``#1''}%
\providecommand \bibnamefont  [1]{#1}%
\providecommand \bibfnamefont [1]{#1}%
\providecommand \citenamefont [1]{#1}%
\providecommand \href@noop [0]{\@secondoftwo}%
\providecommand \href [0]{\begingroup \@sanitize@url \@href}%
\providecommand \@href[1]{\@@startlink{#1}\@@href}%
\providecommand \@@href[1]{\endgroup#1\@@endlink}%
\providecommand \@sanitize@url [0]{\catcode `\\12\catcode `\$12\catcode
  `\&12\catcode `\#12\catcode `\^12\catcode `\_12\catcode `\%12\relax}%
\providecommand \@@startlink[1]{}%
\providecommand \@@endlink[0]{}%
\providecommand \url  [0]{\begingroup\@sanitize@url \@url }%
\providecommand \@url [1]{\endgroup\@href {#1}{\urlprefix }}%
\providecommand \urlprefix  [0]{URL }%
\providecommand \Eprint [0]{\href }%
\providecommand \doibase [0]{https://doi.org/}%
\providecommand \selectlanguage [0]{\@gobble}%
\providecommand \bibinfo  [0]{\@secondoftwo}%
\providecommand \bibfield  [0]{\@secondoftwo}%
\providecommand \translation [1]{[#1]}%
\providecommand \BibitemOpen [0]{}%
\providecommand \bibitemStop [0]{}%
\providecommand \bibitemNoStop [0]{.\EOS\space}%
\providecommand \EOS [0]{\spacefactor3000\relax}%
\providecommand \BibitemShut  [1]{\csname bibitem#1\endcsname}%
\let\auto@bib@innerbib\@empty
\bibitem [{\citenamefont {Castro-Alvaredo}\ \emph
  {et~al.}(2018{\natexlab{a}})\citenamefont {Castro-Alvaredo}, \citenamefont
  {De~Fazio}, \citenamefont {Doyon},\ and\ \citenamefont
  {Sz{\'e}cs{\'e}nyi}}]{smref:CastroAlvaredo2018PRL}%
  \BibitemOpen
  \bibfield  {author} {\bibinfo {author} {\bibfnamefont {O.~A.}\ \bibnamefont
  {Castro-Alvaredo}}, \bibinfo {author} {\bibfnamefont {C.}~\bibnamefont
  {De~Fazio}}, \bibinfo {author} {\bibfnamefont {B.}~\bibnamefont {Doyon}},\
  and\ \bibinfo {author} {\bibfnamefont {I.~M.}\ \bibnamefont
  {Sz{\'e}cs{\'e}nyi}},\ }\href
  {https://doi.org/10.1103/PhysRevLett.121.170602} {\bibfield  {journal}
  {\bibinfo  {journal} {Phys. Rev. Lett.}\ }\textbf {\bibinfo {volume} {121}},\
  \bibinfo {pages} {170602} (\bibinfo {year} {2018}{\natexlab{a}})},\ \Eprint
  {https://arxiv.org/abs/1805.04948} {arXiv:1805.04948 [cond-mat.stat-mech]}
  \BibitemShut {NoStop}%
\bibitem [{\citenamefont {Castro-Alvaredo}\ \emph
  {et~al.}(2018{\natexlab{b}})\citenamefont {Castro-Alvaredo}, \citenamefont
  {De~Fazio}, \citenamefont {Doyon},\ and\ \citenamefont
  {Sz{\'e}cs{\'e}nyi}}]{smref:CastroAlvaredo2018JHEP}%
  \BibitemOpen
  \bibfield  {author} {\bibinfo {author} {\bibfnamefont {O.~A.}\ \bibnamefont
  {Castro-Alvaredo}}, \bibinfo {author} {\bibfnamefont {C.}~\bibnamefont
  {De~Fazio}}, \bibinfo {author} {\bibfnamefont {B.}~\bibnamefont {Doyon}},\
  and\ \bibinfo {author} {\bibfnamefont {I.~M.}\ \bibnamefont
  {Sz{\'e}cs{\'e}nyi}},\ }\href {https://doi.org/10.1007/JHEP10(2018)039}
  {\bibfield  {journal} {\bibinfo  {journal} {JHEP}\ }\textbf {\bibinfo
  {volume} {2018}}\bibfield  {number} {\bibinfo  {number} { (10)},\ \bibinfo
  {pages} {039}},\ }\Eprint {https://arxiv.org/abs/1806.03247}
  {arXiv:1806.03247 [hep-th]} \BibitemShut {NoStop}%
\bibitem [{\citenamefont {Castro-Alvaredo}\ \emph {et~al.}(2019)\citenamefont
  {Castro-Alvaredo}, \citenamefont {De~Fazio}, \citenamefont {Doyon},\ and\
  \citenamefont {Sz{\'e}cs{\'e}nyi}}]{smref:CastroAlvaredo2019JMP}%
  \BibitemOpen
  \bibfield  {author} {\bibinfo {author} {\bibfnamefont {O.~A.}\ \bibnamefont
  {Castro-Alvaredo}}, \bibinfo {author} {\bibfnamefont {C.}~\bibnamefont
  {De~Fazio}}, \bibinfo {author} {\bibfnamefont {B.}~\bibnamefont {Doyon}},\
  and\ \bibinfo {author} {\bibfnamefont {I.~M.}\ \bibnamefont
  {Sz{\'e}cs{\'e}nyi}},\ }\href {https://doi.org/10.1063/1.5098892} {\bibfield
  {journal} {\bibinfo  {journal} {J. Math. Phys.}\ }\textbf {\bibinfo {volume}
  {60}},\ \bibinfo {pages} {082301} (\bibinfo {year} {2019})},\ \Eprint
  {https://arxiv.org/abs/1904.02615} {arXiv:1904.02615 [math-ph]} \BibitemShut
  {NoStop}%
\bibitem [{\citenamefont {Zhang}\ and\ \citenamefont
  {Rajabpour}(2021)}]{smref:Zhang2021JSTAT}%
  \BibitemOpen
  \bibfield  {author} {\bibinfo {author} {\bibfnamefont {J.}~\bibnamefont
  {Zhang}}\ and\ \bibinfo {author} {\bibfnamefont {M.~A.}\ \bibnamefont
  {Rajabpour}},\ }\href {https://doi.org/10.1088/1742-5468/ac1f28} {\bibfield
  {journal} {\bibinfo  {journal} {J. Stat. Mech.}\ }\textbf {\bibinfo {volume}
  {2021}},\ \bibinfo {pages} {093101} (\bibinfo {year} {2021})},\ \Eprint
  {https://arxiv.org/abs/2010.16348} {arXiv:2010.16348 [cond-mat.stat-mech]}
  \BibitemShut {NoStop}%
\bibitem [{\citenamefont {Chen}\ \emph {et~al.}(2020)\citenamefont {Chen},
  \citenamefont {Dai},\ and\ \citenamefont {Pang}}]{smref:ChenDaiPang:1503.01766}%
  \BibitemOpen
  \bibfield  {author} {\bibinfo {author} {\bibfnamefont {J.-W.}\ \bibnamefont
  {Chen}}, \bibinfo {author} {\bibfnamefont {S.-H.}\ \bibnamefont {Dai}},\ and\
  \bibinfo {author} {\bibfnamefont {J.-Y.}\ \bibnamefont {Pang}},\ }\href
  {https://doi.org/10.1016/j.nuclphysb.2019.114892} {\bibfield  {journal}
  {\bibinfo  {journal} {Nucl. Phys. B}\ }\textbf {\bibinfo {volume} {951}},\
  \bibinfo {pages} {114892} (\bibinfo {year} {2020})},\ \Eprint
  {https://arxiv.org/abs/1503.01766} {arXiv:1503.01766 [hep-th]} \BibitemShut
  {NoStop}%
\bibitem [{\citenamefont {Fujimura}\ \emph {et~al.}(2023)\citenamefont
  {Fujimura}, \citenamefont {Nishioka},\ and\ \citenamefont
  {Shimamori}}]{smref:FujimuraNishiokaShimamori:2309.11889}%
  \BibitemOpen
  \bibfield  {author} {\bibinfo {author} {\bibfnamefont {H.}~\bibnamefont
  {Fujimura}}, \bibinfo {author} {\bibfnamefont {T.}~\bibnamefont {Nishioka}},\
  and\ \bibinfo {author} {\bibfnamefont {S.}~\bibnamefont {Shimamori}},\
  }\href@noop {} {\bibinfo {title} {Entanglement r\'enyi entropy and
  boson-fermion duality in massless thirring model}} (\bibinfo {year} {2023}),\
  \Eprint {https://arxiv.org/abs/2309.11889} {arXiv:2309.11889 [hep-th]}
  \BibitemShut {NoStop}%
\bibitem [{\citenamefont {Wolff}(2004)}]{smref:hr:Wolff2004}%
  \BibitemOpen
  \bibfield  {author} {\bibinfo {author} {\bibfnamefont {U.}~\bibnamefont
  {Wolff}},\ }\href {https://doi.org/10.1016/S0010-4655(03)00467-3} {\bibfield
  {journal} {\bibinfo  {journal} {Comput. Phys. Commun.}\ }\textbf {\bibinfo
  {volume} {156}},\ \bibinfo {pages} {143} (\bibinfo {year} {2004})},\ \bibinfo
  {note} {erratum: Comput. Phys. Commun. 176, 383 (2007)},\ \Eprint
  {https://arxiv.org/abs/hep-lat/0306017} {arXiv:hep-lat/0306017} \BibitemShut
  {NoStop}%
\bibitem [{\citenamefont {Christ}\ \emph {et~al.}(2024)\citenamefont {Christ},
  \citenamefont {Eranki},\ and\ \citenamefont
  {Kelly}}]{smref:ChristErankiKelly2024Pvalues}%
  \BibitemOpen
  \bibfield  {author} {\bibinfo {author} {\bibfnamefont {N.~H.}\ \bibnamefont
  {Christ}}, \bibinfo {author} {\bibfnamefont {R.}~\bibnamefont {Eranki}},\
  and\ \bibinfo {author} {\bibfnamefont {C.}~\bibnamefont {Kelly}},\
  }\href@noop {} {\bibinfo {title} {Bootstrap-determined p-values in lattice
  qcd}} (\bibinfo {year} {2024}),\ \Eprint {https://arxiv.org/abs/2409.11379}
  {arXiv:2409.11379 [hep-lat]} \BibitemShut {NoStop}%
\bibitem [{\citenamefont {Edwards}\ \emph {et~al.}(1998)\citenamefont
  {Edwards}, \citenamefont {Heller},\ and\ \citenamefont
  {Klassen}}]{smref:EdwardsHellerKlassen1998}%
  \BibitemOpen
  \bibfield  {author} {\bibinfo {author} {\bibfnamefont {R.~G.}\ \bibnamefont
  {Edwards}}, \bibinfo {author} {\bibfnamefont {U.~M.}\ \bibnamefont
  {Heller}},\ and\ \bibinfo {author} {\bibfnamefont {T.~R.}\ \bibnamefont
  {Klassen}},\ }\href {https://doi.org/10.1016/S0550-3213(98)80003-5}
  {\bibfield  {journal} {\bibinfo  {journal} {Nucl. Phys. B}\ }\textbf
  {\bibinfo {volume} {517}},\ \bibinfo {pages} {377} (\bibinfo {year}
  {1998})},\ \Eprint {https://arxiv.org/abs/hep-lat/9711003}
  {arXiv:hep-lat/9711003 [hep-lat]} \BibitemShut {NoStop}%
\end{thebibliography}
\end{document}